\documentclass[preprint]{aastex}

\usepackage{multirow}
\usepackage{color}
\usepackage[normalem]{ulem}

\def\deg{\ifmmode {^\circ}\else {$^\circ$}\fi}

\def\linebreak{\hfil\break}

\def\nl{\hfil\linebreak}
\def\
{\hfil\linebreak}

\newcommand{\dex}[1]{\times 10^{#1}}

\def\degree{\ifmmode {^\circ}\else {$^\circ$}\fi}
\def\mum{\ifmmode {\rm \mu {\rm m}}\else $\rm \mu {\rm m}$\fi}
\def\arcsec{\ifmmode ^{\prime \prime}\else $^{\prime \prime}$\fi}

\def\inch{\ifmmode ^{\prime \prime}\else $^{\prime \prime}$\fi}
\def\arcmin{\ifmmode ^{\prime}\else $^{\prime}$\fi}

\def\msun{\ifmmode {\rm M_{\odot}}\else $\rm M_{\odot}$\fi}
\def\mearth{\ifmmode {\rm M_{+\mskip-14.6muO\,}}\else $\rm M_{+\mskip-14.6muO\,}$\fi}
\def\mearth{\ifmmode {\rm M_{\earth}}\else $\rm M_{\earth}$\fi}
\def\mmdot{\ifmmode {\rm M}_{*}\dot{\rm M}_{disk}\else {M$_{*}\dot{\rm M}_{disk}$}\fi}
\def\msunsqyr{\ifmmode {\rm M}^{2}_{\odot}{\rm /yr}\else {M$^{2}_{\odot}$/yr}\fi}
\newbox\grsign \setbox\grsign=\hbox{$>$} \newdimen\grdimen \grdimen=\ht\grsign
\newbox\simlessbox \newbox\simgreatbox
\setbox\simgreatbox=\hbox{\raise.5ex\hbox{$>$}\llap
     {\lower.5ex\hbox{$\sim$}}}\ht1=\grdimen\dp1=0pt
\setbox\simlessbox=\hbox{\raise.5ex\hbox{$<$}\llap
     {\lower.5ex\hbox{$\sim$}}}\ht2=\grdimen\dp2=0pt

\begin{document}


\title{SED analysis of \\
    class~I and class~II FU Orionis stars}


\author{Luciana V. Gramajo}
\affil{Observatorio Astron\'omico, Universidad Nacional de C\'ordoba, Argentina,\\ Laprida 854, 5000 C\'ordoba, Argentina}
\email{e-mail: luciana@oac.uncor.edu}

\author{Javier A. Rod\'on}
\affil{European Southern Observatory, Alonso de C\'ordova 3107, \\ Vitacura, Casilla 19001, Santiago 19, Chile}
\email{e-mail: jrodon@eso.org}

\and

\author{Mercedes G\'omez}
\affil{Observatorio Astron\'omico, Universidad Nacional de C\'ordoba, Argentina,\\ Laprida 854, 5000 C\'ordoba, Argentina}
\email{e-mail: mercedes@oac.uncor.edu}

\begin{abstract}
FU Orionis stars (FUORS) are eruptive pre-main sequence objects thought to represent quasi-periodic or recurring
stages of enhanced accretion during the low-mass star-forming process. We characterize the sample of known and
candidate FUORS in an homogeneous and consistent way, deriving stellar and circumstellar parameters for
each object. We emphasize the analysis in those parameters that are supposed to vary during the FUORS stage. We
modeled the SEDs of 24 of the 26 currently known FUORS, using the radiative transfer code of \citet{whitney2003b}.
We compare our models with those obtained by \citet{robitaille2007} for Taurus class~II and I sources in
quiescence periods, by calculating the cumulative distribution of the different parameters.
 FUORS have more massive disks: we find that $\sim80\%$ of the disks in FUORS are more massive than any 
Taurus class~II and I sources in the sample.  Median values for the disk mass accretion rates
are  $\sim$ 10$^{-7}$ $\msun$/yr vs $\sim$ 10$^{-5}$ $\msun$/yr for 
standard YSOs (young stellar objects) and FUORS, respectively.  
While the distributions of envelope mass accretion rates for class I FUORS and for standard class I objects are similar, FUORS, on average, 
have higher envelope mass accretion rates than standard class~II and class~I sources.
Most FUORS ($\sim$ 70\%) have envelope mass accretion rates above $10^{-7}\,\msun$/yr.
In contrast, 60\% of the classical YSO sample have accretion rates below this value.
Our results support the current scenario in which changes experimented by the circumstellar disk
explain the observed properties of these stars. However, the increase in the disk mass accretion rate
is smaller than theoretically predicted \citep{frank1992,hartmann1996a}, though in good agreement with
previous determinations.
\end{abstract}


\section{Introduction}
FU Orionis stars (FUORS) are a class of variable young stellar objects that show brightness variations of the eruptive type \citep{herbig1977}.
The main feature observed in these variables is a sudden increase in brightness  ($ 3-6\,$mag in the optical),
in an elapse of time of a few months. This episode is known as the ``outburst'', after which the object remains
bright for years or decades, and then fades in a few centuries back to the pre-outburst stage. The outburst,
however, occurs in a different way for each FUORS (see, e.g., \citealt{hartmann1996a,clarke2005a}).

These stars exhibit several indicators of youth, such as the presence of the lithium $ 6707\,$\AA~ line in optical spectra, and the association with reflection nebulae and infrared excesses originating from dust grains in circumstellar disks. Moreover, they are spatially and kinematically related to known star-forming regions and in some cases, FUORS have high extinction values in the optical suggesting that they are still embedded in the parent cloud material (see, e.g., \citealt{hartmann1996a}). 

FUORS show several properties that strongly suggest the presence of a circumstellar disk,
such as broad spectral energy distributions (SEDs, \citealt{kenyon1988b}),
stellar spectral types that become progressively colder with increasing wavelength \citep{hartmann1985,kenyon1988b},
spectral linewidths that increase with decreasing wavelength \citep{hartmann1987a,hartmann1987b},
double-peaked line profiles in high-resolution optical and near-infrared (NIR) spectra \citep{hartmann1985,kenyon1988b}
as well as P-Cygni profiles with no evidence for redshifted emission or absorption \citep{kenyon1988b,hartmann1995},
and finally, deep, broadened infrared CO bands in absorption \citep{kenyon1988a,reipurth1997b}. 

Another class of eruptive variables are the so-called EXOR stars, named after EX Lup, the progenitor
of the class \citep{herbig1989,herbig2008}. Their optical brightness increases by $ 1-4 $\,mag on time scales
of weeks or months, then fading back during a few months to its original state after. During its low activity
stage they exhibit T~Tauri-like characteristics, while during the outburst stage they usually display emission
in the optical spectrum as well as in the infrared CO bandheads (e.g., \citealt{aspin2010}).

To reproduce the SEDs of FUORS, modelers use dusty disks and infalling envelopes (e.g.,
\citealt{hartmann1985,kenyon1988b,calvet1991,hartmann1995,calvet1998,whitney2003a,whitney2003b}).
Indeed the presence of a circumstellar disk is essential to explain the FU Orionis phenomenon.
The disk is where material coming from the surrounding infalling envelope accumulates, heats-up,
and finally destabilizes the structure of the disk itself, causing a thermal
\citep{frank1992, bell1994}  and/or a gravitational 
\citep{zhu2009,zhu2010,vorobyov2005,vorobyov2006,vorobyov2010}
 instability that eventually leads to the characteristic outburst.

During this episode an increase of the brightness takes place, affecting mainly the optical wavelengths
since the excess emission comes from the inner regions of the disk, which are heated by the viscous dissipation
released after the instability has triggered an increase in the disk accretion rate.
\citet{frank1992} suggests that the central objects of the FUORS systems alternate between low ($10^{-7}\,$\msun/yr) and high ($10^{-4}\,$\msun/yr) mass accretion rates. The former corresponds to a low activity, quiescent state, while the latter corresponds to periods of high activity. Furthermore, the transformations undergone by the disk are what cause  the observed phenomenon.
Similarly, EXOR events are also attributed to thermal instabilities in the inner disks \citep{aspin2011a}.

Currently, 26 FUORS have been identified and classified as class~I or class~II
objects according to the shape of their SEDs \citep{lada1987}.
This includes the ``confirmed'' FUORS, for which the sudden increase in brightness has been recorded, and the ``candidate'' FUORS, which share many, if not all, of the properties of bona-fide FUORS but for which an outburst has not been observed. 
In this paper we model and analyze the SEDs of 24 confirmed and candidate FUORS to determine
the physical and geometrical parameters of the star and the disk.

In Section 2 we present the sample, describe the adopted model and outline the procedure used in the SED modeling.
In Section 3 we analyze the individual sources. Our results are presented in Section 4. Finally in Section 5
we summarize the results and conclusions.

\section{Radiative transfer SED modeling}

\subsection{The sample}

Our sample includes 24 of the 26 known and candidate class~I and II FU Orionis stars.
The FUORS CaI~136, in NGC 3372 in the Carina nebula \citep{tapia2006} and V733~Cep in
the dark cloud L1216 \citep{reipurth2007,peneva2010}, both
identified as FUORS candidates, were not included as not enough
fluxes to construct the SED were available in the literature. In particular, for these objects
only fluxes in the near-infrared wavelengths have been published.

For some of the 24 remaining objects, only fluxes in the optical, NIR and mid-infrared (MIR)
wavelengths are available. The lack of measurements in the far-infrared (FIR) and sub-mm spectral
regions makes it only possible to derive reliable parameters for the protostar and the inner disk, since
the outer regions of the disk and the envelope emit mostly at FIR and sub-mm wavelengths.

Table \ref{t:sample} presents our sample and summarizes the main properties of each source, such as luminosity, optical extinction (A$_{\rm V}$), variation in the K band ($\Delta$K), year of the outburst (if registered), association with molecular outflows/jets, SED class, spectral type, distance, and whether the central star is a binary.

We classified the sources according to their observational properties. In particular,
we used the CO band at $2.3\,\mu$m to classify the objects as confirmed FUORS if it appears in absorption
(e.g., \citealt{reipurth1997b,hartmann2004}), or as a FUORS candidate otherwise
(i.e., if the band is in emission or absent, e.g., \citealt{reipurth2004b,aspin2011a}).
 The FUORS in our sample were subsequently classified as class~II visible pre-main sequence stars
or as embedded class~I objects.  Sources with $\dot{M}< 10^{-7}\,\msun$/yr 
are class~II stars surrounded by disks, whereas objects with $\dot{M}> 10^{-7}\,\msun$/yr are class~I 
objects embedded in infalling envelopes. This criterion is based on the best-fit values for the
envelope mass accretion rates listed in Table 1 from \cite{robitaille2007}. 

To construct the SED of each object we compiled all their fluxes available in the literature between $\sim 0.3\,\mu$m and $\sim 3\,$mm.  Table \ref{t:fluxes} lists the fluxes compiled for each object. The uncertainties for the fluxes are not always
provided in the literature. In those cases we adopted ``typical'' errors. 
In addition we include the MIR spectra published by \citet{green2006} and \citet{quanz2007b}, when available.

The spectra published by \citet{green2006} for the sources FU~Ori, V1515~Cyg, V1057~Cyg, BBW~76, and V346~Nor were obtained with the Spitzer infrared spectrograph (IRS) in the $5-36\,\mu$m range in December 2003 and May 2004.  \citet{quanz2007b} published data obtained with both the Spitzer and the Infrared Space Observatory (ISO) telescopes. In this work we include the Spitzer-IRS spectra for RNO~1B, RNO~1C, Par~21, and L1551~IRS5, observed in the $5-14\,\mu$m range between December 2003 and March 2004.
Only in the case of the source Re~50~N~IRS1 we use the spectrum obtained with ISO short wavelength spectrometer (SWS) in the $5-15\,\mu$m range, which was obtained in October 2007.  We have not included all of the ISO spectra because their signal to noise ratio is low for most of the
targets. 

\subsection{Procedure}

To analyze the SEDs we used the code developed by \citet{whitney2003a,whitney2003b} and
the grid of models computed by \citet{robitaille2006}. Briefly, the dust radiative transfer model of \citet{whitney2003b} reproduces a complete protostellar system comprised by a central source emitting photons and a circumstellar disk and envelope. The code uses a Monte Carlo radiative transfer scheme that follows photon packets emitted by the central star as they are scattered, absorbed, and re-emitted throughout the disk and envelope.

The geometry of the protostellar system (star$+$disk$+$infalling envelope) is highly parametrized.
The model has 15 modifiable parameters that can be divided into three groups. The central source parameters,
i.e., stellar mass (M$_{*}$), stellar radius (R$_{*}$), and stellar temperature (T$_{*}$); the infalling
envelope parameters comprising the envelope mass accretion rate ($\dot{M}$), envelope outer radius (R$ _{\rm max}$),
cavity density ($\rho_{\rm cav}$), cavity opening angle ($\rm\theta$)\footnote{$\rm\theta$ is measured from the axis
of rotation at the outer radius of the envelope.}; and the disk parameters disk mass (M$_{\rm disk}$),
disk outer -or centrifugal- radius ($R_{\rm c}$)\footnote{The centrifugal radius and the outer disk radius are usually related. The envelope material falls to $R_{\rm c}$ at the disk equatorial plane, providing an indication of the extension of the disk (see \citealt{whitney2003a,robitaille2006}).}, disk inner radius (R$_{\rm min}$), disk mass accretion rate ($\dot{M}_{\rm disk}$), disk radial density exponent (A), disk scale height  exponent (B),
and disk fiducial scale height ($z_{01}$)\footnote{The disk fiducial scale height, $z_{01}$, is the scale height of the inner disk at R$_{*}$ in units of R$_{*}$.}.
The ambient density surrounding the young stars ($\rho_{\rm amb}$) is included as the 15th parameter.
In addition to these parameters, the inclination angle to the line of sight ($i$) is one of the most critical parameters when modeling the SEDs. In our analysis we adopted a value of  $z_{01}=0.03\,$R$_{*}$, as it remains practically unchanged for this type of objects.

We used the \citet{robitaille2006} grid (also called ``the SED fitting tool'') of young stellar object (YSO) models\footnote{The grid is available at \url{http://caravan.astro.wisc.edu/protostars}.} to obtain an initial model for each source. \citet{robitaille2006} used the \citet{whitney2003a} code to compute a grid of 20,000 axisymmetric radiation transfer models of YSOs at 10 viewing angles, resulting in a total of 200,000 SEDs in the wavelength range $0.36-1000\,\mu$m that cover the ``typical'' range of values of physical and geometrical parameters that characterize YSOs.
The SED fitting tool offers the advantage that data in a wide range of wavelengths can be used simultaneously, without losing information. In addition, this tool allows the evaluation of the uniqueness or the goodness of the fit, calculating the $\chi^2$-per-data point value of each model in the grid, following Eq.~6 in \cite{robitaille2007}.

For each object analyzed we selected the best model from the grid of \citet{robitaille2006} corresponding to the minimum value of $\chi^2$ that reproduces the SED, and at the same time, gives reasonable values for the known parameters according to previous determinations from the literature. In other words for each source we select the model with the smallest $\chi^2$ that provides parameter values in agreement with those already published. Spectra were not used in this procedure. However in all cases the selected models reproduce the spectra well enough.
These initial models were used as starting points for a detailed analysis with the \citet{whitney2003b} code. At this step we included in the modeling the available MIR spectra. The direct application of the \citet{whitney2003b} code allows, among other things, to introduce discrete step variations in the values of the parameters.
In this way a refined model, i.e., a fit with a better $\chi^2$ can be obtained, since in the \citet{robitaille2006} grid step variations are fixed.

As mentioned before, the sample we analyze consists of class~I and class~II objects. The parameters for which a more accurate value can be obtained from the SED modeling are related to the envelope in the first case, and to the disk in the second. This is particularly relevant to characterize the FU Orionis phenomenon.

Tables \ref{t:resclass2} and \ref{t:resclass1} list the parameters of the models that best fit the
SEDs of the 24 analyzed objects. Table \ref{t:resclass2} corresponds to class~II FUORS, and
Table \ref{t:resclass1} to class~I FUORS. In both cases we provide as an additional parameter
the envelope mass (M$_{\rm {env}}$). Although not independent, this parameter has been determined
by other authors and thus provide a direct way to compare our results with others.  In the last
column we give sample values for class~I and II parameters \citep{robitaille2007}.
Figures \ref{f:v1515cyg} to \ref{f:v2775} show the best fit obtained in each case. Uncertainties
in fluxes available in the literature are indicated with error bars, except when  smaller than the size of the symbols. 

Since FUORS are variable stars, when constructing their SEDs we attempted to use contemporary data whenever possible. This was particularly the case in the optical and NIR wavelengths, where the variations are larger.
Five of the analyzed objects (V1515~Cyg, V1057~Cyg, L1551~IRS5, RNO~1B, and V1331~Cyg) show a significant dispersion in the observed fluxes, due to the variability of the source during the period of time covered by the data.
For this reason, fluxes at different times were selected to construct individual SEDs. 
In the case of RNO~1B and V1331~Cyg we defined two SEDs that were modeled independently 
(Figures \ref{f:rno1b} and \ref{f:v1331cyg}). On the other hand for V1515~Cyg, V1057~Cyg, and L1551~IRS5 the identification of fluxes corresponding to different observing periods was not useful to reduce their dispersion.  
Therefore, we chose to model the fluxes contemporary to the observed MIR spectra
(Figures \ref{f:v1515cyg}, \ref{f:v1057cyg}, and \ref{f:l1551}).
For five of the analyzed sources (V1647~Ori, OO~Ser, V2492~Cyg, HBC~722 and V2775~Ori) we can
clearly distinguish two epochs, before and after the outburst. Consequently, two SEDs were modeled
(Figures \ref{f:v1647oripre} and \ref{f:v1647oripost}, \ref{f:ooser}, \ref{f:v2492cyg}, \ref{f:hbc722}, and \ref{f:v2775}, respectively). 
In general, fluxes from the literature were obtained with different aperture sizes. For this reason, for sources V1515~Cyg, FU~Ori, V1057~Cyg, AR~6B, V346~Nor, and Re~50~N~IRS we show models corresponding to different apertures (Figures \ref{f:v1515cyg}, \ref{f:fuori}, \ref{f:v1057cyg}, \ref{f:ar6b}, \ref{f:v364nor}, and \ref{f:re50nirs1}, respectively).

\section{Analysis of individual sources}

\subsection{Class~II FU Orionis stars }

\subsubsection{V1515 Cygni} 

The outburst of V1515 Cygni, one of the three prototypes of the FUOR class, was detected in the optical in 1950 and it has remained in an outburst stage since then \citep{herbig1977}. 
\citet{goodrich1987} suggested the presence of a molecular outflow associated with this object (see also \citealt{evans1994}).  Furthermore, they argued that the inclination angle of this source to the line of sight should be close to zero, according to the shape of the large-scale nebula associated.
\citet{weintraub1991} detected V1515 Cygni at 450, 800 and 850 $\mu$m, and the central source has an estimated luminosity between $77$\,L$_{\odot}$ and $200\,$L$_{\odot}$ \citep{sandell2001,green2006}.
\citet{kospal2011a} obtained $^{13}$CO maps of V1515~Cyg that shows an arc-shaped emission structure.

The SED of V1515~Cyg (Figure \ref{f:v1515cyg}) has two peaks, at $\sim 1.5\,\mu$m and $\sim 60\,\mu$m.
The Spitzer-IRS spectrum in the $5-36\,\mu$m range shows the presence of silicate in emission at $\sim 9.7 \,\mu$m \citep{green2006}.
The observed fluxes from $4\,\mu$m to $200\,\mu$m show a relatively large dispersion. This is mainly caused by the difference in the time when observations were obtained, which can be seen in Figure~\ref{f:v1515cyg}, where the crosses correspond to the fluxes observed between 2003 and 2004 and diamonds to the observations obtained between 1983 and 1996.

The best fit for V1515~Cyg (Figure~\ref{f:v1515cyg}, Table~\ref{t:resclass2}) corresponds to the fluxes obtained in 2003-2004 (crosses in Figure \ref{f:v1515cyg}), which are contemporaneous to the Spitzer spectrum. The 1983-1996 data (diamonds in Figure \ref{f:v1515cyg}) have a large dispersion, which makes the modeling more difficult. We adopted the published spectral type for this source (G2--G5, see Table~\ref{t:sample} in \citealt{kolotilov1983}), which corresponds to a temperature in the $5860-5770\,$K range \citep{kenyon1995}.

Figure~\ref{f:v1515cyg} shows our best model for two different apertures, $60\arcsec$ (solid line) and $11\arcsec$ (dashed line). The latter aperture value is similar to that used in the extraction of the spectrum \citep{green2006}.

\citet{sandell2001} determined an upper limit for the disk mass of V1515~Cyg of $\sim 0.13\,\msun$, from observations at 1.3 mm.
\citet{lodato2001} estimated an opening angle for the cavity $\theta\sim 20\degr-28\degr$, a disk mass accretion rate $\dot{\rm M}_{disk}= 1.0\times 10^{-5}\,\msun$/yr, and a disk mass M$_{disk}=0.9-1.5\,\msun$.
\citet{green2006} modeled the NIR and MIR SED as well as the Spitzer spectrum and derived a maximum temperature for the central source of $7710\,$K, and $ \mmdot=3.5\times 10^{-5}\,\msunsqyr $.
\citet{zhu2008} estimated an inclination angle $i=0\degr$ and a central star mass M$_{*}=0.3\,\msun$,
a stellar radius R$_{*}=2.8\,$R$_{\odot}$, a disk inner radius R$_{min}=0.25\,$AU and a value of $\mmdot=1.3\times 10^{-5}\,\msunsqyr$.

Table \ref{t:resclass2} lists our best fit parameters for V1515~Cyg. The stellar temperature is lower than the maximum determined by \citet{green2006}, but consistent with the spectral type. 
The disk mass ($0.13\,\msun$) is also lower than that obtained by \citeauthor{lodato2001} (\citeyear{lodato2001}, M$_{disk}=0.91-1.52\,\msun$).
Furthermore, from our disk mass accretion rate $\dot{\rm M}_{disk}=1.0\times 10^{-5}\,$\msun/yr and stellar mass, M$_{*}= 0.3\,\msun$, we derived $\mmdot=1.1\dex{-5}\,\msunsqyr$.
This value is lower than previous estimations by others authors, nevertheless the stellar mass agrees with the determination of \citet{zhu2008} and the disk mass-accretion rate matches the estimation of \citet{lodato2001}.
The values of $\theta=25\degr$ and R$_*=2.0\,$R$_{\odot}$ listed in Table \ref{t:resclass2} are in good agreement with those determined by \citet{lodato2001} and \citet{zhu2008}, respectively.

\subsubsection{BBW 76}

BBW 76, also known as BRAN 76 and IRAS~07486$-$3258 was identified by \citet{reipurth1985a} as a 
FU Orionis star. Later, \citet{eisloeffel1990} confirmed this identification based on high resolution optical spectra, in particular by the P-Cygni Balmer line profiles and absorption line widths, similar to FUORS prototypes. In addition, \cite{reipurth2002a} identified several other observational properties such as the change of the spectral type toward later types with increasing wavelength, common to well known FU Ori stars.
BBW~76 is not associated with any known molecular outflow \citep{sandell2001}, and \citet{green2006} suggested that BBW~76 might be a class~I object.

Figure \ref{f:bbw76} shows the SED of BBW~76, which presents a maximum at around $1.5\,\mu$m.
The $5-36\,\mu$m Spitzer-IRS spectrum shows a strong silicate absorption at $\sim 9.7\,\mu$m \citep{green2006}.
The best model obtained for BBW 76 is shown in Figure \ref{f:bbw76} (solid line), and its parameters in Table~\ref{t:resclass2}. This model successfully reproduces the observed fluxes, however, at $\sim 10\,\mu$m the observed fluxes do not match the Spitzer spectrum and also show a moderate dispersion. In the SED modeling more weight was given to spectrum than to the individual flux values.

\citet{sandell2001} determined a disk mass of $0.15\,\msun$ from sub-mm observations.
\citet{green2006} estimated a maximum temperature of $7710\,$K, a disk inner radius R$_{min}=3.9\,$R$_{\odot}$, a luminosity L$_{*}=1.8$\,L$_{\odot}$, and $\mmdot=7.2\dex{-5}\,\msunsqyr$.
\citet{zhu2008} obtained R$_{*}=4.6\,$R$_{\odot}$, R$_{min}=0.64\,$AU, and $\mmdot=8.1\dex{-5}\,\msunsqyr$, with an inclination angle $i=50\degr$.

The stellar radius ($3.0\,$R$_{\odot}$) and the disk inner radius (R$_{min}=0.42\,$AU) we derive agree with the values obtained by \citet{zhu2008}, however our modeled disk mass ($0.08\,\msun$) is less than that determined by \citeauthor{sandell2001} (\citeyear{sandell2001}, M$_{disk}\sim 0.15\,\msun$).
The disk mass accretion rate is $\dot{\rm M}_{disk}=1\dex{-5}\,\msun$/yr, then $\mmdot=8\dex{-7}\,\msunsqyr$, is two orders of magnitude lower than that obtained by other authors.

\subsubsection{V1735~Cygni}

V1735~Cygni is located in the IC~5146 stellar cluster in L1031, at a distance of $900\,$pc \citep{hilton1995}, and it is also known as Elias 1-12.
\citet{elias1978} identified V1735~Cygni as a FU Orionis type variable. Its outburst took place between $\sim 1957$ and 1965 \citep{hartmann1996a}.
V1735 Cyg is associated with a high-mass molecular outflow \citep{levreault1983}, and has a luminosity of $25\,$L$_{\odot}$.

The observed SED is shown in Figure \ref{f:v1735cyg}, and it reveals two peaks, one around $1\,\mu$m and the other in the $60-100\,\mu$m range. 
Although the observed fluxes cover the spectral range from the infrared to close to $100\,\mu$m, data at longer wavelengths are scarce, resulting in an uncertain behavior of the SED in that spectral range. This difficult the determination of reliable parameters for the envelope.
Nevertheless, the best model obtained (solid line, Figure \ref{f:v1735cyg}) reproduces satisfactorily well the observed SED.

Table \ref{t:resclass2} shows the parameters corresponding to the model presented in Figure \ref{f:v1735cyg}.  From observations in the mm, \citet{sandell2001} derived a mass of $0.42\,\msun$, that they associate with the disk. However, our model disk mass M$_{disk}=0.20\,\msun$ is roughly half of that value.
The envelope mass we obtain is M$_{env}=0.9\,\msun$, which suggests that the envelope might contribute to the mass value determined by \citet{sandell2001}.
The disk mass accretion rate $\dot{M}_{disk}=1.4\dex{-5}\,\msun$/yr agrees with previous estimates for other class~II FU Orionis.

\subsubsection{V883~Orionis}

\citet{strom1993} reported V883~Orionis as a FU Orionis object in the IC~430 nebula of the Orionis region. It has a luminosity of $400\,$L$_{\odot}$, and is not associated with any known molecular outflow \citep{sandell2001}.  
Figure \ref{f:v883ori} shows its observed SED. The dispersion of the fluxes is small and it is relatively well covered in the region beyond $100\,\mu$m. However, few fluxes are available in the NIR region.
This SED is rather flat and without any distinguishable feature. 

Table \ref{t:resclass2} lists the parameters corresponding to the model in Figure \ref{f:v883ori}. The disk mass M$_{disk}=0.3\,\msun$ is consistent with the $ 0.39\,\msun $ estimated by \citet{sandell2001} from mm observations.
The disk mass accretion rate $\dot{M}_{disk}=1\dex{-5}\,\msun$/yr matches the values derived for the other class~II FU Orionis objects.

\subsubsection{RNO~1B}

This object, also as known as V710~Cas, was identified as a FUORS by \citet{staude1991}. It is located in the L1287 dark cloud at a distance of $850\,$pc \citep{yang1991}, and constitutes a binary system with RNO~1C (also a FUORS, see following section), for which \citet{quanz2006} estimated a separation of $\sim 5000\,$AU.
According to \cite{snell1990} and \cite{yang1991}, RNO~1B is associated with a high-mass molecular outflow. However, \citet{mcmuldroch1995} identified RNO~1C as the driving source of the outflow. 

Figure \ref{f:rno1b} shows the observed SED of RNO 1B. The fluxes have a relatively large dispersion, which can be attributed to the different epochs of observation. For this reason, the observations are divided into two periods, until 1995 (crosses) and after 1996 (asterisks).
The SED of this object includes the Spitzer-IRS $5-14\,\mu$m spectrum \citep{quanz2007b}. The solid line in Figure~\ref{f:rno1b} indicates the model for pre-1995 fluxes, and with a dotted-dashed line the post-1996 model.

In Table~\ref{t:resclass2} we list the parameters corresponding to the data obtained before 1995, and in brackets we indicate the values obtained from the fluxes observed after 1996 whenever they differ.
Comparing the results from the two epochs, we see that the disk mass, the temperature and the disk mass accretion rate have all decreased. Particularly the disk mass has decreased by a factor 20.
The decrease in the disk mass accretion rate might be an indicator that the central star would be in its way to enter the T Tauri or class~II evolutionary stage. However, this should be confirmed by more detailed determinations of $ \dot{\rm M}_{disk} $.
It is worthwhile to mention that $ \dot{\rm M}_{disk}\sim 1.0\dex{-5}\,\msun$/yr at the time of highest brightness, which is the same order of magnitude as for other class~II FUOR sources.

\subsubsection{RNO~1C}
 
RNO~1C was identified as a FU Orionis type star by \citet{kenyon1993b}. As mentioned before, RNO~1C and RNO~1B form a binary system in which both stars are FU Orionis variables.
Figure~\ref{f:rno1c} show the observed SED for RNO~1C and the $5-14\,\mu$m Spitzer-IRS spectrum obtained by \citet{quanz2007b}.
There are no observed fluxes around $100\,\mu$m available in the literature, thus the behavior of the SED in that spectral region is very uncertain.
Furthermore, the model reproduces well the shape of the Spitzer spectrum, but fails to reproduce individual fluxes around these wavelengths.

The parameters derived from this model are shown in Table \ref{t:resclass2}. The outer radius value R$_{max}=6000$\,AU is in agreement with the $\sim 5000\,$AU determined by \citet{mcmuldroch1995} for the size of the envelope using CS molecular line observations. 
The disk mass accretion rate $\dot{\rm M}_{disk}=8.0\dex{-6}\,\msun$/yr agrees with the expected value for these type of objects.

\subsubsection{PP~13S}

PP~13S is a protostar \citep{tapia1997, sandell2001, tsukagoshi2005} embedded in the small dark cloud L1473, at a distance of $ 350\,$pc \citep{cohen1983a}. This source is associated with a bipolar molecular outflow traced by the CO$(2-1)$ and CO$(1-0)$ transitions \citep{sandell1998,tsukagoshi2005}. 
\citet{sandell1998} identified PP~13S as a FU Orionis object from the broad and deep shape of the CO absorption band at $2.3\,\mu$m.

The SED (see Figure \ref{f:pp13s}) has a maximum around $8\,\mu$m. However, the lack of data between $10\,\mu$m and $200\,\mu$m makes its shape uncertain.
Nevertheless, our best SED model (Figure \ref{f:pp13s}, solid line) reproduces satisfactorily the observed fluxes.

From NIR images, sub-mm continuum, and CO line observations \citet{sandell1998} suggested the existence of a disk associated with PP~13S with an inclination of $40\degr$ with respect to the line of sight.
\citet{tsukagoshi2005} estimated for PP~13S an envelope mass M$_{env}\sim 0.27\,\msun$, a mass accretion rate $\dot{\rm M}\sim 5\dex{-6}\,\msun$/yr, and an inclination $i\sim 59\degr$, from mm continuum data and C$^{18}$O$(1-0)$ observations.

Table \ref{t:resclass2} presents the parameters derived from the model of the SED shown in Figure \ref{f:pp13s}. 
The inclination angle $i = 50\degr$ is in agreement with previous determinations. The disk mass accretion rate value of $8\dex{-6}\,\msun$/yr, as well as the envelope mass M$_{env}=0.12\,\msun$, agree with the values determined by \citet{tsukagoshi2005}.

\subsubsection{V1647~Orionis}

This source is located in the Lynds~1630 dark cloud in M78, at a distance of $400\,$pc, and illuminates the McNeil's reflection nebula \citep{lis1999}. Two outbursts have been registered for this source. The first one occurred between October 2003 and February 2004, for which pre- and post- outburst observations are available \citep{briceno2004,abraham2004b,reipurth2004a,mcgehee2004,andrews2004,walter2004}. 
The second outburst took place in 2008-2009 \citep{itagaki2008,kun2008}.

\citet{aspin2008} obtained optical, NIR and MIR observations for V1647 Ori after the first outburst. Based on the relative long outburst and the detection of the CO overtone in absorption, these authors suggested its classification as a FUOR. 
Later, \citet{aspin2009} observed a very weak CO overtone bandhead absorption when the star was experimenting a second brightness increase in August 2008.

\citet{aspin2011b} observed this source in the NIR and noticed the star remained in an outburst state during the 2008-2011 period, supporting the hypothesis of relatively long outbursts and thus the FUORS classification.
However, \citet{aspin2006} suggested that V1647~Ori may be an EXOR variable. In particular a NIR spectrum, taken after the second outburst, shows the CO overtone bandheads in emission in addition to
other emission lines in the optical and NIR \citep{aspin2010,aspin2011b}, which are features commonly found in EXOR variables.
In summary, V1647~Ori shows photometrical properties similar to FUORS stars, and spectroscopic characteristics common to EXOR variables \citep{aspin2011b,semkov2012a}.

In Figures \ref{f:v1647oripre} and \ref{f:v1647oripost} we show the SED before the 2003-2004 outburst and the SED for the post-outburst period between 2004 and 2008. For the period after the second outburst in 2008 there are currently not enough data to construct an SED.
Table \ref{t:resclass2} lists the parameter values for both modeled SEDs, pre- and post- first outburst.

\citet{reipurth2004b} obtained an inclination angle $i=30\degr$ and an opening angle $\theta=60$\degr, from the analysis of Gemini images. 
\citet{muzerolle2005} modeled the SEDs before and after the 2003-2004 outburst, adopting a flat accretion disk (i.e., without flaring), a stellar mass M$_{*}=0.5\,\msun$, and a stellar radius R$_{*}= 2.0\,$R$_{\odot}$. 
They obtained a mass accretion rate $\dot{\rm M}\sim 10^{-6}\,\msun$/yr and an envelope total mass of M$_{env}=3\dex{-3}\,\msun$ for the SED before the outburst.
From the post-outburst SED, on the other hand, they derived a disk mass accretion rate $\dot{\rm M}_{disk}\sim 10^{-5}\,\msun$/yr, assuming the bolometric luminosity is dominated by the accretion luminosity.
Pre- and post-first-outburst mass accretion rates derived by \citet{acosta-pulido2007} are 
$\dot{\rm M}=5\dex{-7}\,\msun$/yr and $\dot{\rm M}=1-7\dex{-6}\,\msun$/yr, respectively, and an inclination angle $i\sim 61\degr$.

\citet{aspin2008} used optical, NIR, and MIR observations after the outburst and estimated a T$_{eff} \sim 3800\,$K and a R$_* \sim 5$\,R$_{\odot}$, together with a stellar mass value M$_* \sim 0.8\,\msun$, this time from the position of V1647~Ori in the HR diagram.
They also estimated a value $\dot{\rm M}_{disk}=1.0\pm 0.5\dex{-6}\,\msun$/yr for the disk mass accretion rate, and $\theta=65\degr$, $i=30\degr$ for the cavity opening angle and the inclination angle to the line of sight, respectively.

For the second outburst period (i.e., 2008-2009), \citet{aspin2011b} derived a disk mass accretion rate $\dot{\rm M}_{disk}=4\pm 2\dex{-6}\,\msun$/yr, similar to that obtained by \citet{aspin2008} for the first outburst. 

When comparing our results for the pre- and after-outburst SEDs, we see that the stellar temperature and the mass accretion rates of the disk and the envelope all increased during the outburst (see Table~\ref{t:resclass2}). In particular, $ \dot{\rm M}_{disk} $ increased by an order of magnitude from $0.1\dex{-6}\,\msun$/yr to $5\dex{-6}\,\msun$/yr.
On the other hand, the stellar mass and the envelope mass, as well as the geometrical parameters
$i=60\deg$ and $\theta=7\deg$, remain unchanged.

Our results derived from the SEDs analysis before and after the 2003-2004 outburst are, in general, in agreement with those obtained by \citet{muzerolle2005} and \citet[][see Figures \ref{f:v1647oripre} and \ref{f:v1647oripost}, and Table \ref{t:resclass2}]{acosta-pulido2007}.
In particular, our values for the inclination angle and the disk mass accretion range for both pre- and post-outburst models agree very well with those derived by \citet{acosta-pulido2007}.
Furthermore, our determination for the disk mass accretion rate is also comparable to the value obtained by \citet{muzerolle2005} for this source after of the first outburst.

The stellar mass derived from our analysis is similar to the value obtained by \citet{aspin2008}, and our value of $5.0\dex{-6}\,\msun$/yr for the disk mass accretion rate after the outburst is on the same order as the $1.0\pm0.5\dex{-6}\,\msun$/yr they derived.
However, the inclination angle ($i=60\degr$ vs $30\degr$) as well as the opening angle ($\theta=7\degr$ vs $65\degr$) differ.
Moreover, \citet{aspin2011a} derived a disk mass accretion for the second outburst that is similar with our determination. 

Our model parameters derived from the SED after the outburst and the results obtained by other authors agree with those expected for a class~II FUORS, with exception of the disk mass accretion rate that turned out smaller than expected ($\dot{\rm M}_{disk}=2\dex{-5}\,\msun$/yr vs $10^{-4}\,\msun$/yr; \citealt{hartmann1996a}).
However, as pointed out by \citet{aspin2011b} this star shows several observational properties common to EXOR variables.

\subsection{Class~I FU Orionis stars}

\subsubsection{FU Orionis}

This source is one of the three prototypes of the class. The outburst was observed in 1937 with a luminosity of \mbox{$340\,$L$_\odot$} \citep{sandell2001}. FU Ori is a binary system with a separation of $217\,$AU \citep{malbet2005,quanz2006}, and has no associated optical jet or molecular outflow \citep{evans1994}.

Its SED, shown in Figure~\ref{f:fuori}, present two peaks. One at $\sim 1.5\,\mu$m and the other at $\sim 100\,\mu$m. The Spitzer-IRS ($5-36 \,\mu$m) spectrum displays a silicate emission between $10\,\mu$m and $18\,\mu$m \citep{green2006}.
While the Spitzer spectrum gives a good constraint for the MIR region of the SED, fluxes at sub-mm and mm wavelengths are scarce and show a high dispersion.

Figure~\ref{f:fuori} shows the best model obtained for FU~Ori. For this source we adopted $T_{*} \sim 6030\,$K, in agreement with its G0 spectral type (see Table~\ref{t:sample} in \citealt{kenyon2000}), and 
the calibration of \citep{kenyon1995}. We plot the models corresponding to two apertures, $60\arcsec$ (continuous line), and $20\arcsec$ (dashed line).  This last aperture is similar to that used by \citet{green2006} to extract the spectrum.

Several authors have analyzed this object. \citet{kenyon1988b} adopted a stationary accretion disk model and reproduced both the SED and the observed line profiles. They derived a stellar mass of $0.37\,\msun$ and a temperature of $7200\,$K. In addition they estimated M$_{*}\dot{\rm M}_{disk}=0.5-4.0\times10^{-4}\,$M$^2_{\odot}$/yr for $\cos i = 0.5$.
\citet{popham1996} fixed the value of the stellar mass at $0.7 \,\msun$ and used an accretion disk with a boundary layer to model both optical spectra and line profiles. For $\cos i = 0.5$ they derived M$_{*}\dot{\rm M}_{disk}=1.4\times10^{-4}\,$M$^2_{\odot}$/yr.

More recently, \citet{sandell2001} used observations in the sub-mm to estimate an upper limit for the disk mass of $0.02 \,\msun$.
\citet{lodato2001}, in turn, modeled the SED of FU Orionis using a self-gravitant accretion disk and obtained M$_{*}\dot{\rm M}_{disk}=5.2\times10^{-5}\,$M$^2_{\odot}$/yr for M$_{*}=1\,\msun$, $\cos i = 0.65$, and R$_{min}= 8\,$R$_{\odot}$. Subsequently, \citet{lodato2003} modeled the lines profiles in addition to the SED, and derived M$_{*}\dot{\rm M}_{disk}=10^{-4}\,$M$^2_{\odot}$/yr, for M$_{*}=0.7\,\msun$ and $\cos i=0.5$.

\citet{malbet2005} used interferometric data in the NIR and determined $\dot{\rm M}_{disk}=6.5\times 10^{-5}\,\msun$/yr. \citet{green2006} modeled the SED and the Spitzer infrared spectrum with an accretion disk. They obtained M$_{*}\dot{\rm M}_{disk}=1.0\times 10^{-4}\,$M$^2_{\odot}$/yr, for R$_i=0.58\,$AU and R$_c=70\,$AU, adopting M$_*=0.3\,\msun$ and a maximum stellar temperature of $7710\,$K. 
\citet{zhu2008} modeled the Spitzer-IRS spectrum of FU~Ori and derived $i= 55^{\circ}$, R$_*=5$R$_{\odot}$, and M$_{*}\dot{\rm M}_{disk} = 7.4\times 10^{-5}\,$M$^2_{\odot}$/yr.

Our model of the SED of FU~Orionis provides the parameters listed in Table \ref{t:resclass1}.  The central stellar mass M$_*= 0.7\,\msun$ agrees with previous determinations, specially with the value determined by \citet{lodato2003}.
The disk mass accretion rate we obtain is $\dot{\rm M}_{disk}=10^{-5}\,\msun$/yr for a stellar mass M$_{*}= 0.7 \,\msun$, therefore, M$_{*}\dot{\rm M}_{disk}=0.7\times 10^{-5}\,$M$^2_{\odot}$/yr.
This value is similar to those determined by \citet{kenyon1988b}, \citet{lodato2003} and \citet{zhu2008}. However, it is lower than \citet{lodato2001}. The disk mass is $0.01\,\msun$, in agreement with the estimation of \citet{sandell2001}. Other parameters,
such as R$_*$ ($5.0$\,R$_{\odot}$), R$_{min}$ ($0.47\,$AU) and R$_c$ ($70\,$AU), are consistent with previous determinations
by \citet{green2006} and \citet{zhu2008}.

\subsubsection{V1057 Cygni}

\citet{welin1971} noticed that V1057~Cygni increased $\sim6\,$mag in brightness in less than a year (1969-1971), which was reflected in a change of the spectral type of this object from M to early-A \citep{herbig1977}. Then, it gradually declined, fading by about 6 magnitudes in the following six years from the outburst.
V1057~Cyg has an estimated luminosity between $170\,$L$_{\odot}$ and $370\,$L$_{\odot}$ \citep{sandell2001,green2006}, and it is associated with a molecular outflow \citep{evans1994} and surrounded by an envelope \citep{kospal2011a}.

Figure \ref{f:v1057cyg} shows our best model for V1057 Cygni. Since the data display a large scatter, in a first approximation we divided the observations in four epochs. We chose to model the fluxes contemporaneous to the Spitzer spectrum (diamonds in Figure \ref{f:v1057cyg}) because older flux values show significant differences with the spectrum and do not provide a complete coverage in wavelength to well constrain our modeling attempt.
The model for V1057 Cyg in Figure \ref{f:v1057cyg} is plotted for two apertures, $60\arcsec$ (solid line) and $11\arcsec$ (dotted-dashed line). The last aperture is similar to that of the \citep{green2006} spectrum. 
In the region around $10\,\mu$m the fluxes show dispersion independently of the observing epoch and the aperture used. For this reason, in our modeling we give more value to the spectrum than to the individual flux values. In addition, we adopted a spectral type corresponding to the time when the observations were obtained (F7/G3 I/II, \citealt{herbig2003}), and derived a temperature ${\rm T}\sim 5900-6500\,$K, in concordance with \citet{kenyon1995}.

\citet{kenyon1988b} derived a maximum value for the inclination angle of $i=30\degr$, a lower limit for the stellar mass of M$_{*}>0.1\,\msun$, a radius R$_{*}\sim 4\,$R$_{\odot}$ and $\mmdot\sim0.5-3\dex{-4}\,\msunsqyr$.
\citet{popham1996} modeled both the SED and the line profiles and estimated  $\dot{\rm M}_{disk}=1.0\dex{-4}$\,\msun/yr for M$_{*}=0.5\,\msun$ and R$_{*}=5.03\,$R$_{\odot}$, with an inclination angle of $30\degr$.
\citet{lachaume2004} modeled the SED of V1057~Cyg and obtained $\mmdot=2\dex{-5}\,\msunsqyr$, and R$_{min}=2\,$R$_{\odot}$. Finally, \cite{green2006} adopted an inclination of $i=0\degr$, and determined a maximum temperature T$_{*}<6590\,$K, a R$_{min}=3.7\,$R$_{\odot}$, and $\mmdot=4.5\dex{-5}\,\msunsqyr$.

Table \ref{t:resclass1} lists the best SED model parameters for V1057~Cygni. The derived stellar temperature is less than the maximum estimated by \citet{green2006}, while the stellar mass ($0.5\,\msun$) is in agreement with \citet{kenyon1988b} and \citet{popham1996}. Our disk mass accretion rate is $\dot{\rm M}_{disk}=1.4\dex{-4}$\,\msun/yr, thus we derive $\mmdot=7.0\dex{-5}\,\msunsqyr$, in agreement with previous determinations.

\subsubsection{Z CMa}

This object is a close binary with a separation of $0.1\arcsec$ \citep{koresko1991,thiebaut1995,leinert1997}, consisting of the two young stars Z~CMa~NW and Z~CMa~SE.

Z~CMa~NW is a Herbig Be star surrounded by a dusty cocoon with a hole \citep{szeifert2010,canovas2012}. This component has a mass of $12\,\msun$ and a B8 spectral type \citep{vandenacker2004,alonso-albi2009}.
Z~CMa~SE has been classified as a FU Orionis object by \citet{hartmann1989a}, based on the detection of a blueshifted $2\,\mu$m CO first-overtone $\nu' -\nu'' = 2-0$ and double-peaked optical absorption lines, with a velocity difference of about $100\,$kms$^{-1}$.
This source has a luminosity of $420\,$L$_{\odot}$ \citep{sandell2001}, a stellar mass of $1.1\,\msun$ \citep{pfalzner2008}, and a F5 spectral type \citep{kenyon1989}.
\citet{canovas2012}, using optical polarimetric images, found that the Z~CMa system is surrounded by a common circumbinary envelope.

Z~CMa was associated with a CO bipolar molecular outflow by \citet{evans1994}. More recently, \citet{whelan2010} obtained adaptive-optics-assisted [Fe II] spectro-images that show the presence of two jets.  In addition, observations carried out with OSIRIS at Keck revealed a parsec-scale wiggling outflow emanating from the Herbig Be star (Z~CMa~NW), suggesting that the central source may be double.
Z~CMa~SE is, on the other hand, associated with a micro or small-scale jet (see also \citealt{canovas2012}). 

Z~CMa has shown outburst events of less than one visual magnitude in a 5--10 years time scale in 1987, 2000, and 2004 \citep{vandenacker2004,grankin2009}, typical of EXORS variables.
In January 2008, the brightness of Z~CMa increased by about two visual magnitudes \citep{grankin2009}.  Based on spectropolarimetric observations, \citet{szeifert2010} concluded that the outburst is associated with the Herbig Be component (Z~CMa~NW), which is embedded in a dusty cocoon. Moreover, the dynamical time scale of the wiggling outflow emanating from Z~CMa~NW (4--8 years) agrees with the timescale between the outburst \citep{whelan2010}.

Figure \ref{f:zcma} shows the SED of Z~CMa. The observed fluxes cover reasonably well all the spectral range, and the scatter in the observed fluxes is low. The best fit obtained (solid line) reproduces satisfactorily well the shape of the observed SED, with the exception of the NIR region.
This is likely due to the binarity, since being a multiple system, the NIR portion of the SED has contributions from more than one of the stellar photospheres.
For the modeling of this source we used a temperature value appropriated to the
associated spectral type (${\rm T}\sim 6440\,$K, \citealt{kenyon1995}).

Table \ref{t:resclass1} shows the best model parameters obtained for Z~CMa.
The stellar mass has a value M$_{*}=0.8 \,\msun$, similar to that determined by \citet{pfalzner2008}. The disk mass accretion rate $\dot{\rm M}_{disk} = 2\dex{-5}\,\msun$/yr is on the same order as what was derived by these authors ($\dot{\rm M}_{disk}=7.9\dex{-5}\,\msun$/yr).

\subsubsection{AR 6A/6B}

These stars are FU Orionis variables that lie in the NGC~2264 star-forming region, at a distance of $800\,$pc. They form a binary system with a separation of $ \sim2200\,$AU ($2.8\arcsec$, \citealt{aspin2003}).
The source AR~6A has, in addition, a third companion AR~6C discovered by \citet{aspin2003}, with a separation of $ \sim 700\, $AU ($0.85\arcsec$).
\citet{moriarty-schieven2008} detected a molecular flow associated with AR 6A/6B.

Figures \ref{f:ar6a} and \ref{f:ar6b} show the SEDs of AR~6A and AR~6B, respectively. Fluxes compiled from the literature cover just the $1\,\mu$m to $20\,\mu$m range.
The fluxes associated with AR~6B exhibit a large dispersion (Figure \ref{f:ar6b}), since they were obtained with two different apertures. The best models for each of these sources reproduce well
the observed SEDs. For AR~6B, the model is plotted for two apertures, $60\arcsec$ (solid line), and  $30\arcsec$ (dotted-dashed line).

Table \ref{t:resclass1} lists the parameter values derived for the SEDs of AR~6A and AR~6B.
For the masses and temperatures of the central sources, we derived M$_{*}=0.80\,\msun$ and M$_{*}=0.87\,\msun$
for AR~6A and AR~6B, respectively, while for both we obtain T$_{*}\sim 4100\,$K.
Parameters associated with the inner disk should be reasonably
well constraint in the $1-20\,\mu$m wavelength region. On the contrary, the external
disk and the envelope parameters are poorly constraint in the models in Table \ref{t:resclass1} and
Figures \ref{f:ar6a} and \ref{f:ar6b}, since as mentioned before no fluxes for wavelength $> 20\,\mu$m
are available for these sources.

\subsubsection{L1551~IRS5}

This object, also known as IRAS 04287$+$1801, is a young protostellar binary system with a separation of 45 AU \citep{rodriguez1998}, associated with a bipolar outflow seen in the optical and NIR \citep{snell1980,mundt1983,moriarty-schieven1988,stocke1988,davis1995b}. L1551~IRS5 shows an  optical spectrum characteristic of FU Orionis objects \citep{looney1997}, for which it has been suggested that L1551~IRS5 belongs to this class.
\citet{sandell2001} estimated a mass of $0.23\,\msun$ for the disk from observations in the mm.

Figure \ref{f:l1551} displays the observed SED of L1551~IRS5. Fluxes obtained from the literature cover well the spectral range between $1\,\mu$m and $1200\,\mu$m, but have a modest dispersion around $100\,\mu$m.  In the observed SED we include the Spitzer spectrum published by \citet{quanz2007b}, which shows a silicate absorption at $9.7\,\mu$m typical of class~I objects. It also shows CO$_{2}$ in absorption at $6.85\,\mu$m. The best fit we obtained reproduces satisfactorily well the shape of the SED in the infrared region, as well as the spectrum around $10\,\mu$m.
However, for wavelengths beyond $100\,\mu$m the fit is relatively poorer.

From the modeling of low resolution NIR images, \citet{whitney1997} obtained a mass accretion rate $\dot{\rm M}=5\dex{-6}\,\msun$/yr, $R_c=30\,$AU, $\theta=20^\circ$, and $i \sim 70\deg-90\deg$. 
On the other hand, \citet{osorio2003} determined $\dot{\rm M}=7\dex{-5}\,\msun$/yr, and $R_c=150\,$AU from the SED model of this source, while \citet{robitaille2007} estimated an envelope mass accretion rate in the range between $5.5\dex{-6}\,\msun$/yr and $3.0\dex{-4}\,\msun$/yr.
At the same time, \citet{gramajo2007} analyzed images in the K and L bands, and obtained an inclination angle of $i = 72\deg-77\deg$, an envelope mass accretion rate $\dot{\rm M}\sim 5\dex{-6}\,\msun$/yr, a centrifugal radius $R_c = 40-100\,$AU, and an opening angle $\theta =20\deg$.

Table \ref{t:resclass1} lists the parameters for the best fit for L1551~IRS5 (see Figure \ref{f:l1551}).
In general, these parameters are consistent with those determined by other authors. In particular, the inclination angle of $70\degr$ agrees with that derived by \citet{gramajo2007} and \citet{whitney1997}.
However, the centrifugal radius is somewhat larger ($R_c=200\,$AU vs $40-100\,$AU), while the envelope mass accretion rate is an order of magnitude higher than that estimated by \citet[][ $\dot{\rm M}=10^{-5}\,\msun$/yr vs $5\dex{-6}\msun$/yr]{gramajo2007}.
Furthermore, the envelope mass accretion rate is less than that derived by \citet{whitney1997}, but it is consistent with the range of values estimated by \citet{robitaille2007}. 
On the other hand, the disk mass obtained from our SED modeling (M$_{disk}= 0.2\msun$), is in good agreement with that obtained by \citet{sandell2001}.

\subsubsection{V900~Mon}
\label{s:v900mon}

V900~Mon, also known as 2MASS~06572222$-$0823176, was initially recognized as an eruptive variable \citep{thommes2011,reipurth2012} in the L1656 small cloud located in a filamentary bridge between the Mon R2 complex and the CMa OB1 clouds, at a distance of $\sim1100\,$pc \citep{gregorio-hetem2008,lombardi2011}. This object is deeply embedded in a large cool envelope, with an estimated extinction of $ A_{V}~13\,$mag \citep{reipurth2012}.

The same authors note that the spectra of V900~Mon has a striking resemblance to those of the prototype of the class, FU~Ori, and by extension to the whole FUORs class. The NIR spectra of V900~Mon shows prominent CO bandhead absorption as well as large H$ _{2} $O broadband absorptions, suggesting a very late spectral type. In the optical, on the other hand, it shows characteristics suggesting an earlier than mid- to late-K spectral type. The appearance of the classic P~Cigny profile in lines such as H$ \alpha $ and the $ \lambda8662 $ Ca\,\textsc{ii}, as well as the $\lambda6497$ Ba\,\textsc{ii} feature further support the inclusion of V900~Mon into the FUORs class \citep{reipurth2012}.

The photometric history of this object suggests that V900~Mon started its brightening sometime before the 1970's and is still ongoing. \citet{reipurth2012} note that the brightness increase of this source is more consistent with that of the class-prototype V1515~Cyg.
From \emph{Spitzer} photometry, \citet{reipurth2012} suggest that V900~Mon is a Class~I source bordering the Class~II sources, and the outburst appears to have occurred at an earlier evolutionary stage when the star was still partly embedded.

In Figure \ref{f:v900} we show the SED of V900~Mon covering the $1-200\,\mu$m spectral range. There are no sub-mm flux measurements available.
The best models obtained reproduce satisfactorily the observed SEDs. The parameters for the model in Figure \ref{f:v900} are listed in Table \ref{t:resclass1}. We note that the lack of sub-mm fluxes means that the envelope parameters of the model are not well constrained.

From our modeling, V900~Mon appears as a Class~I source with an envelope mass accretion rate $ \dot{M} \sim 4.0\dex{-6}\,\msun$/yr. The values we obtain for the disk mass $ (0.1\,\msun) $ and disk mass accretion rate $ (2.0\dex{-6}\,\msun{\rm /yr}) $ are comparable with those derived for other Class~I sources. Furthermore, the results we obtain are similar to those of V1647~Ori after its outburst (V1647~Ori~(post) in Table~\ref{t:resclass2}), in agreement with the predictions of \citet{reipurth2012}.

\subsubsection{ISO-ChaI~192}

This class~I protostar in the Chamaeleon~I dark cloud is also known as GM~Cha,
[CCE98]2-41\footnote{[CCE98] from Cambresy, Copet, Epchtein et al. (1998).},
DENIS-P~J1109.5$-$7633, [PMK99]~ISOCAM~ChaI-Na2\footnote{[PMK99]
from Persi, Marenzi, Kaas et al. (1999).}, and [PMK99]~IR~ChaI-Na1
\citep{cambresy1998,persi1999,gomez2003b}, and it is associated with a
CO molecular outflow \citep{mattila1989,persi2007}. 

The SED of ISO-ChaI~192 is shown in Figure \ref{f:isochai192}. For wavelengths greater than $30\,\mu$m only a single flux value is available, at $70\,\mu$m.
This affects the reliability of the parameters associated with the disk and the envelope.

\citet{persi2007} modeled the SED of ISO-ChaI 192 using the code of \citet{whitney2003b}, and obtained an envelope mass accretion rate of $\dot{\rm M}=1-3\dex{-6}\,\msun$/yr, a disk
mass accretion rate of $\dot{\rm M}_{disk}= 1-7 \dex{-7}\,\msun$/yr, a centrifugal radius $R_c =5-20\,$AU, an opening angle $\theta=5\degr-30\degr$, and an inclination angle $i=35\deg-45\degr$.
These authors adopted fixed values for the stellar parameters (M$_*=0.55\,\msun$, R$_* =2.5\,$R$_{\odot}$, and T$_* = 3600\,$K), for the disk mass (M$_{disk} = 0.15 \,\msun$), and the disk inner radius (R$_{min} = 5.5\,$R$_{*}$).  

The parameters for the SED model in Figure \ref{f:isochai192} are listed in Table \ref{t:resclass1}. The values derived for the inclination angle ($i= 50\degr$) and the opening angle ($\theta= 20\degr$) are consistent with those obtained by \citet{persi2007}. 
In addition, other parameters such as the disk mass accretion rate $\dot{\rm M}_{disk}= 1\dex{-7}\,\msun$/yr and the envelope mass accretion rate $\dot{\rm M}= 5\dex{-6}\,\msun$/yr agree with those derived by these authors. Nevertheless, the stellar parameters we obtain for the central source correspond to a more massive star (M$_{*}=1.2\,\msun$ vs $0.55\,\msun$, R$_*=6.1\,$R$_{\odot}$ vs $2.5\,$R$_{\odot}$, and T$_* = 5000\,$K vs $3600\,$K) than that adopted by \citet{persi2007}.

\subsubsection{V346 Norma}

This object has a luminosity of $\sim 135\,$L$_{\odot}$ \citep{sandell2001}, and was discovered in 1983 by \citet{graham1983} in the dark cloud Sa~187 in Norma. V346~Nor shows FUOR-like characteristics \citep{reipurth1985b,graham1985,frogel1983}, and is associated with a bipolar molecular outflow \citep{reipurth1997a, sandell2001}.
This protostar is located near the YSO Reipurth~13, consequently the outflow of V346~Nor may be affected by the presence of this other young star \citep{prusti1993}.

In Figure~\ref{f:v364nor} we present the observed SED of V346~Norma, which includes the spectra obtained with Spitzer-IRS in the $5-35\,\mu$m range \citep{green2006}. The absorption around $10\,\mu$m clearly seen in this spectrum is probably due to silicates.
The observed fluxes have a relatively large scatter in the $10-100\,\mu$m region, likely due to the different apertures used. 

Table \ref{t:resclass1} lists the parameters corresponding to the SED model shown in Figure \ref{f:v364nor}. This figure displays the SED model for three values of apertures, $60\arcsec$ (solid line), $30\arcsec$ (dotted-dashed line), and $11\arcsec$ (dashed line).
The value obtained for the mass of the envelope, M$_{env}= 0.3 \,\msun$, agrees with the M$_{env}\sim 0.5\,\msun$ derived by \citet{sandell2001} and is greater than the value estimated for the disk mass (M$_{disk}=0.05\,\msun$), however it has to be noted that the fit at mm wavelengths is somewhat poor.

\subsubsection{OO Ser}

OO~Ser, previously known as DEOS (Serpens Deeply Embedded Outburst Star), is an embedded class~I source \citep{enoch2009}, located in the Serpens star-forming region at a distance of $311\,$pc \citep{delara1991}. The outburst probably occurred in 1995 \citep{hodapp1996}, however its nature or membership to the class~I FUORS is somewhat uncertain.
After of the outburst, the source has been declining in brightness \citep{kospal2007}, and \citet{hodapp2012} suggested that this tendency has already stopped.
In Figure \ref{f:ooser} we show the SED of OO Ser divided in two epochs, the outburst (1995-1996) period and the post-outburst stage, after 1996.

The fluxes cover the spectral range between $\sim 1-60\,\mu$m, with only two flux measurements available in the sub-mm range. Fluxes around $2\,\mu$m have a relatively large dispersion.
The parameters for the model in Figure \ref{f:ooser} are listed in Table \ref{t:resclass1}. The disk mass accretion rate value $\dot{\rm M}_{disk}= 5\dex{-5}\,\msun$/yr is the highest of all class~I FUORS in our sample.

\subsubsection{Re~50~N~IRS1}

Re~50 was discovered in the L1641 molecular cloud by \citet{reipurth1985a}. Later observations of the source IRAS~05380$-$0728, located $1.5\arcmin$ north of Re~50 (i.e., Re~50~N), allowed the identification of Re~50~N~IRS1, which has a stellar counter-part observed at $3.6\,\mu$m \citep{casali1991}.
Re~50~N~IRS1 is an embedded class~I object, located at a distance of $460\,$pc \citep{geers2009,sandell2001} and associated with a bipolar molecular outflow \citep{reipurth1986}.
\citet{strom1993} proposed Re~50~N~IRS1 as a FU Orionis type object.

In Figure \ref{f:re50nirs1} we present the SED and model of Re~50~N~IRS1. The observed fluxes obtained from the literature do not completely cover the range between the NIR and the $2000\,\mu$m.  However, the ISO-SWS spectrum provides a good coverage of the $5-15\,\mu$m range \citep{quanz2007b}.
The dispersion of the observed fluxes, both at $10\,\mu$m and at longer wavelengths is relatively significant.  In a first attempt to model these data we tested different sets of observations, according to the period of time in which they were obtained.  However, given the small number of fluxes available, this turned out to be inconvenient. Furthermore, different apertures sizes were used in the fluxes determinations, which is the likely cause of the large dispersion. 

Table \ref{t:resclass1} lists the parameters obtained from the model. The disk mass accretion rate $\dot{\rm M}_{disk}=1.3\dex{-6}\,\msun$/yr and the envelope mass accretion rate $\dot{\rm M}=1.24\dex{-5}\,\msun$/yr are higher than values for class~I protostars \citep{whitney2003b}. However, $\dot{\rm M}_{disk}$ is on the order of those obtained for the other class~I FUORS in our sample, which strengthens its classification as a FUORS.

\subsubsection{V2492 Cygni}
\label{sec-v2492cyg}

This object, also known as PTF10NVG, IRAS~20496$+$4354 and VSXJ205126.1$+$440523, is a class~I object located to the South-East of the Pelican North Nebula \citep[$d=550\,$pc, ][]{straizys1989,bally2003,covey2011}, at an angular distance of about $ 2\deg $ from the protostellar object and FUORS HBC~722. Furthermore, V2492~Cyg may be associated with an outflow \citep{bally2003,covey2011}.

\citet{covey2011} observed the outburst in 2010, while conducting a monitoring survey of the North American Nebula region, registering a brightness increase of $\sim 5\,$mag in the optical and NIR. 
They associated this outburst with a FU Orionis event, since V2492~Cygni presents NIR spectroscopic characteristics similar to the FUORS V1647~Ori. Among these similarities are the weak P~Cygni profiles in the Balmer and CaII lines \citep{covey2011,aspin2011a}.
In addition, V2492~Cygni shows similar outburst variations as V1647~Ori in the optical. However, the time scale between outbursts is more similar to that of EXOR variables than of bona fide FUORS \citep{kospal2011a}.

V2492~Cygni also displays other characteristics typical of FUORS stars, such as having an FG-supergiant spectral type in the optical and an M-supergiant in the NIR.
Moreover, V2492 Cyg has Na I D, K I and He I blueshifted absorptions associated
with strong outflows \citep{covey2011}.

\citet{aspin2011a} observed that the CO overtone bandheads are strongly in emission.
They also noted that the properties of V2492 Cyg during its outburst are similar to the 2008 outburst of V1647 Ori, which in turn is more similar to an EX Lupi (the progenitor the EXor class) event than to a FUOR event.
In summary, whether V2492 Cygni is a FUORS or an EXOR is still under debate \citep{covey2011,kospal2011a}.

In Figure \ref{f:v2492cyg} we show the SEDs of V2492 Cyg corresponding to three epochs, before the 2010 outburst, and in two outburst periods, September and November 2010.
Pre-outburst fluxes (before 2010) cover the spectral range between $\sim 0.1-1300\,\mu$m, with fluxes around $80\,\mu$m presenting a large dispersion. On the other hand, both SEDs for the September and November 2010 outbursts only cover the $\sim 0.1-3 \,\mu$m spectral range, which difficult the determination of reliable parameters for the external disk and envelope.
The best models obtained for the different epochs reproduce satisfactorily the observed SEDs.  The parameters for the models in the Figure \ref{f:v2492cyg} are listed in Table \ref{t:resclass1}. 

The derived parameters for the V2492~Cygni SED before the outburst agree with parameters for class~I sources \citep{whitney2003b}, and are also in concordance with the values obtained by \citet{aspin2011a} using the grid of \citet{robitaille2006} and \citet{robitaille2007}.
The disk mass accretion rate ($\dot{\rm M}_{disk}=0.1\dex{-6}\,\msun$/yr vs $0.4\pm 0.5\dex{-6}\,\msun$/yr), and the stellar parameters (R$_{*}= 2.5\,$R$_{\odot}$ vs $2.8-3.0\,$R$_{\odot}$ and T$_{*}=5000\,$K vs $6100-6500\,$K) increase during the outburst event in a similar way as  for other class~I FU Orionis objects in our the sample.

\subsubsection{V1331 Cygni}

This protostar, also known as LkHa~120 and IRAS~20595$+$5009, is located in the L988 complex at a distance of $\sim 550\,$pc (see \citealt{herbig2006} for a summary), and is associated with a bipolar molecular outflow \citep{levreault1988b,mundt1998}.
\citet{biscaya1997} suggested the presence of a circumstellar disk with a mass of $\sim 0.5\,\msun$ 
surrounded by a gaseous envelope.
\cite{mcmuldroch1993} observed in CO synthesis maps an external expanding gas ring, and \cite{quanz2007a} detected two circumstellar rings of dust separated by a gap.

V1331 Cyg shares several characteristics with FU Orionis stars, and it has been classified as a pre-outburst FUORS \citep{welin1976,herbig1989}.  However, its nature still remains uncertain \citep{biscaya1997,sandell2001}.
Figure \ref{f:v1331cyg} shows the SED of V1331 Cyg. The fluxes obtained from the literature show a relatively large dispersion,  likely due to the different periods in which they were obtained.  

For this reason, the observed fluxes have been divided into two intervals of time. The first corresponds to the observations before 1991 and the second to the post-2001 period. The corresponding modeled SEDs are displayed with solid and dashed lines, respectively.
Table \ref{t:resclass1} shows the values of the parameters corresponding to the SED defined by the data obtained before 1991. In brackets we list the values of the model parameters for the post-2001 data when they differ from the pre-1991 ones.
In our modeling we took into account the different spectral types corresponding to the two periods indicated,  F0/F4 and G5, respectively \citep{chavarria-k1981,hamann1992,herbig2003}. 
Using the spectral type calibration of \citet{kenyon1995} we derived a temperature ${\rm T}\sim6600\,$K  and ${\rm T}\sim5770\,$K, respectively.

The two models have different disk masses (${\rm M}_{disk}=0.1\,\msun$ vs $0.02\,\msun$).
In general, the parameters values for the second SED are lower, although this difference can only be considered marginal. An exception is the disk mass accretion rate, which decreased by an
order of magnitude between 1991 and 2001 ($\dot{\rm M}_{disk}=2.0\dex{-6}\,\msun$/yr vs $0.1\dex{-6}\,\msun$/yr).
All this suggests that V1331 Cyg has entered into a post-outburst stage.

\subsubsection{HBC~722}
\label{sec:hbc722}

HBC~722, also known as V2493~Cygni and PTF~10qpf, is an YSO located at $\sim 520\,$pc in the North America/Pelican Nebula (e.g., \citealt{laugalys2006}).
The outburst of this star, occurred in March -- August 2010, was detected independently by \citet{munari2010}, from low resolution spectra, and by \citet{semkov2010}, from (BVRI) photometry and optical spectroscopy. Based on the similarity of the light curve with those of FU~Ori and V1057~Cyg, \citet{semkov2010} suggested that HBC 722 was a  FUOR-like object.
More recently, this suggestion was confirmed by \citet{miller2011} from infrared photometry and spectroscopy of this star as well as high and low resolution optical spectroscopy. In particular, they reported an increase in brightness of $4\,$mag, an optical spectrum consistent with a G supergiant and a NIR spectrum resembling those of late K–M giants/supergiants.
Nevertheless, \cite{kospal2011b} argued against the bone fide FUOR classification of this source, based on its fast fading rate. \citet{semkov2012b}, however, used the shape of the long-term light curve to confirm the FUOR nature of HBC~722.  
\citet{dunham2012} analyzed sub-mm continuum and molecular line emission indicating that HBC 722 is associated with a outflow.

Figure \ref{f:hbc722} shows the pre-outburst and the post-outburst SEDs. The observed fluxes for both SEDs cover the spectral range from optical to $10\,\mu$m particularly well. On the contrary, only a few data points are available for wavelengths around $100\,\mu$m in the case of the post-outburst SED, whereas no data for $\lambda>10\,\mu$m have been found for the pre-outburst SED. Fluxes for $\lambda<10\,\mu$m allow a reliable identification of the outburst event, however the scarcity of fluxes above $10\,\mu$m, particularly in the pre-outburst SED, rendered uncertain the determination of the envelope parameters.  
Nevertheless, the best models obtained for the outburst SED (solid line, Figure \ref{f:hbc722}) and the pre-outburst SED (dotted-dashed line) reproduce satisfactorily well the SEDs observed in the optical and infrared spectral regions.

Table \ref{t:resclass1} lists the parameters from the modeling of the post-outburst SED, and in brackets we indicate the values obtained for the pre-outburst SED when they differ.
Notable differences are the stellar temperature T$_{\rm *}$ and the disk mass accretion rate $\dot{\rm M}_{disk}$. T$_{\rm *}$ increases from $5600\,$K to $7100\,$K, while $\dot{\rm M}_{disk}$ does it from $0.4\dex{-6}\,\msun$/yr to $4\dex{-6}\,\msun$/yr.
Post-outburst parameters agree with previous estimates for other class~I FU Orionis.

\subsubsection{Parsamian 21}

Par~21, also known as IRAS~19266$+$0932, was discovery by \citet{parsamian1965} and classified as a FUORS by \citet{staude1992} based on an optical spectrum and infrared properties of the central star that illuminates a cometary nebula. These authors also associated this protostar with a small bipolar HH flow aligned along the polar axis of the nebula. 
\cite{allen2004} classified the star as a class~II object, and \cite{kospal2008} resolved a circumstellar envelope with a polar cavity and an edge-on disk on their high-resolution NIR direct and polarimetric images.
However, \citet{quanz2007b} suggest that this source is a post-AGB star, based on the detection of PAH emission features on a $4-5\,\mu$m infrared spectrum obtained with Spitzer. 

Figure \ref{f:par21} shows the observed SED for Par 21, including the $5-14\,\mu$m Spitzer-IRS spectrum \citep{quanz2007b}.
The model follows the shape of the Spitzer spectrum, but fails to reproduce individual fluxes at these wavelengths.
In addition the model underestimates the fluxes around 4 $\mu$m by a factor of $\sim$ 3 and overestimates the fluxes
around 50 $\mu$m by a factor of $\sim$ 10.  Consequently, the disk parameters, in particular the disk outer -or centrifugal- radius
($R_{\rm c}$) and the disk inner radius (R$_{\rm min}$), are likely to be poorly determined by our model.
Table \ref{t:resclass1} presents the parameters corresponding to the model shown in Figure \ref{f:par21}. The distance to this source is uncertain (see Table \ref{t:sample}), but for our SED model we adopted a distance of $1800\,$pc, as estimated by \citet{sandell2001}. The disk mass we derived (M$_{disk}=0.30\,\msun$) agrees with that determined by \citet{sandell2001}.

\citet{kospal2008} analyzed this source using a SED model that includes an optically thick and geometrically thin accretion disk with an optically thin envelope with no cavity. In addition, no central source is simulated, since during the outburst stage the inner disk contribution overwhelm the star flux \citep{hartmann1996a}. Moreover, they adopted an edge-on disk that obscures the central star. 
In general, the model parameters derived by \cite{kospal2008}, including the envelope mass (M$_{env}=0.22\,\msun$), roughly agree with our determinations. They also suggest that Par~21 has an edge-on disk, consistent with our model inclination angle ($i=79\deg$).
However, our inner disk radius is larger than what they derived.

More recently, \citet{liu2011b} analyzed the SEDs of a sample of Herbig Ae/Be stars, including Par 21, using the \citet{robitaille2006} grid. They derive M$_{\rm *}=3.74\,\msun$, R$_{\rm *}=5.68\,$R$_{\odot}$, and T$_{\rm *}=8511\,$K. Both the stellar mass and radius are higher than our estimations (see Table \ref{t:resclass1}), but we are in a good agreement on the stellar temperature. 
The inclination angle $i=87.13\deg$ obtained by \citet{liu2011b} is consistent with our result of $i=79\deg$. However, we find our disk mass and disk mass accretion rate differ from their determinations. Their estimations of ${M}_{disk}=0.04\,\msun$ and $\dot{\rm M}_{disk}=4\dex{-6}\,\msun$/yr correspond rather to a classical or inactive class~II object than to an FU Orionis star. 
For the disk mass and mass accretion rate we derived ${M}_{disk}=0.3\,\msun$ and
$\dot{\rm M}_{disk}=4\dex{-6}\,\msun$/yr, respectively, in good  agreement with previous
estimates for other class~I FU Orionis.

\subsubsection{V2775~Ori}
\label{s:v2775ori}

This object, also known as 2MASS~J05424848$-$0816347, was first reported by \citet{carattiogaratti2011} as a FU Orionis star in the L1641 region of the Orion molecular cloud, at a distance of $ 420\,$pc \citep{sandstrom2007,menten2007,kim2008}.
V2775~Ori is suspected to be part of a wide binary system with a separation of $\sim17300\,$AU ($\sim0.08\,$pc), associated to precessing jets \citep{carattiogaratti2011}. \citet{fischer2012} observed this source in the near-IR and concluded that its spectra are consistent with a FU Orionis object. They observe CO absorption lines in the K-band, broad H$_{2}$O absorption, strong and wide blueshifted He~I, and a lack of atomic hydrogen emission.

In Figure \ref{f:v2775} we show the SED of V2775~Ori divided in two epochs, the pre-outburst (before 2005) period and the post-outburst stage, after 2006.
The outburst fluxes cover the $\sim 1-900\,\mu$m spectral range, however the fluxes in the pre-outburst period cover only the spectral range between $\sim 1-70\,\mu$m,  without any measurements available in the sub-mm range.
The best models obtained for the different epochs reproduce satisfactorily the observed SEDs. The parameters for the model shown in Figure \ref{f:v2775} are listed in Table \ref{t:resclass1}.

When comparing our results for the pre- and post-outburst SEDs, we see that the stellar temperature and the disk mass and mass accretion rate all increased during the outburst (see Table~\ref{t:resclass1}). In particular, $ \dot{\rm M}_{disk} $ increased by about two orders of magnitude from $0.6\dex{-6}\,\msun$/yr to $10\dex{-6}\,\msun$/yr.
On the other hand, the stellar mass, as well as the geometrical parameters $i=60\deg$ and $\theta=7\deg$, remain unchanged, while the envelope mass accretion rate decreases by one order of magnitude from $ 3\dex{-6}\,\msun$/yr to $7\dex{-7}\,\msun$/yr.

\citet{carattiogaratti2011} find that this source is one of the lowest mass YSOs presenting a strong outburst. Based on the features of its spectra (strong CO bandheads, H$ _{2} $O broadband absorption, Br$ \gamma $ in emission) they adopted an M spectral type. In comparisons of their NIR spectra with AMES-DUSTY models \citep{allard2001} they find a reasonable match for $T_{\rm eff} \sim3200\,$K. Furthermore, applying the main-sequence models of \citet{siess2000} they derived a stellar mass of $\sim0.24\,\msun$.

\citet{fischer2012} modeled this source and obtained that the disk accretion rate increased in around one order of magnitude ($\sim2\dex{-6}\,\msun$/yr to $\sim10^{-5}\,\msun$/yr) and that the envelope mass accretion rate remained constant with a value of $7\dex{-7}\msun$/yr and suggested that V2775~Ori is approaching the end of the envelope dominated phase. For their best model they adopted R$ _{*} = 2.09\,$R$_{\odot}$, $ {\rm T}=4000\,$K and M$ _{*} =0.5\,\msun$, and with an outer radius of $ 10000\,$AU they derived an envelope mass of $ 0.09\,\msun $, an opening angle of $ 25\deg $ and inclination of $ 49\deg $.

From our models we obtain M$ _{*}=0.5\,\msun$, in good agreement with \citet{carattiogaratti2011}. However, the stellar temperature estimated before and after the outburst ($ 5600\,$K and $ 6800\,$K, respectively) is significantly higher than what \citet{carattiogaratti2011} and \citet{fischer2012} obtained.
Geometrical parameters as the opening angle and inclination are similar, although somewhat higher than the those derived by \citet{fischer2012}. On the other hand, the increase in the disk mass-accretion rate we obtain is one order of magnitude larger than what they derive.

Based on the increase of its envelope mass-accretion rate from $ 3\dex{-6}\,\msun$/yr to $ 7\dex{-7}\,\msun$/yr during the outburst, we suggest that before the outburst this was a Class I object, but now during the outburst it is in the late stages of that class. This is in agreement with \citet{fischer2012}.

\section{Results}
\label{s:results}

In this section we analyze the results derived from the modeling of the 24 known and candidate FUORS in our sample. We compare parameters values for these objects with those of standard class~II and class~I objects that are not in an eruptive phase, obtained from the sample of YSOs in Taurus modeled by \citet{robitaille2007}. Table \ref{t:averages} lists the average values for each parameter of the class~II and class~I FUORS. For sources RNO~1B, V1647~Ori, OO~Ser, V2492~Cyg, HBC~722, V2775~Ori, and V1331~Cyg we obtained more than one solution, corresponding to different periods of observations. In those cases, we choose the parameters corresponding to the outburst stage for computing the average. Table \ref{t:averages} also gives average values of these parameters for standard class~II and class~I YSOs from \citet{robitaille2007}.
From Table~\ref{t:averages} we see that class~II and class~I FUORS disks have smaller inner radii (R$_{min}$) in comparison with the average values for the standard class~II and class~I from \citet{robitaille2007}. In addition, for FUORS the centrifugal radii of the circumstellar disks are also smaller than for classical YSOs.

To better describe and compare how the different model parameters change between the FUORS during the outburst and the standard class~II and class~I objects in quiescence stage, we analyze the cumulative distribution of the disk mass, the disk mass accretion rate, the envelope mass accretion rate, and the stellar temperature. Those parameters were selected since they show the largest variations.
For the typical class~II and class~I objects, we use the parameters from the models of \citet{robitaille2007} for standard YSOs in Taurus. The resulting distributions are shown in Figure~\ref{f:cumul}. 
We apply the Kolmogorov-Smirnov (KS) test to compare these distributions for each of the four parameters selected.
Table~\ref{t:median} lists the median values for both groups as well as the KS test results, i.e., the maximum difference $D$ between the distributions and the significance or confidence level $s$. 
The cumulative distributions for both groups (standard class~II and class~I objects in Taurus and FUORS) are different with a high level of confidence\footnote{If both distributions were identical, then $s=1$.}. 
We also see that within each parameter, the values for each group are distributed in different ways. 

On average, FUORS disks are more massive and have higher accretion rates than standard class~II and class~I disks. 
None of the disks in standard class~II and class~I objects have masses above $0.06\,\msun$ while $\sim80\%$ of
the disks in FUORS have masses $\ge0.10\,\msun$.
 Nevertheless, the mass distributions for classical YSOs and FUORS objects have similar spreads. 
Standard class~II and class~I objects span at least two orders of magnitude in mass from $2.5\dex{-4}\,\msun$ to $6\dex{-2}\,\msun$
(see also \citealt{andrews2005}), while FUORS disks span a mass range from $ 0.01\,\msun$ to $0.37\,\msun $.
The values listed on Table~\ref{t:averages} indicate that class~II and class~I FUORS show increases in the disk mass
by one order of magnitude (see also Table~\ref{t:median}). 

\begin{figure}[htb]
\includegraphics[angle=0,width=0.5\textwidth]{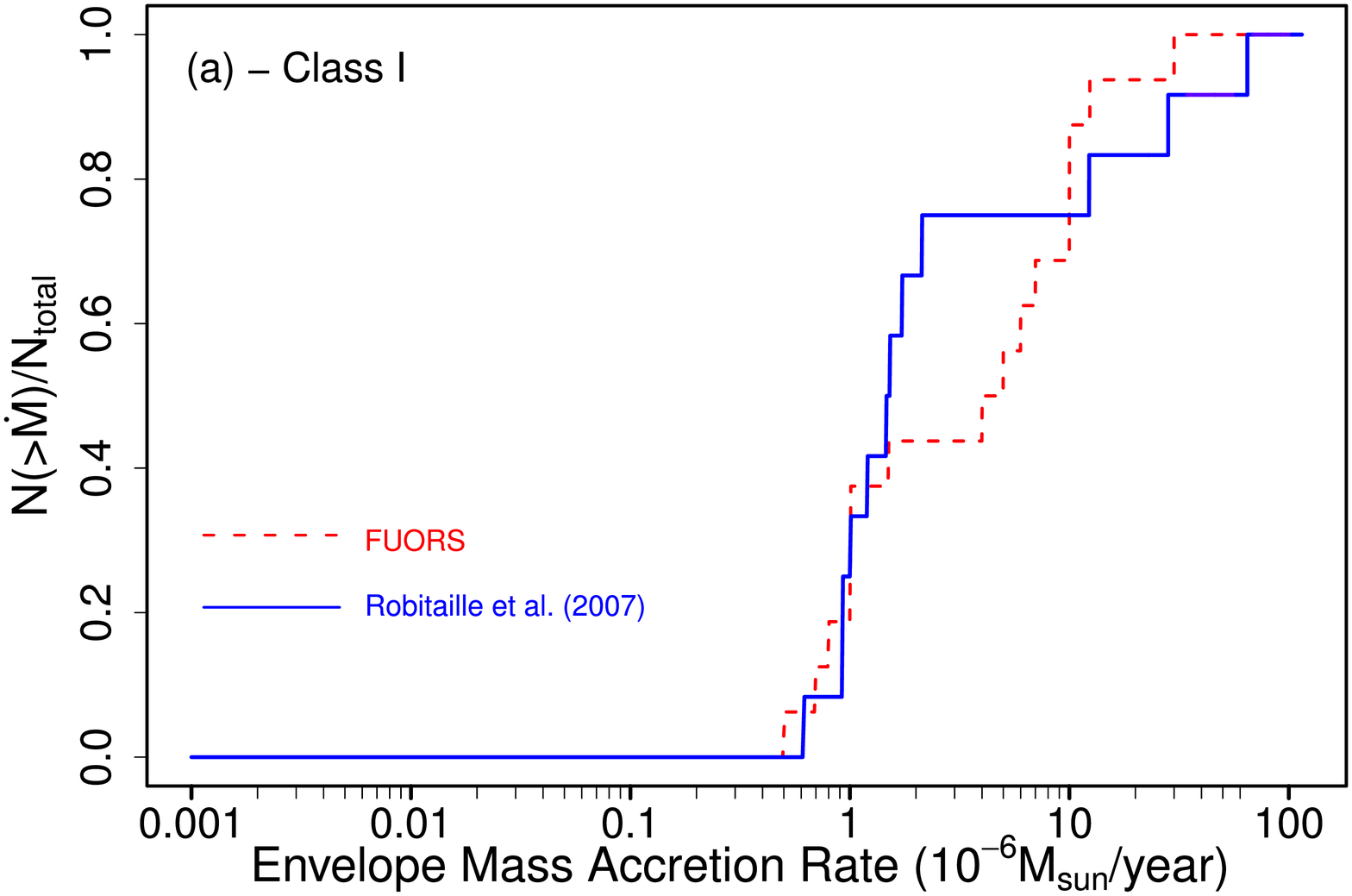}
\includegraphics[angle=0,width=0.5\textwidth]{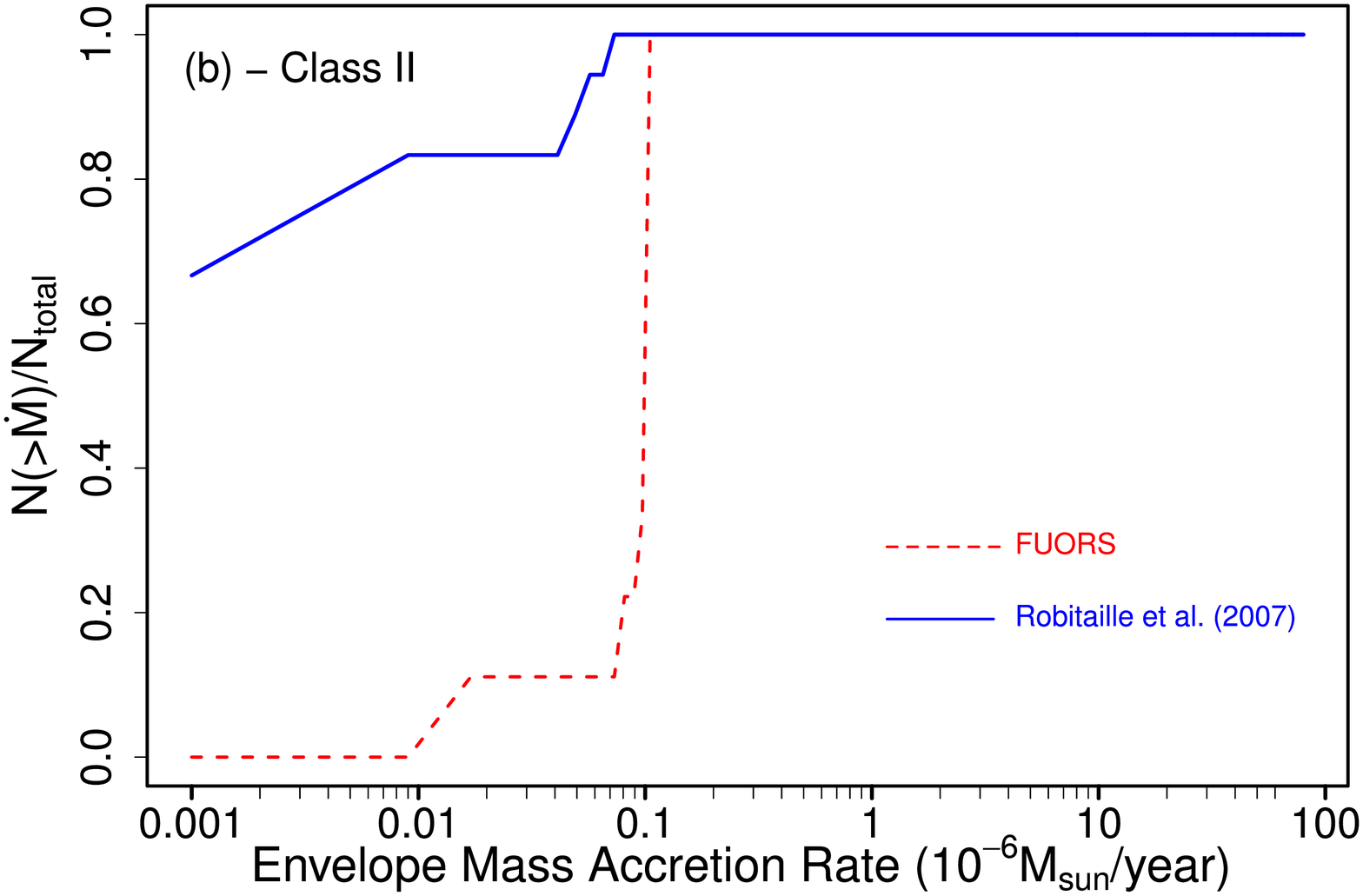}
\caption{Cumulative distributions for the envelope mass accretion rate. 
In panel~$a$, we compare the class I FUORS distribution (red dashed line)
to the standard class~I objects distribution (blue solid line).  Panel~$b$ shows the
class II FUORS distribution (red dashed line) vs the standard class~II objects distribution
(blue solid line). FUORS have been analyzed in this contribution. Standard
class I and class II objects have been modeled by \cite{robitaille2007}.}
\label{f:Menv}
\end{figure}

Regarding the disk mass accretion rate, $\sim 90\%$ of the FUORS have $\dot{\rm M}_{disk}>10^{-6}\,$M$_{\odot}$/yr, while $\sim95\%$ of the standard class~II and class~I objects have $\dot{\rm M}_{disk}<10^{-6}$\,M$_{\odot}$/yr. The median mass accretion rate for FUORS is $\sim10^{-5}\,$M$_{\odot}$/yr, in contrast with $\sim10^{-7}$\,M$_{\odot}$/yr for classical YSOs (see Table~\ref{t:median}). {Despite of being unusual or rare, a significant amount of mass can be accumulated
onto the central star during relatively short periods of time (the FUORS events),
contributing to its final mass.

The comparison of the behavior of the envelope mass accretion rate for both
distributions shows that a large fraction of FUORS ($\sim70\%$) have accretion
rates $>10^{-7}\,\msun$/yr. In contrast, 60\% of the classical YSOs have accretion
rates below this value. On average, FUORS have higher envelope mass accretion rates
than standard class~II and class~I sources ($\sim10^{-6}\,\msun$/yr vs $\sim10^{-8}\,\msun$/yr,
respectively, see Table~\ref{t:median}).
For class~I FUORS, the envelope mass accretion rate remains practically unchanged during
the FU Orionis stage (see Table \ref{t:averages}). In the cases where we had an SED before and after the outburst,
this parameter remained unchanged for both SED models (see Tables \ref{t:resclass2}, and \ref{t:resclass1}).
 Figure~\ref{f:Menv} shows the envelope mass accretion rate, analized per YSO class, i.e., for the class~I
FUORS sample vs standard class~I objects (left panel) and for the class~II FUORS group vs classical class~II YSOs
(right panel). The distributions of class~I FUORS and standard class~I objects are similar ($s=0.15$, $D=0.31$), while class~II FUORS
and classical class~II are different ($s=1.5\dex{-4}$, $D =0.89$).}

The case for the stellar temperature is different (see Figure \ref{f:cumul}, bottom right panel). The distributions for both groups are similar in shape, only shifted by about 2000 K to higher temperatures for class~II and class~I FUORS, which reflects the observed rise in stellar luminosity during the outburst event. The higher stellar temperature would also account for the hotter or earlier spectral type. 

For 7 of the stars in our sample, 2 class II (RNO 1B, V1647 Ori) and 5 class I (OO Ser, V2492 Cyg, HBC 722, V2775 Ori, V1313 Cyg), we had modeled two SEDs (see Tables~\ref{t:resclass2} and \ref{t:resclass1}), during the outburst and at the quiescence stage.  Envelope parameters such as mass, radius, and mass accretion rate do change. This suggests that the outbursts are triggered by an instability after a long build-up phase. In general, the remaining disk and stellar parameters change significantly.

\section{Summary and Discussion}
\label{s:summary}

In this work we present the modeling of the SEDs of a sample of 24 class~II and class~I FU Orionis stars.
These SEDs were constructed from fluxes obtained from the literature (Table~\ref{t:fluxes}), including Spitzer-IRS infrared spectra in the $5-35\,\mu$m range for V1515~Cyg, BBW~76, FU~Ori, V346~Nor and V1057~Cyg, and in the $5 -14\,\mu$m range for RNO~1B, RNO~1C, L1551~IRS5 and Par~21 \citep{green2006,quanz2007b}.
For Re~50~N~IRS1, we used an ISO-SWS spectrum in the $5 -15\,\mu$m range obtained by \citet{quanz2007b}.

\begin{figure}
\epsscale{.65}
\plotone{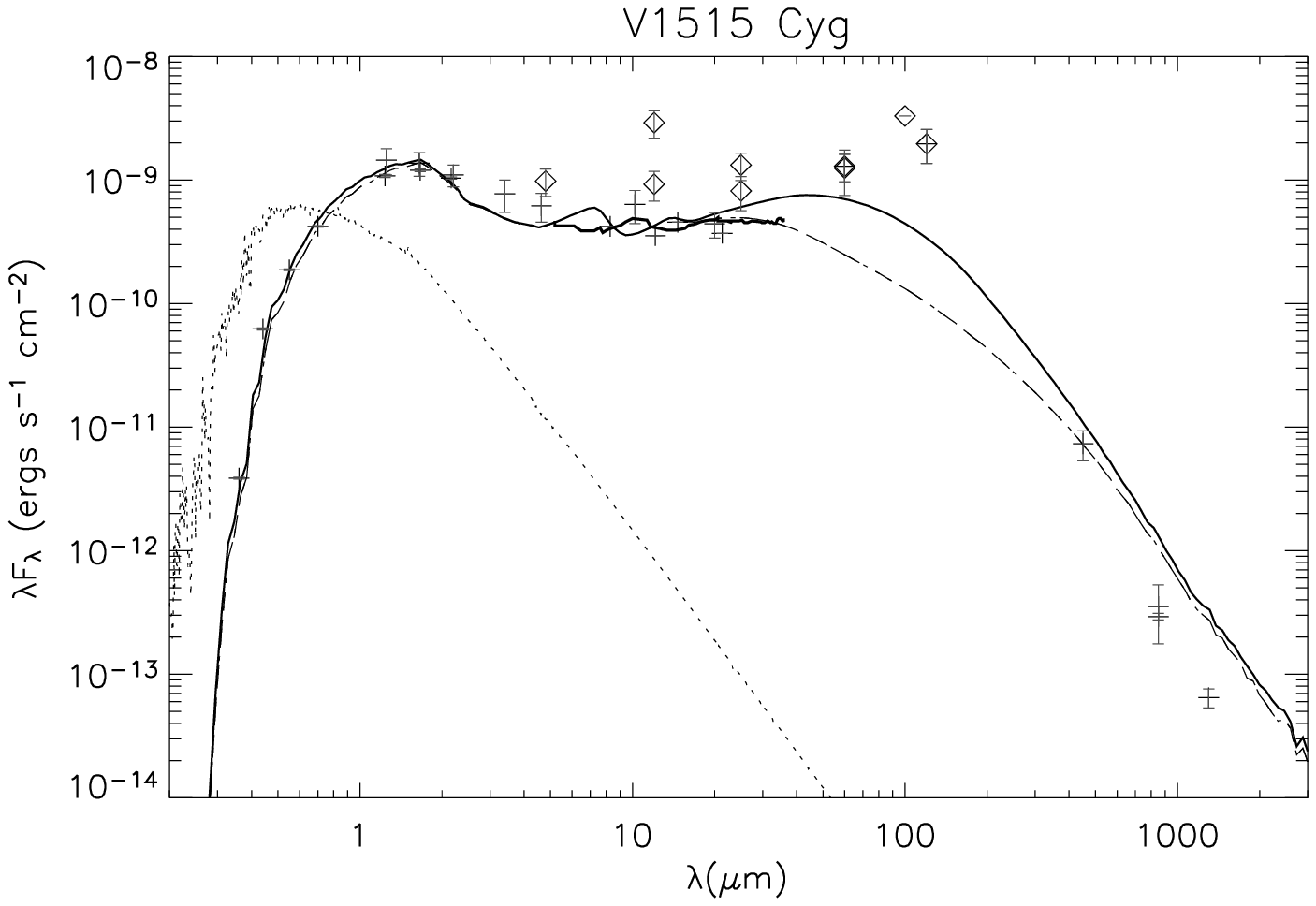}
\caption{SED and model for V1515~Cyg. 
The crosses correspond to the observed fluxes in the $2003-2004$ period, diamonds in the $1983-1996$ period, and the bars represent the uncertainties for the fluxes. In some cases error bars are not seen because they are smaller than the size of the points. In thick solid line we include the Spitzer $5-36\,\mu$m spectrum from \citet{green2006}. The dotted line shows the Kurucz model for the stellar photosphere. Solid and dashed lines are the SED models for aperture sizes of $ 60\arcsec $ and $ 11\arcsec $, respectively.}
\label{f:v1515cyg}
\end{figure}

\begin{figure}
\epsscale{.65}
\plotone{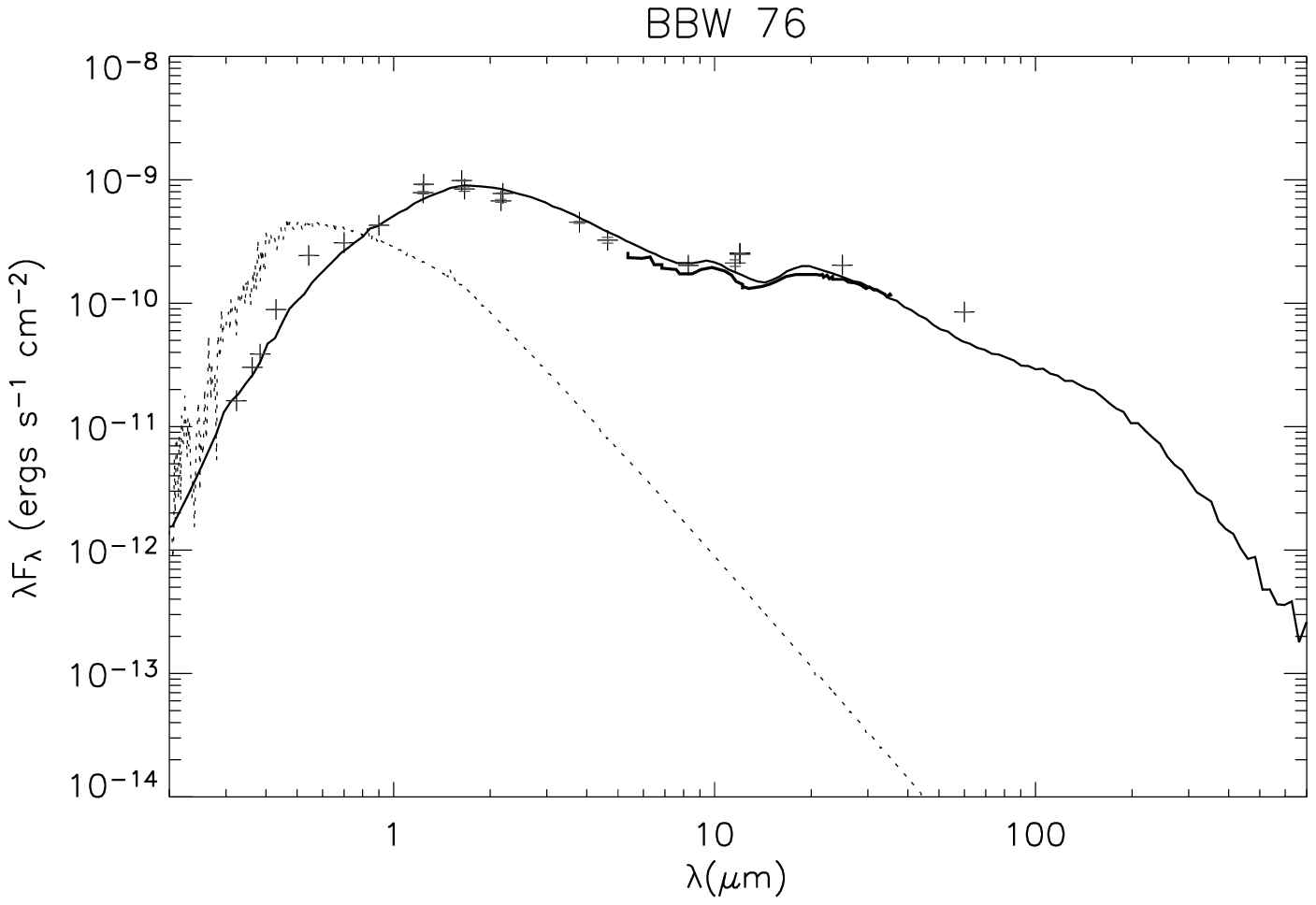}
\caption{SED and model for BBW~76.  The crosses correspond to the observed fluxes and the uncertainties are represented with bars. In some cases error bars are not seen because they are smaller than the size of the points. In thick solid line we include the Spitzer $5-36\,\mu$m spectrum from \citet{green2006}. The dotted line shows the Kurucz model for the stellar photosphere. Solid line is the SED model.}
\label{f:bbw76}
\end{figure}

\begin{figure}
\epsscale{.65}
\plotone{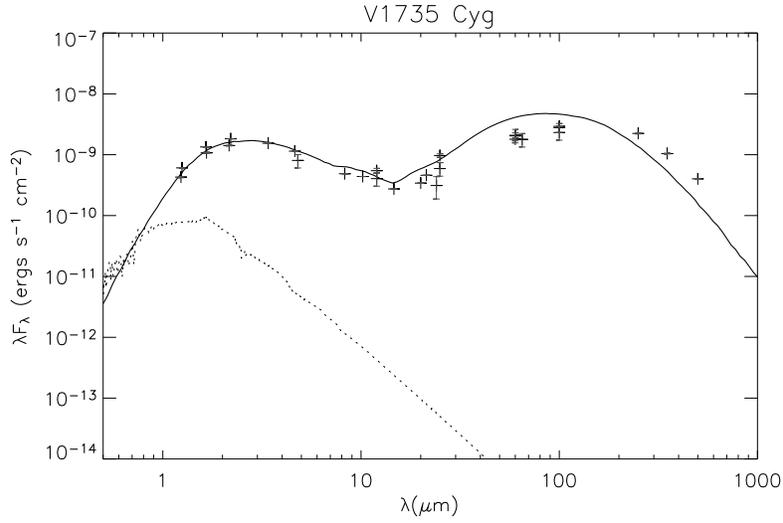}
\caption{SED and model for V1735~Cyg. The crosses correspond to the observed fluxes and the uncertainties are represented with bars. In some cases error bars are not seen because they are smaller than the size of the points. The dotted line shows the Kurucz model for the stellar photosphere. Solid line is the SED model.} 
\label{f:v1735cyg}
\end{figure}

\begin{figure}
\epsscale{.65}
\plotone{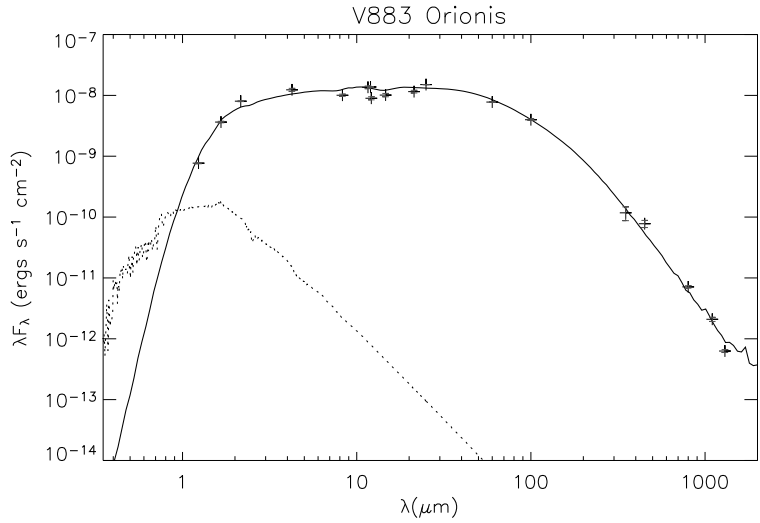}
\caption{Same as Figure~\ref{f:v1735cyg} but for V883~Ori.}
\label{f:v883ori}
\end{figure}

\begin{figure}
\epsscale{.65}
\plotone{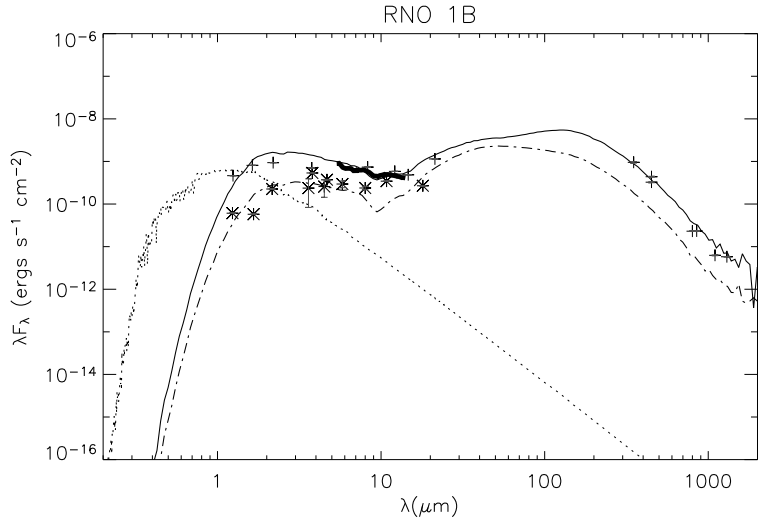}
\caption{SED and model for RNO~1B. Crosses are observed fluxes in the period before 1995 and asterisks in the period after 1996. The bars
represent the uncertainties for the fluxes, that in some cases are not seen because they are smaller than the size of the points. We include the Spitzer $5-14\,\mu$m spectrum from \citet{quanz2007b}. The dotted line shows the Kurucz model for the stellar photosphere. Solid and dashed lines are the SED models for each period, respectively.} 
\label{f:rno1b}
\end{figure}

\begin{figure}
\epsscale{.65}
\plotone{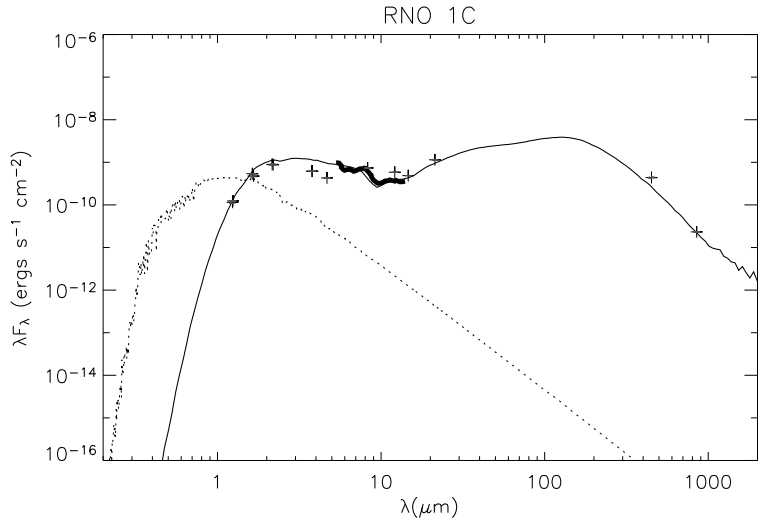}
\caption{Same as Figure~\ref{f:v1735cyg} but for RNO~1C. We include the Spitzer $5-14\,\mu$m spectrum from \citet{quanz2007b}.} 
\label{f:rno1c}
\end{figure}

\begin{figure}
\epsscale{.65}
\plotone{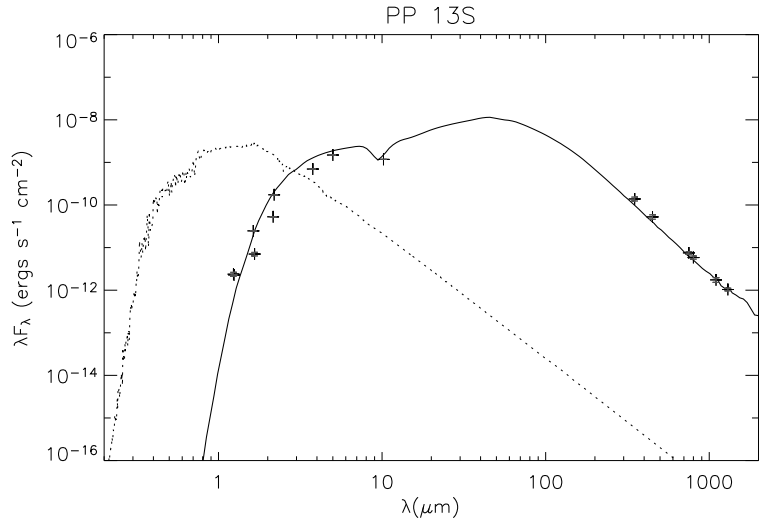}
\caption{Same as Figure~\ref{f:v1735cyg} but for PP~13S.}
\label{f:pp13s}
\end{figure}

\begin{figure}
\epsscale{.65}
\plotone{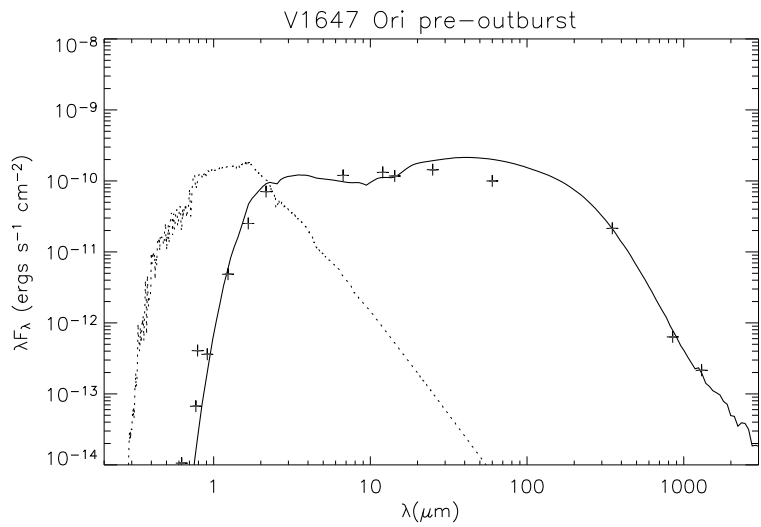}
\caption{Same as Figure~\ref{f:v1735cyg} but for V1647 Ori before its first outburst.}
\label{f:v1647oripre}
\end{figure}

\begin{figure}
\epsscale{.65}
\plotone{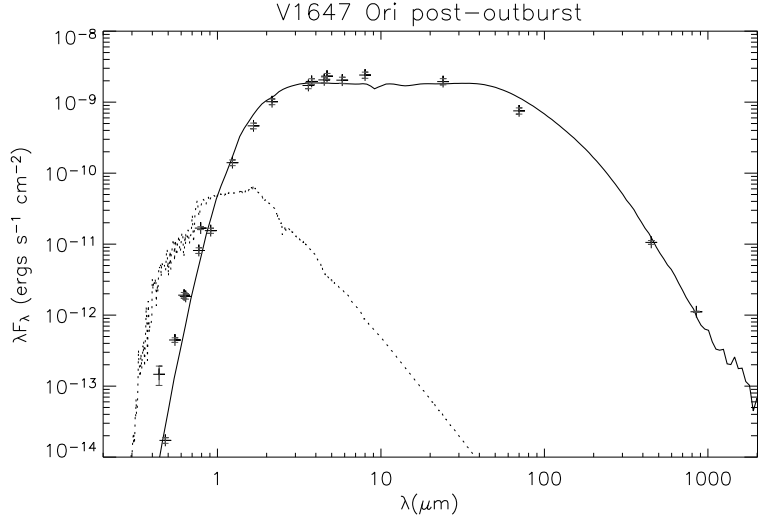}
\caption{Same as Figure~\ref{f:v1735cyg} but for V1647 Ori after its first outburst.}
\label{f:v1647oripost}
\end{figure}

\begin{figure}
\epsscale{0.65}
\plotone{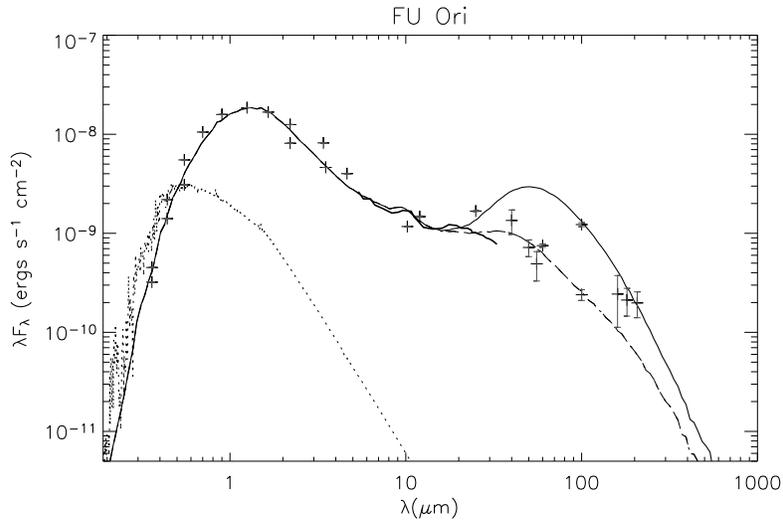}
\caption{SED and model for FU Ori. The crosses correspond to the observed fluxes, and the bars represent the uncertainties for the fluxes. In some cases error bars are not seen because they are smaller than the size of the points. In thick solid line we include the Spitzer $5-36\,\mu$m spectrum from \citet{green2006}. The dotted line shows the Kurucz model for the stellar photosphere.
Solid and dashed lines are the SED models for aperture sizes of $ 60\arcsec $ and $ 20\arcsec $, respectively.} 
\label{f:fuori}
\end{figure}

\begin{figure}
\epsscale{.65}
\plotone{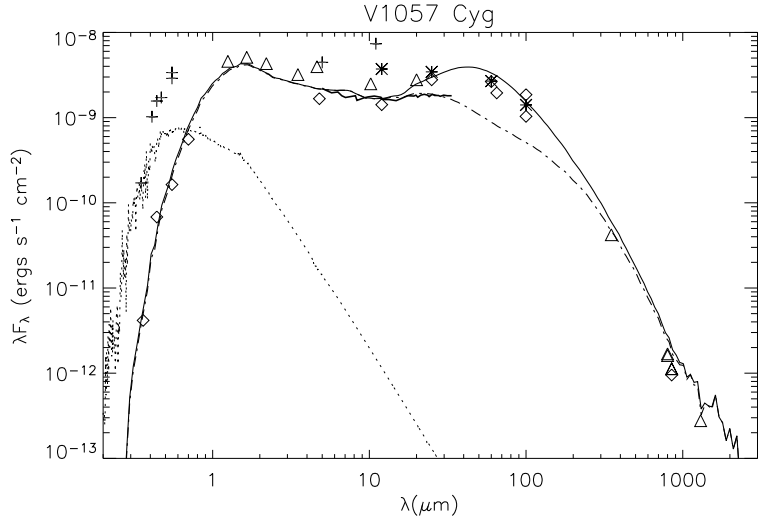}
\caption{SED and model for V1057~Cyg. The crosses are observed fluxes in the period before 1971, asterisks in 1983, triangles in 1989, and diamonds in the $1995-1998$ period. The bars represent the uncertainties for the fluxes, that in some cases are not seen because they are smaller than the size of the points. In thick solid line we include the Spitzer $5-36\,\mu$m spectrum from \citet{green2006}. The dotted line shows the Kurucz model for the stellar photosphere. Our best model closely reproduces the most recent data, including the Spitzer spectrum.  Solid and dot-dashed lines are the SED models for aperture sizes of $ 60\arcsec $ and $ 11\arcsec $, respectively.}
\label{f:v1057cyg}
\end{figure}

\begin{figure}
\epsscale{.65}
\plotone{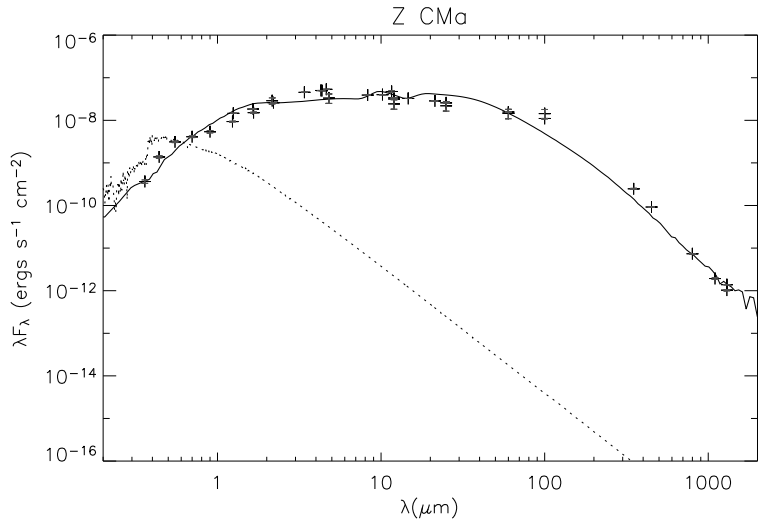}
\caption{Same as Figure~\ref{f:v1735cyg} but for Z~CMa.}
\label{f:zcma}
\end{figure}

\begin{figure}
\epsscale{.65}
\plotone{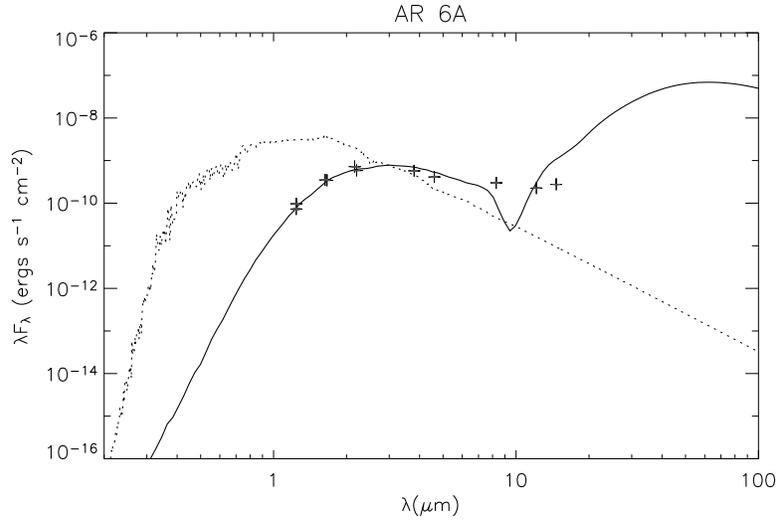}
\caption{Same as Figure~\ref{f:v1735cyg} but for AR~6A. Only NIR and MIR fluxes are available (see text).} 
\label{f:ar6a}
\end{figure}

\begin{figure}
\epsscale{.65}
\plotone{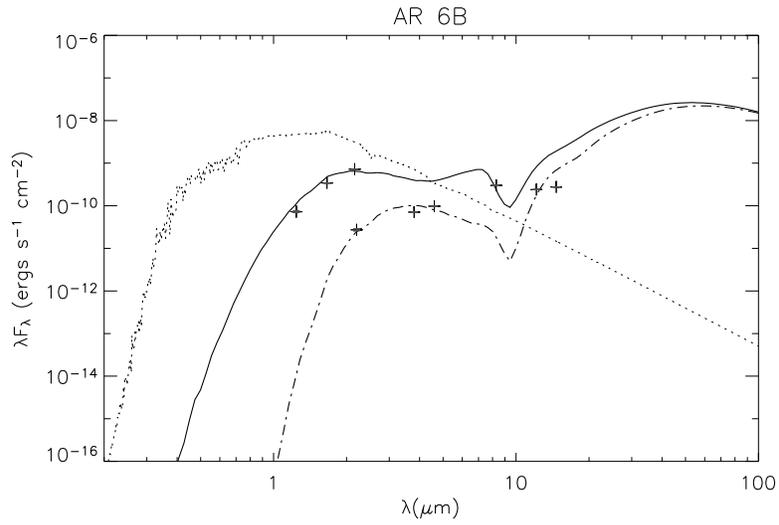}
\caption{Same as Figure~\ref{f:v1735cyg} but for AR~6B.
Solid and dashed lines are the SED models for aperture sizes of $ 60\arcsec $ and $ 30\arcsec $, respectively. Only NIR and MIR fluxes are available (see text).} 
\label{f:ar6b}
\end{figure}

\begin{figure}
\epsscale{.65}
\plotone{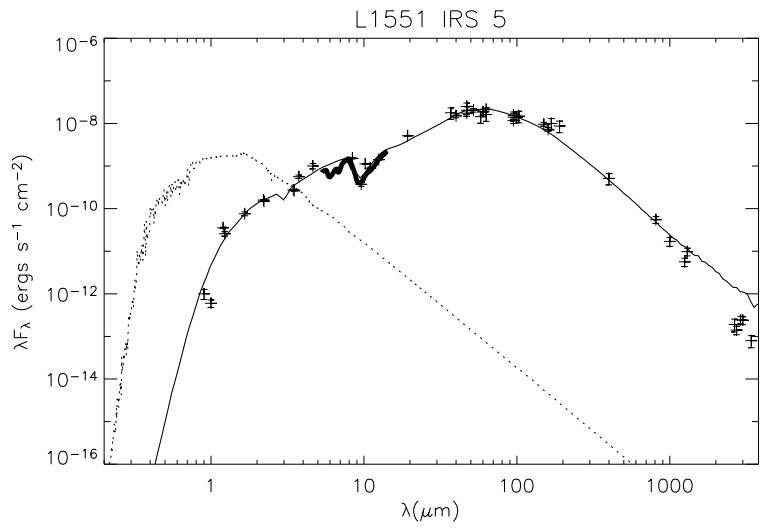}
\caption{Same as Figure~\ref{f:v1735cyg} but for L1551~IRS5. We include the Spitzer spectrum from \citet{quanz2007b}.} 
\label{f:l1551}
\end{figure}

\begin{figure}
\epsscale{.65}
\plotone{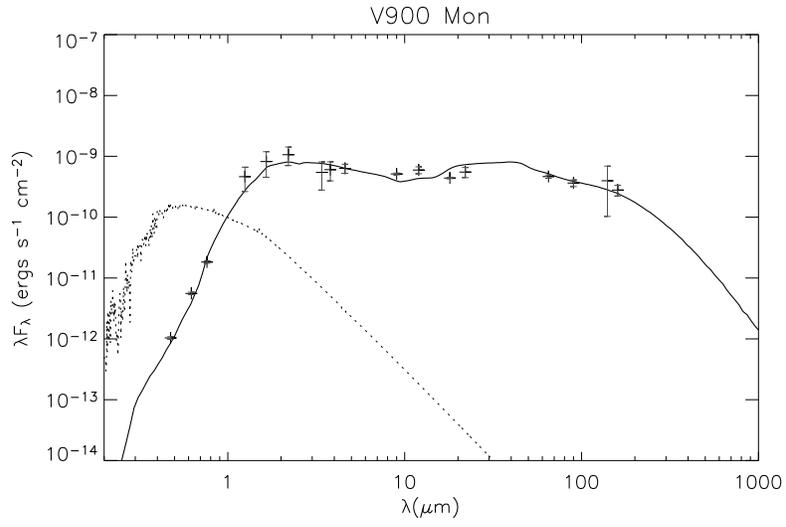}
\caption{Same as Figure~\ref{f:v1735cyg} but for V900~Mon.} 
\label{f:v900}
\end{figure}

\begin{figure}
\epsscale{.65}
\plotone{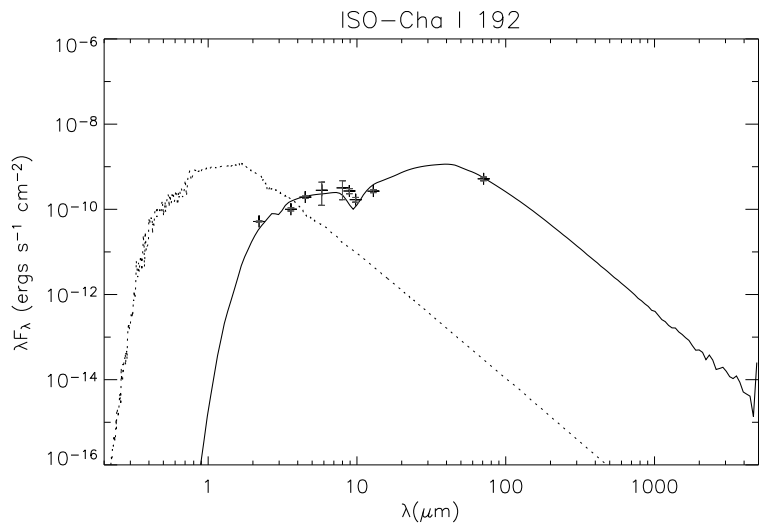}
\caption{Same as Figure~\ref{f:v1735cyg} but for ISO-Cha I 192.}
\label{f:isochai192}
\end{figure}

\begin{figure}
\epsscale{.65}
\plotone{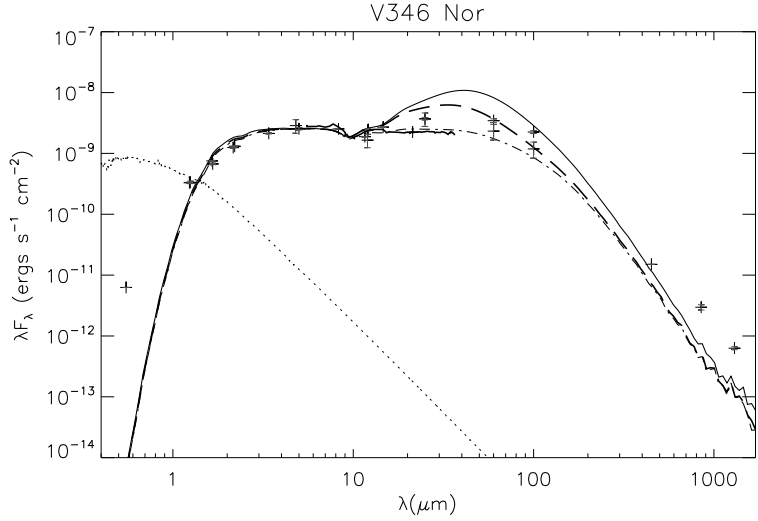}
\caption{Same as Figure~\ref{f:v1735cyg} but for V346~Nor. We include the Spitzer $5-36\,\mu$m spectrum from \citet{green2006}. Solid, dot-dashed, and dashed lines are the SED models for aperture sizes of $ 60\arcsec $, $ 30\arcsec $, and $ 11\arcsec $, respectively. }
\label{f:v364nor}
\end{figure}

\begin{figure}
\epsscale{.65}
\plotone{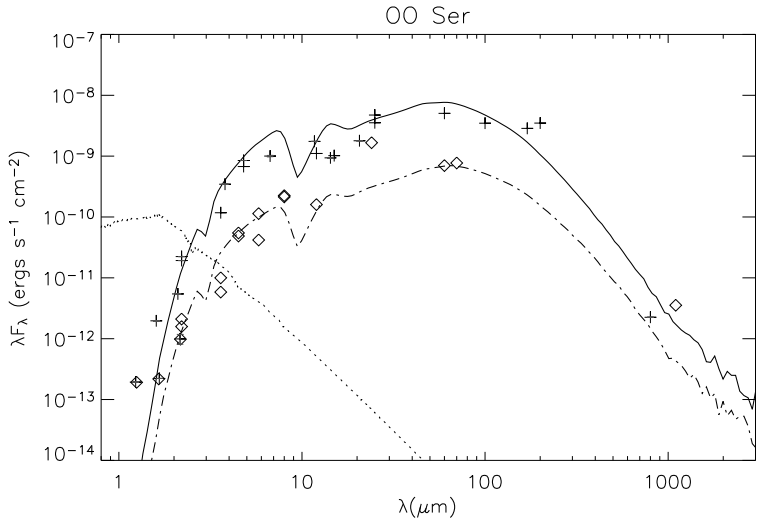}
\caption{SED and model for OO Ser. Crosses are the fluxes observed in the $1983-1999$ period and diamonds in the period after 2004. The bars represent the uncertainties for the fluxes, that in some cases are not seen because they are smaller than the size of the points. The dotted line shows the Kurucz model for the stellar photosphere. Solid and dashed lines are the corresponding SEDs models.} 
\label{f:ooser}
\end{figure}

\begin{figure}
\epsscale{.65}
\plotone{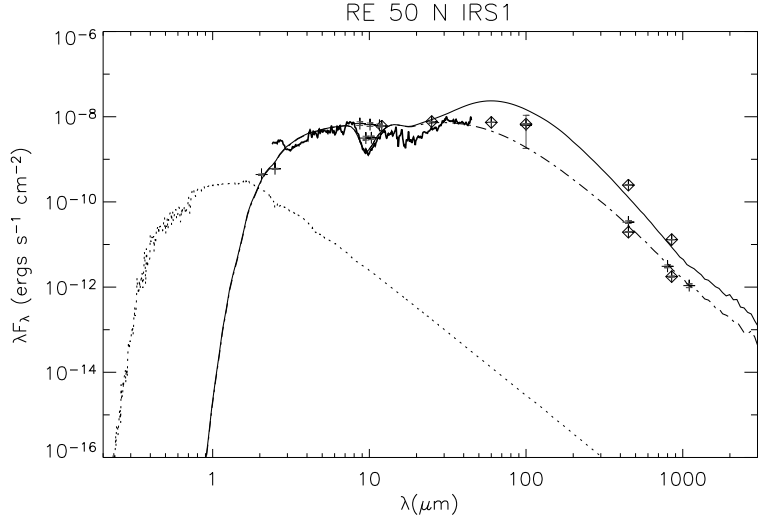}
\caption{SED and model for RE~50~N~IRS1. Crosses are the fluxes observed before 1991 and diamonds after 1992. The bars represent the uncertainties for the fluxes, that in some cases are not seen because they are smaller than the size of the points. We include the ISO-SWS $5-15\,\mu$m spectrum from \citet{quanz2007b}. The dotted line shows the Kurucz model for the stellar photosphere. Solid and dot-dashed lines are the SED models for aperture sizes of $ 50\arcsec $ and $ 11\arcsec $, respectively.} 
\label{f:re50nirs1}
\end{figure}

\begin{figure}
\epsscale{.65}
\plotone{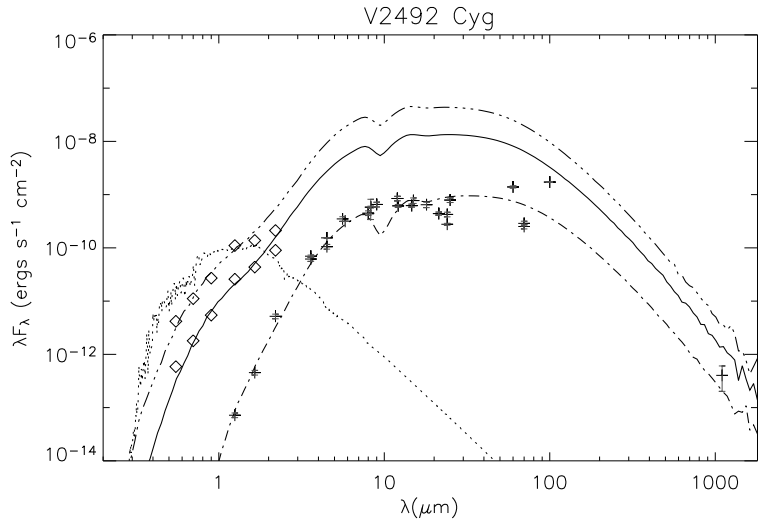}
\caption{SED and model for V2492~Cygni. Crosses are the fluxes observed before the 2010 outbursts, and diamonds during the September and November 2010 outbursts. The bars represent the uncertainties for the fluxes, that in some cases are not seen because they are smaller than the size of the points. 
The dotted line shows the Kurucz model for the stellar photosphere. Dot-dashed, triple-dot dashed, and solid lines are the respective SED models.}
\label{f:v2492cyg}
\end{figure}

\begin{figure}
\epsscale{.65}
\plotone{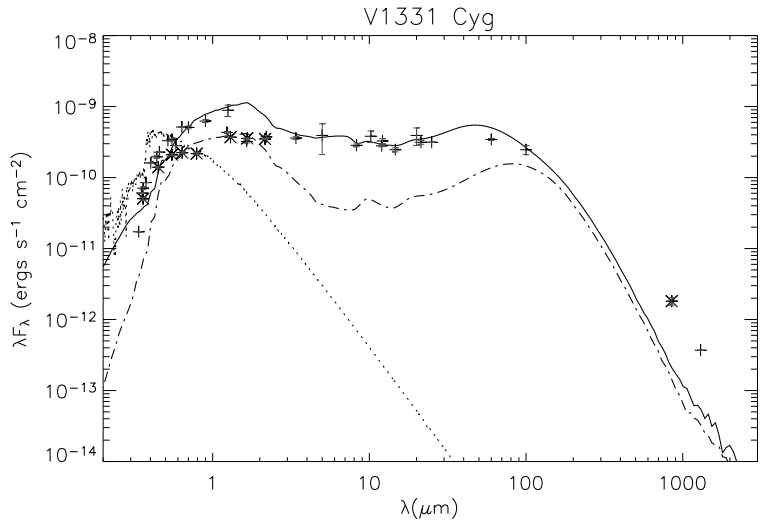}
\caption{SED and model for V1331~Cyg. Crosses are the fluxes observed before 1991 and the asterisks after 2001. The bars represent the uncertainties for the fluxes, that in some cases are not seen because they are smaller than the size of the points. The dotted line shows the Kurucz model for the stellar photosphere. Solid and dashed lines are the corresponding SEDs models.} 
\label{f:v1331cyg}
\end{figure}

\begin{figure}
\epsscale{.65}
\plotone{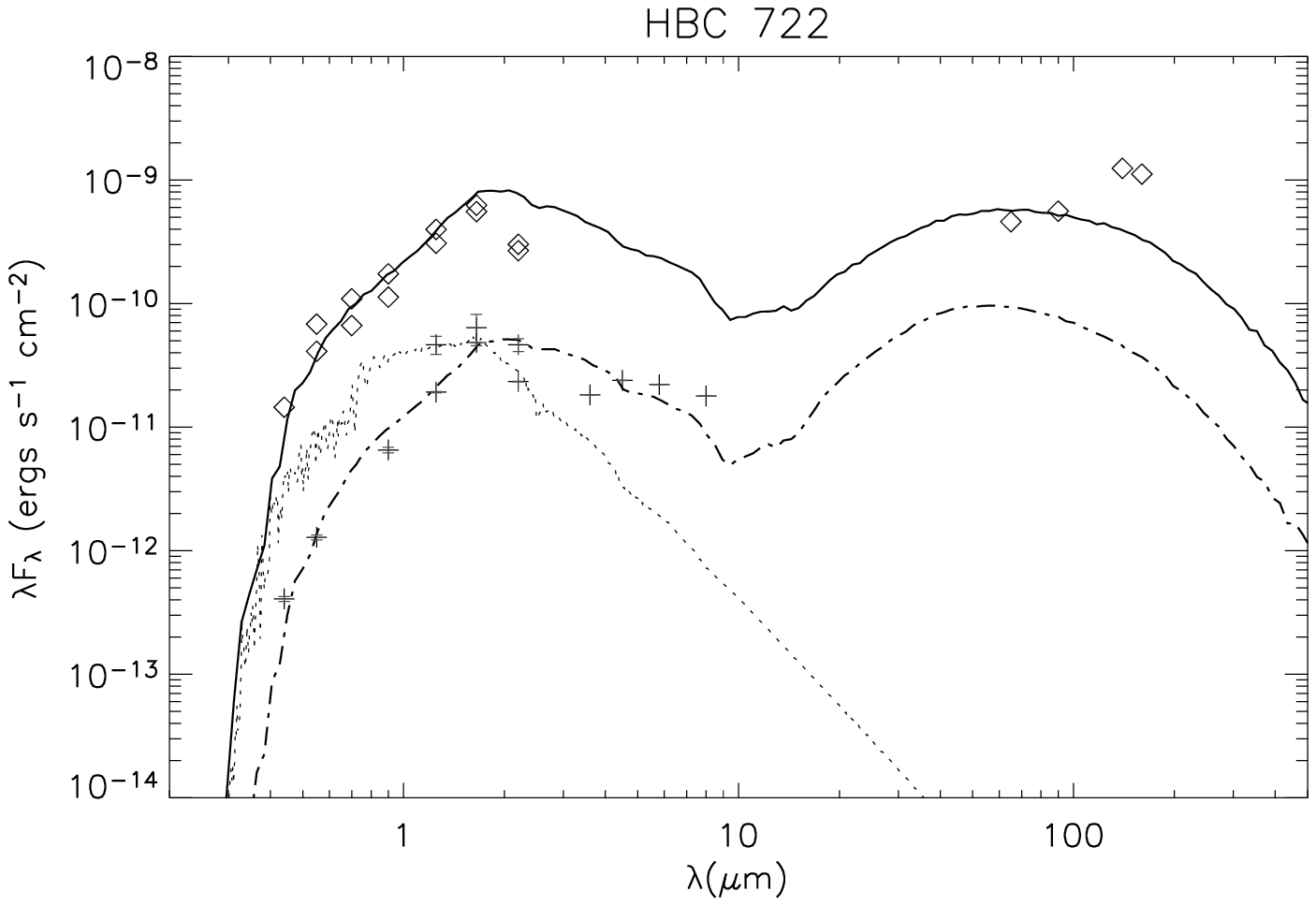}
\caption{SED and model for HBC 722. Crosses are the observed fluxes before the outburst, and diamonds during the outburst. The bars represent the uncertainties for the fluxes, that in some cases are not seen because they are smaller than the size of the points. The dotted line shows the Kurucz model for the stellar photosphere. In dashed and solid lines are the SED models for these periods of time, respectively.}
\label{f:hbc722}
\end{figure}

\begin{figure}
\epsscale{.65}
\plotone{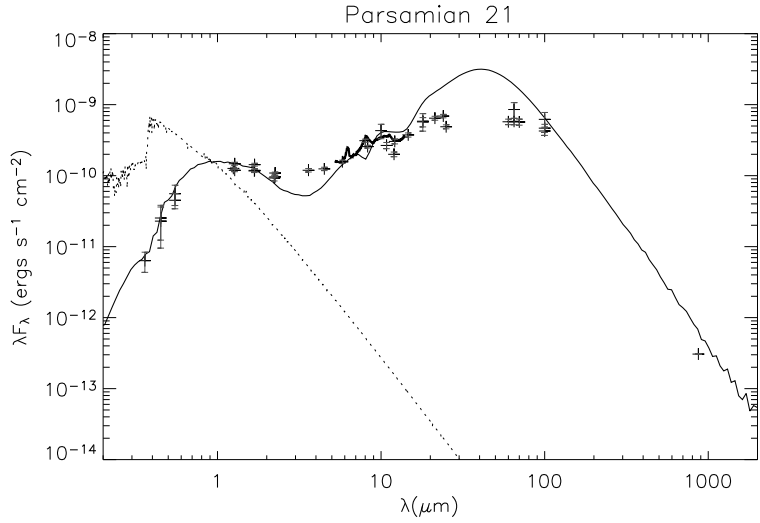}
\caption{Same as Figure~\ref{f:v1735cyg} but for Par 21.
We include the Spitzer $5-14\,\mu$m spectrum from \citet{quanz2007b}.}
\label{f:par21}
\end{figure}

\begin{figure}
\epsscale{.65}
\plotone{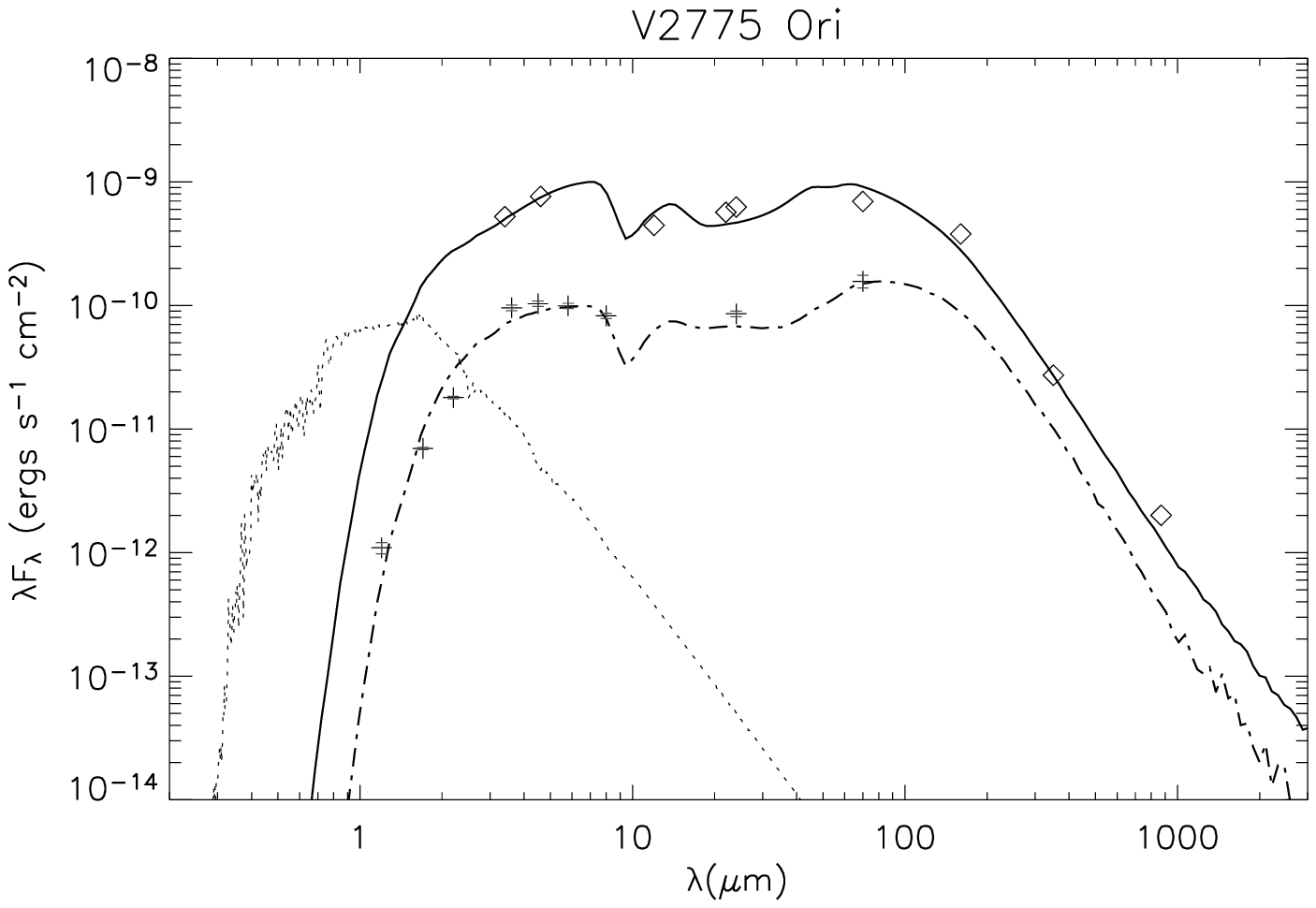}
\caption{SED and model for V2775~Ori. Crosses are the observed fluxes before the outburst, and diamonds during the outburst.  
The bars represent the uncertainties for the fluxes, that in some cases are not seen because they are smaller than the size of the points. The dotted line shows the Kurucz model for the stellar photosphere. In dashed and solid lines are the SED models for these periods of time, respectively.}
\label{f:v2775}
\end{figure}

Initially we modeled each source applying the grid of \citet{robitaille2006}, to later use these models as starting points for a more refined analysis using the code of \citet{whitney2003b}.
The parameters corresponding to the best model fits are given in Tables \ref{t:resclass2} and \ref{t:resclass1} for class~II and class~I FUORS, respectively. Figures \ref{f:v1515cyg} to \ref{f:v2775} show the corresponding SEDs.
For sources V1515~Cyg, BBW~76, PP~13S, V1647~Ori, FU~Ori, V1057~Cyg, Z~CMa, L1551~IRS5, ISO-Cha~192, V2492~Cyg, V1331~Cyg, Par~21, and V2775~Ori we compared our parameters values with those derived by other authors, finding in general a good agreement. For the remaining 11 sources, this is the first time a model of their SED is derived.

\begin{figure}[htb]
\includegraphics[width=0.5\textwidth]{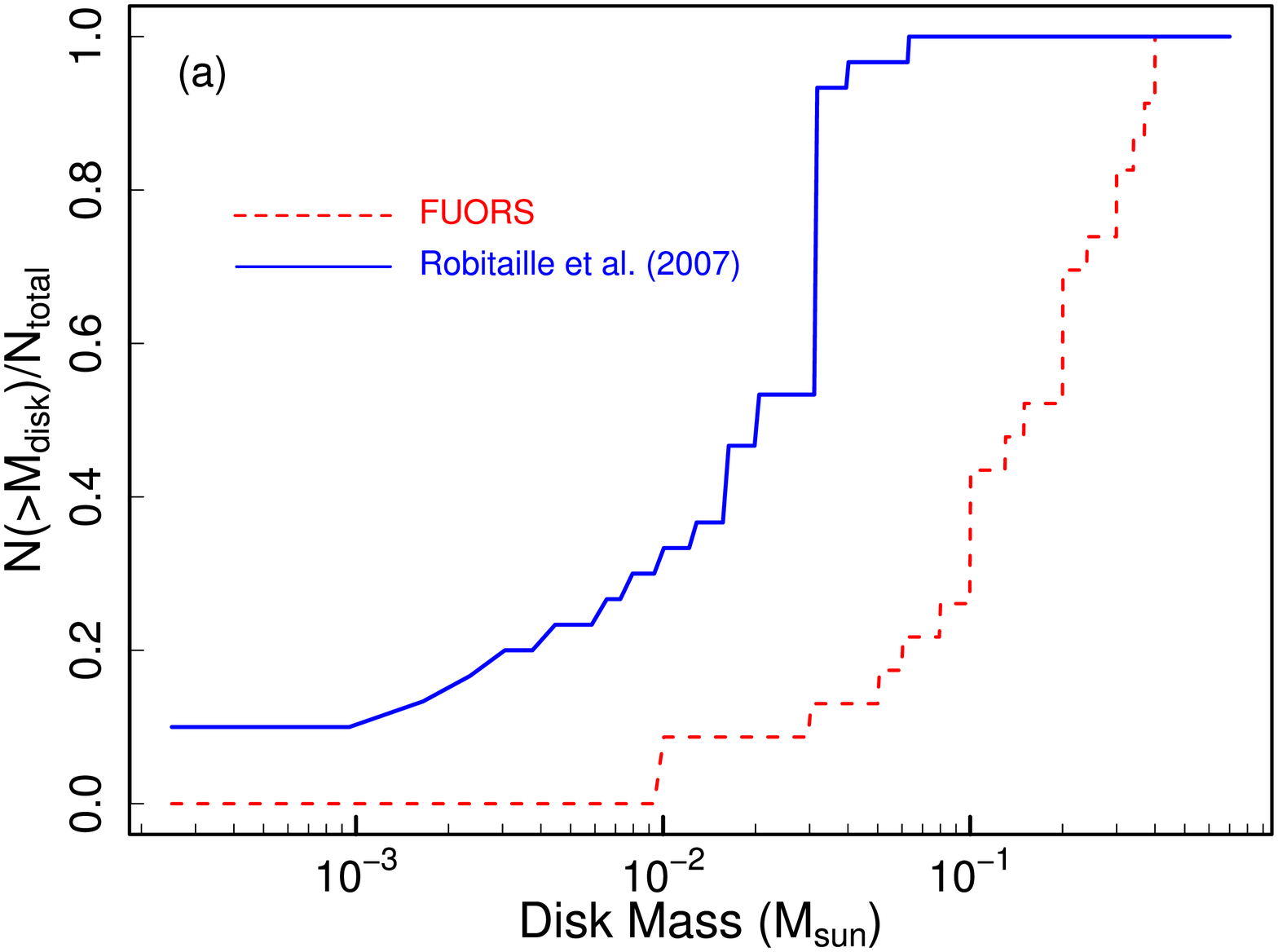}
\includegraphics[width=0.5\textwidth]{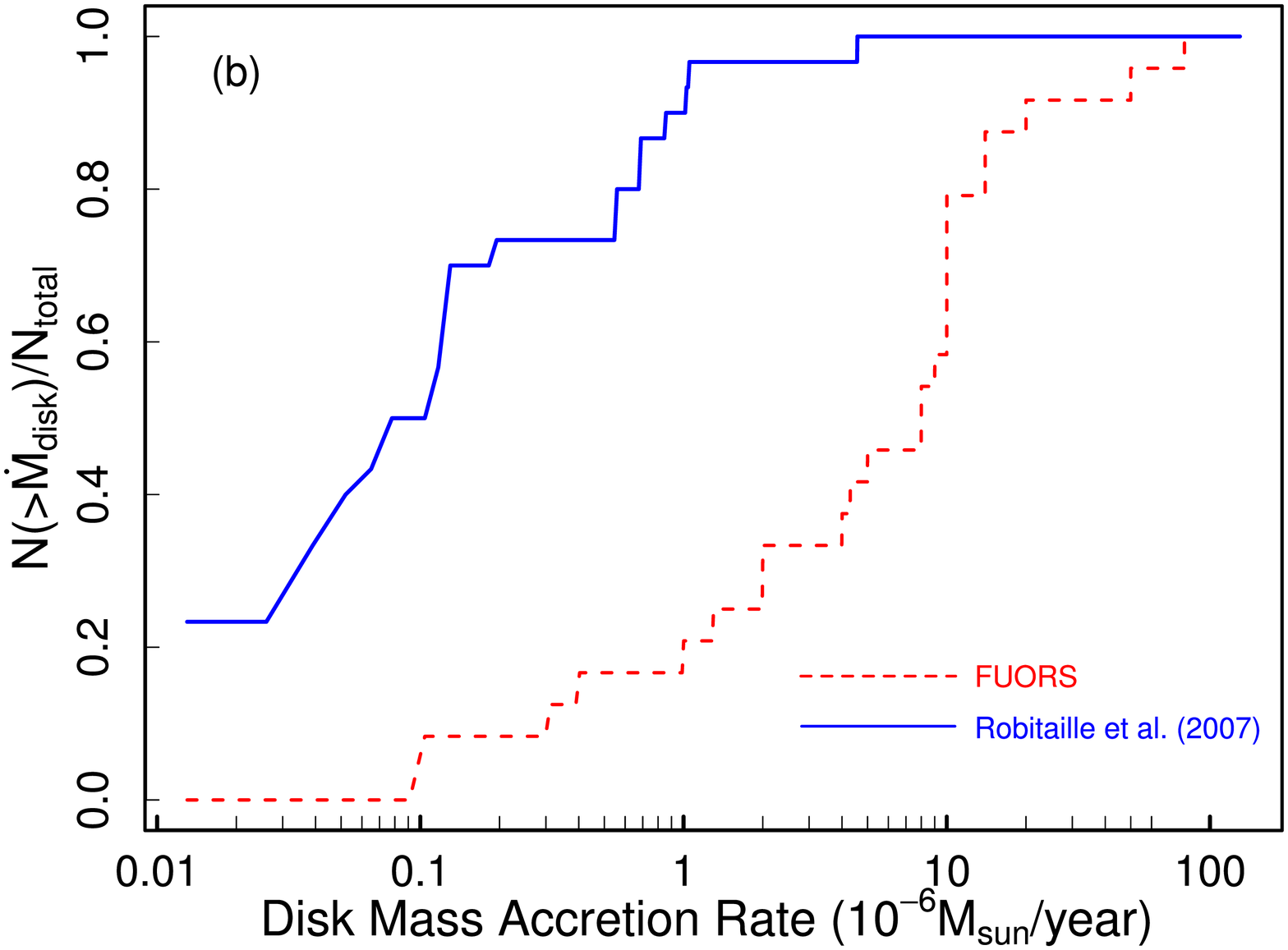}
\includegraphics[width=0.5\textwidth]{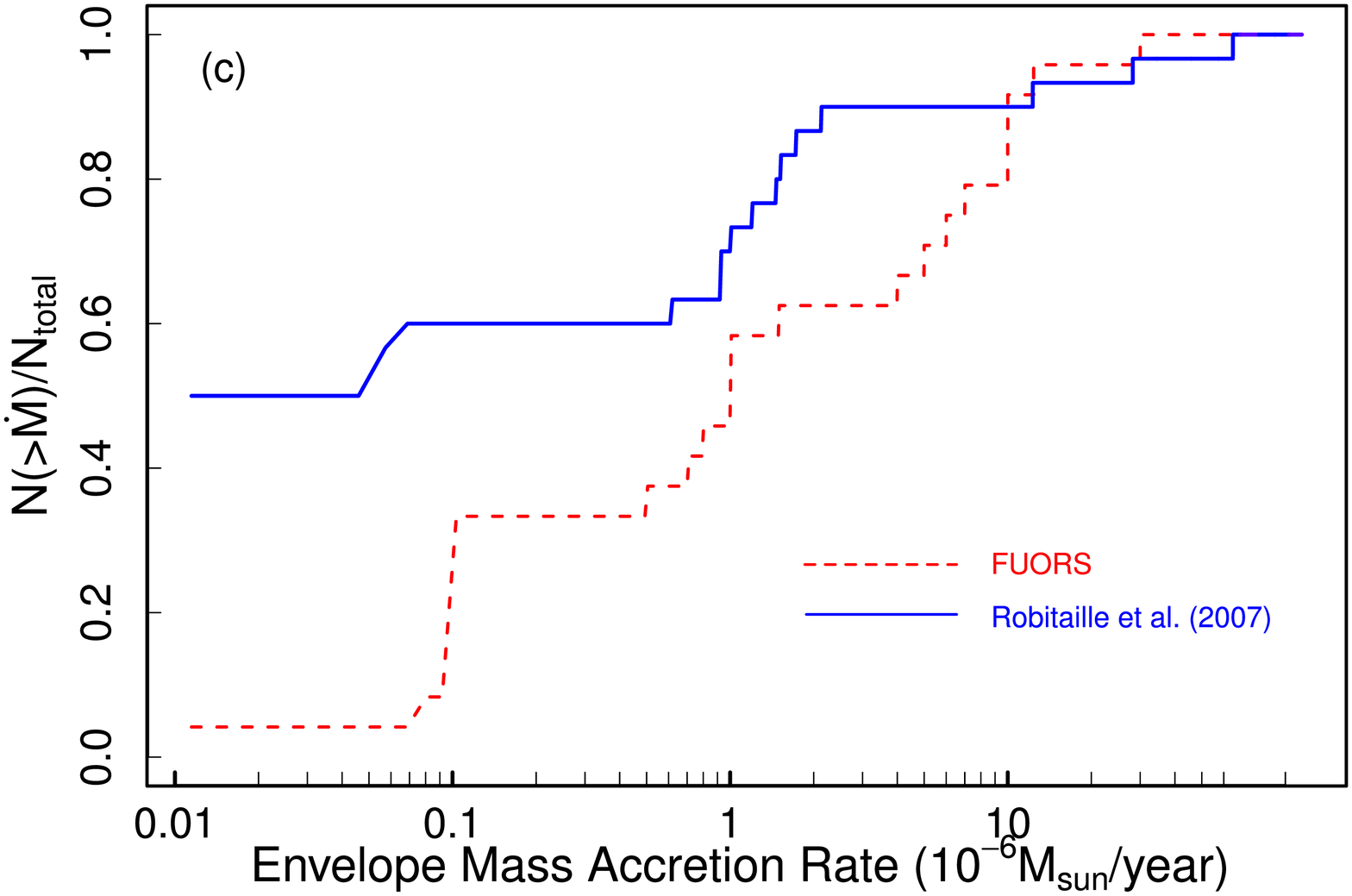}
\includegraphics[width=0.5\textwidth]{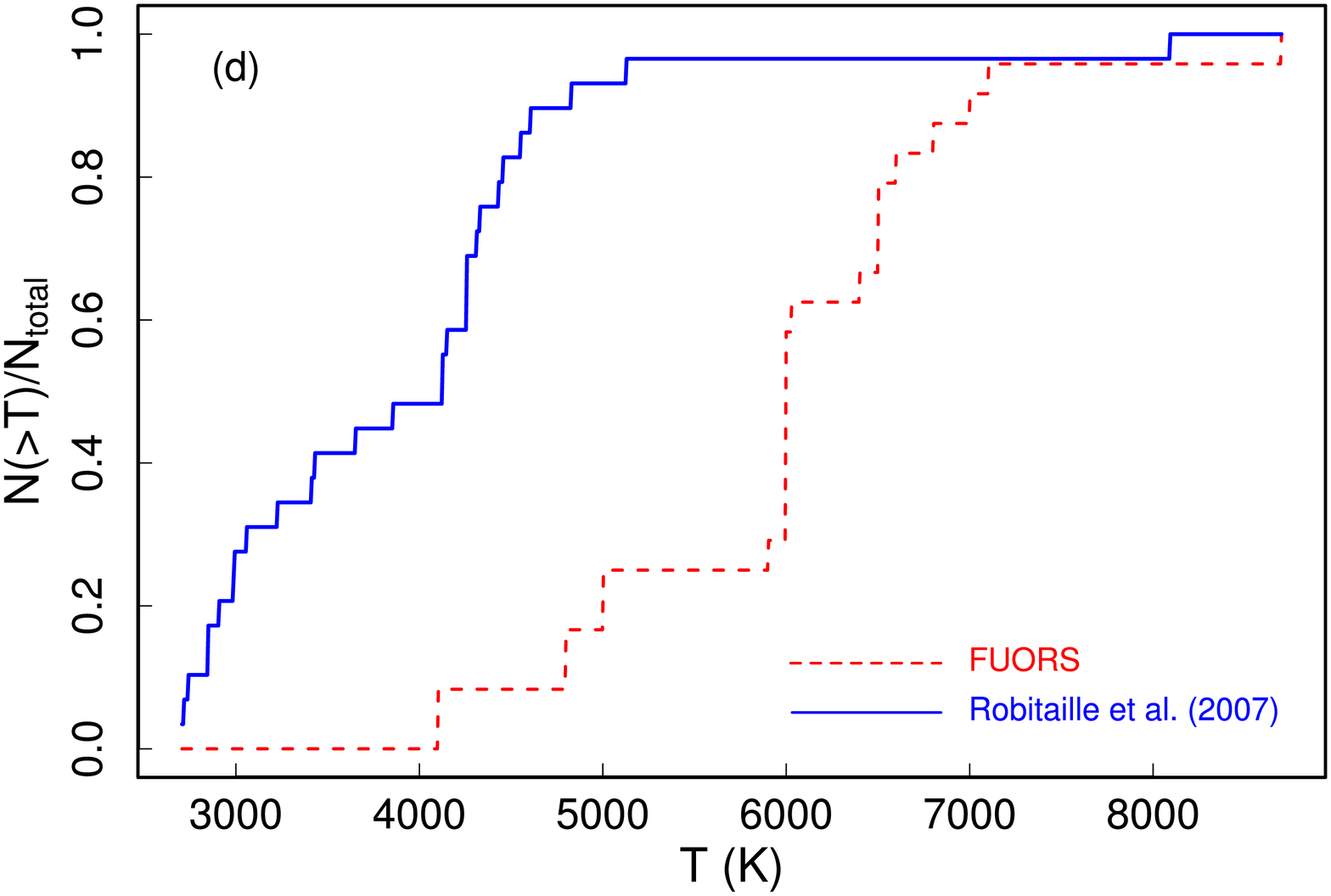}
\caption{Cumulative distribution for the disk mass (panel~$a$), the disk mass accretion rate (panel~$b$), the envelope mass accretion rate (panel~$c$), and the stellar temperature (panel~$d$), for FUORS analyzed in this work (red dashed line) and standard class~I and class~II objects modeled by \citeauthor{robitaille2007} (\citeyear{robitaille2007}, blue solid line).}
\label{f:cumul}
\end{figure}

Figure \ref{f:cumul} shows the accumulative distribution functions of disk masses, disk accretion rates, envelope accretion rates and stellar temperatures of FUORS in our sample and standard class~II and class~I objects in a quiescence state from \citet{robitaille2007}. 
Table \ref{t:median} gives the median values for both groups. The comparison shows that:

\begin{enumerate}

\item
On average FUORS disks are more massive than standard class~II and class~I objects disks. 
About 80\% of FUORS disks have masses $> 0.1\,\msun$, while standard class~II and class~I objects have disk masses $< 0.06\,\msun$.
\item
Disks mass accretion rates are higher for FUORS than for classical YSOS. 
The great majority of FUORS ($\sim90\%$) have $\dot{\rm M}_{disk}>10^{-6}$\,M$_{\odot}$/yr, while $\sim95\%$ of the standard class~II and class~I objects have $\dot{\rm M}_{disk}<10^{-6}$\,M$_{\odot}$/yr. Median disks accretion rates are $\sim 10^{-5}$\,M$_{\odot}$/yr vs $\sim 10^{-7}$\,M$_{\odot}$/yr for FUORS and classical YSOs, respectively. 
\item
The distributions of envelope accretion rates for class I FUORS and standard class I objects are indistinguisable. 
Most FUORS ($\sim70\%$) have envelope accretion rates $> 10^{-7}\,\msun$/yr.
Median envelope accretion rates are $\sim 10^{-6}$\,M$_{\odot}$/yr vs $\sim 10^{-8}$\,M$_{\odot}$/yr for
FUORS and standard YSOs, respectively.
\item
The distribution of stellar temperatures for FUORS and classical YSOs are similar in shape, but the FUORS are shifted $\sim2000\,$K to higher temperatures.

\end{enumerate}

The cumulative distributions for confirmed and candidate FUORS (see Table~\ref{t:sample}) show no significant differences, suggesting that most candidate objects, in fact, belong to the FUORS class. We caution, however, on the small number of objects in each class (14 confirmed and 10 candidate FUORS).

 For the seven objects in our sample, for which we have SEDs both
in the outburst and in the quiescence stage (see Tables~\ref{t:resclass2} and \ref{t:resclass1}), 
2 class II (RNO 1B, and V1647 Ori), and 5 class I (OO Ser, V1313 Cyg, V2492 Cyg, HBC 722, and V2775 Ori),
we note that while the disk and stellar parameters show variations,
the envelope parameters ($\dot{M}$, R$ _{\rm max}$, and M$_{env}$) do not change,
suggesting the outbursts are triggered by an instability after a long build-up phase.

The current scenario of FUORS events states that the circumstellar disk of a YSO builds up material injected from the envelope until it becomes thermally \citep{frank1992, bell1994, hartmann1996a} and/or gravitationally \citep{zhu2009,zhu2010,vorobyov2005,vorobyov2006,vorobyov2010}
unstable. In particular, using the model parameters and disk properties listed in Tables~\ref{t:resclass2} and \ref{t:resclass1} we calculated the Toomre $ Q $ gravitational stability parameter \citep{Toomre1964}.  If $Q < 1$, the disk is unstable.
Most of the models are unstable for $R > 5-20$\,AU. The only exception is L~1551~IRS~5, which is gravitationally unstable at a larger scale ($R \gtrsim 50$\,AU). Nevertheless all disk are unstable well inside the centrifugal radius (see Table \ref{t:averages}).
Consequently, gravitational instabilities may contribute to the outburst eruptions, in addition to thermal instabilities, resulting in an increase of the mass accretion onto the central object.  What we have described so far agrees with this picture. However, the disk mass accretion rate $\dot{\rm M}_{disk}\sim 10^{-5}\,\msun$/yr we obtain (see Table \ref{t:median}) is one order of magnitude lower than the $\sim 10^{-4}\,\msun$/yr predicted by the theory \citep{frank1992,hartmann1996a}.
Nevertheless, previous models of individual FUORS objects obtain $\dot{\rm M}_{disk}$
values consistent with those presented in this work (see, e.g., \citealt{pfalzner2008, aspin2008}). 

Although the average values for the parameters for both groups of FUORS are similar to those theoretically expected, the individual values listed for each object in Tables \ref{t:resclass2} and \ref{t:resclass1} differ significantly. This can be in part attributed to the fact that the group of the FU Orionis stars itself is not an homogeneous sample.
While they all share a particular set of characteristics, those appear in different ways for each object. For instance, while all FUORS show a sudden brightness increase of several magnitudes, followed by a slow decrease to their previous state, the way the brightness jump develops in time is different for each object. A clear example of this is the great diversity in the light curves of the three prototypes of the class \citep{hartmann1996a}. It is therefore reasonable to expect that the values of individual parameters of each member of the group will ultimately differ.

Lastly, we would like to draw attention to three sources in particular. V1647~Ori is a special case on the FUOR sample since it has been well studied before and after the outburst, having SEDs for both epochs. This makes that source a prime candidate for the study of the FU Orionis event, though it has to be approached with care, since its classification as a FUOR or EXOR is still under debate \citep{aspin2006,aspin2011b,semkov2012a}.
The other two particular sources are V2492~Cyg and HBC~722. These objects are, at the moment of writing, the last two for which an FUORS-like outburst has been observed. They show characteristics proper of bona-fide FUORS, as shown in Sections~\ref{sec-v2492cyg} and \ref{sec:hbc722}, however its inclusion in the FUORS class is still not certain. Nevertheless, our SED modeling shows behaviors similar to V1647~Ori, the former newest member of the class.

V1647~Ori and V2492~Cyg also show EXOR characteristics, and from our modeling we see that they do not show a large variation between the outburst and the quiescent phases. For example, their disk masses do not change with the outburst (see Tables~\ref{t:resclass2} and \ref{t:resclass1}), and the variation of the disk mass accretion rate is lower than for other FUORS.
Nevertheless, when compared with other FUORS, the parameter values derived for those sources are still within the range established by the rest of the FUORS sample, and could then be considered FUORS. However, if we had just analyzed only those two sources while taking into consideration that EXOR outbursts are thought to be ``scaled-down'' versions of FUORS outbursts, it is very likely that they would have been considered EXORS.
This shows the uncertainty and difficulty of disentangle the two types of outburst episodes.

Despite sharing common properties, each FUORS or FUORS candidate has its own peculiarities that are not currently well understood. It is therefore of great interest to study the most extreme objects of the class to reach a full understanding of this period of great activity in circumstellar disks.

The work we have presented here is the first compilation of SEDs of the currently known FUORS.
Of the 26 currently known FUORS, 2 do not have enough observations as to construct the SEDs, and thus are not analyzed. For 21 of the remaining 24 we compile in one place the observations taken in all wavelengths, producing the most complete SEDs possible so far. For 3 FUORS (AR 6A, AR 6B, V2492 Cyg), SEDs for only a limited range ($\lambda$ $<$ 20 $\mu$m) were constructed and thus values for the derived parameters are not fully determined.  Finally, for 11 of the 24 FUORS analyzed (V1735 Cyg, V883 Ori, RNO 1B, RNO 1C, AR 6A, AR 6B, V900~Mon,  V346~Nor, OO~Ser, RE 50 N IRS 1 and HBC~722) we provide for the first time a complete SED modeling to determine the physical and geometrical parameters of the star$+$disk$+$envelope system.
Furthermore, this is the first time all the known FUORS with an observed SED are modeled with the same code at the same time, providing an homogeneous set of results. The data we present here will be of great help for future studies in the field.

\acknowledgments
We thank Dr. Barbara A. Whitney for thoroughly reading the manuscript and for 
providing comments and suggestions that greatly improved and clarified the content of the paper.
We are also grateful to Dr. Sascha P. Quanz for sending us Spizer-IRS and ISO-SWS spectra 
and to Dr. Joel D. Green for providing guide to access to Spizer-IRS spectra. We appreciate
the careful revision of the paper done by the referee as well as his/her suggestions that improved
the content and the presentation of this work.

\clearpage

\clearpage

\begin{deluxetable}{lcccclclcc}
\tabletypesize{\scriptsize}
\rotate
\tablecaption{The FU Orionis sample \label{t:sample}}
\tablewidth{0pt}
\tablehead{
\colhead{Source} & \colhead{L($L_{\odot}$)} & \colhead{A$_{\rm V}$(mag)} & \colhead{$\Delta$K} & \colhead{Outburst}& \colhead{CO at 2.3 $\mu$m} & 
\colhead{Outflow/Jet} & \colhead{Spec.Type} & \colhead{d(pc)} & \colhead{refs.} 	
}%
\startdata
\multicolumn{10}{c}{Confirmed FUORS}\\ 
\hline
FU Ori$^{\flat\dagger}$ & $340-466$ & 1.5 &  \ldots & 1937 & absorption & no & G0 & 500 & 1,2,3,4,5,6,7,8,9\\ 
V1515 Cyg$^{\ddagger}$ & $177-200$ & 3.2 &  \ldots& 1950 & absorption & yes? & G2-G5 & 1000 & 1,2,3,7,8,10,11,12,13\\ 
\multirow{2}{*}{V1057 Cyg$^{\dagger}$} & \multirow{2}{*}{$170-370$} & \multirow{2}{*}{$3.0-3.7$} & \multirow{2}{*}{\ldots} & \multirow{2}{*}{1970} & \multirow{2}{*}{absorption} & \multirow{2}{*}{yes} & F5 II/G2 Ib II & \multirow{2}{*}{600} & \multirow{2}{*}{1,2,3,6,8,12,14}\\ 
& & & & & & & F7/G3 I/II & & \\ 
Z CMa$^{\flat\dagger}$ & 420 & 2.8 & \ldots& 2008 & absorption & yes & F5 & 1700 & 1,2,3,11,15,16,17,18,19,20,21\\ 
BBW 76$^{\ddagger}$ & $287-550$ & 2.2 & \ldots& $\sim 1930$ &  \ldots& no & \ldots& 1800 & 2,3,7,8\\ 
V1735 Cyg$^{\ddagger}$ & 250 & 10 & \ldots& $1957-1965$ & \ldots& yes & \ldots& 900 & 1,2,3,15,20,22,24,25\\ 
V883 Ori$^{\ddagger}$ & 400 & \ldots& \ldots& \ldots& absorption & no & \ldots& 460 & 2,3\\ 
RNO 1B$^{\ddagger}$ & 440 & $\sim$9 & \ldots& \ldots& strong absorption & yes? & F8 II & 850 & 3,5,18,26,28,29\\ 
RNO 1C$^{\ddagger}$ & 540 & $\sim$9 & \ldots& \ldots& strong absorption & yes? & \ldots& 850 & 1,18,26,27,28,29\\ 
AR 6A$^{\flat\dagger}$ & 450 & 18 & \ldots& \ldots& \ldots& yes? & G III & 800 & 5,30\\ 
AR 6B$^{\dagger}$ & 450 & 18 & \ldots& \ldots& \ldots& yes? & \ldots& 800 & 2,5,30\\ 
PP 13S$^{\ddagger}$ & 30 & $30-50$ & $\sim$1 & $<1900$ & strong absorption & yes & \ldots& 350 & 2,3,31,32,33,34,35\\ 
\multirow{2}{*}{L1551 IRS5$^{\flat\dagger}$} & \multirow{2}{*}{$\sim30$} & \multirow{2}{*}{$\sim20$} & \multirow{2}{*}{\ldots} & \multirow{2}{*}{\ldots} & \multirow{2}{*}{absorption} & \multirow{2}{*}{yes} & \multirow{2}{*}{K3 V/M3 III} & \multirow{2}{*}{140} & 3,36,37,38,39,40,42\\ 
 & & & & & & & & & 43,44,45,46,48,49,50\\ 
 V900 Mon$^{\dagger}$ & 106 & 13 & \ldots & 1953-ongoing & absorption & yes? & \ldots & 1100 & 98,99,100\\
 \hline
\multicolumn{10}{c}{Candidate FUORS}\\
\hline
V2775 Ori$^{\dagger}$ & \ldots & 18 & 3.8 & 2005/2007 & \ldots & yes & M? & 420 & 101,102,103,104,105\\
ISO-Cha I 192$^{\dagger}$ & 1.5 & 17 & $\sim 2$ & \ldots& \ldots& yes & M3.5-M6.5 & \ldots& 52,53,54,55,56\\ 
V346 Nor$^{\dagger\flat}$ & 135 & 2.7 & \ldots& $\sim1984$ & \ldots& yes & \ldots& 700 & 1,3,51,57,58,59,60\\ 
\multirow{2}{*}{V1331 Cyg$^{\dagger}$} & \multirow{2}{*}{53/60} & \multirow{2}{*}{2.4} & \multirow{2}{*}{\ldots} & \multirow{2}{*}{\ldots} & \multirow{2}{*}{variable emission} & \multirow{2}{*}{\ldots} & \multirow{2}{*}{F0/F4-G5} & \multirow{2}{*}{$550-700$} & 3,23,61,62,63\\ 
 & & & & & & & & & 64,65,66,67,68\\ 
OO Ser$^{\dagger}$ & 15 & \ldots& 4.6 & 1995 & \ldots& no & \ldots& 311 & 70,71\\ 
Re 50 N IRS1$^{\dagger}$ & 50 & $\sim$30 & \ldots& $1960-1970$ & \ldots& yes & \ldots& 460 & 3,57,72,73,74,75\\ 
\multirow{2}{*}{V1647 Ori$^{\ddagger}$} & \multirow{2}{*}{$34-90$} & \multirow{2}{*}{$\sim$10} & \multirow{2}{*}{$\sim$3} & \multirow{2}{*}{2004/2008} & \multirow{2}{*}{\ldots} & \multirow{2}{*}{no} & \multirow{2}{*}{\ldots} & \multirow{2}{*}{400} & 69,76,77,78,79,80\\ 
 & & & & & & & & & 81,82,83,84,85\\ 
HBC 722$^{\dagger}$ & $8.7-12$ & 3.4 & \ldots& 2010 & \ldots& yes? & K0-M7 & 520 & 86,87,88,89,90,91\\ 
Par 21$^{\dagger}$ & 117 & 1.6 & \ldots& \ldots& \ldots& yes & A5e & $400-1800$ & 3,92,93,94\\ 
V2492 Cyg$^{\dagger}$ & $1-6$ & $2-7$ & \ldots& 2010 & \ldots& yes & F-G II / M I & 550 & 95,96,97\\
\enddata
\tablecomments{$^{\dagger}$source classified as a class~I YSO. $^{\ddagger}$source classified as a class~II YSO. $^{\flat}$binary star.}
\tablerefs{
(1) \citet{evans1994};  
(2) \citet{hartmann1996a};     
(3) \citet{sandell2001};    
(4) \citet{kenyon2000};        
(5) \citet{aspin2003};        
(6) \citet{malbet2005};       
(7) \citet{green2006};      
(8) \citet{zhu2008};          
(9) \citet{kenyon2000};
(10) \citet{goodrich1987};  
(11) \citet{terranegra1994}; 
(12) \citet{herbig1977}; 
(13) \citet{kolotilov1983};
(14) \citet{herbig2006};
(15) \citet{lorenzetti2001};   
(16) \citet{herbst1978};  
(17) \citet{quanz2006}; 
(18) \citet{polomski2005};  
(19) \citet{kenyon1989};
(20) \citet{grankin2009}
(21) \citet{hartmann1989a}
(22) \citet{sato1992};
(23) \citet{levreault1988b};
(24) \citet{connelley2007};  
(25) \citet{harvey2008};
(26) \citet{staude1991};    
(27) \citet{kenyon1993b};   
(28) \citet{mcmuldroch1995};    
(29) \citet{greene1996};  
(30) \citet{moriarty-schieven2008}; 
(31) \citet{cohen1983b}; 
(32) \citet{tapia1997};  
(33) \citet{sandell1998}; 
(34) \citet{aspin2000}; 
(35) \citet{aspin2001};    
(36) \citet{strom1976};      
(37) \citet{snell1980};     
(38) \citet{cohen1984};    
(39) \citet{snell1985};  
(40) \citet{doppmann2005};
(41) \citet{prato2009};
(42) \citet{mundt1985};   
(43) \citet{carr1987a}; 
(44) \citet{adams1987};   
(45) \citet{carr1990};   
(46) \citet{davis1995b}; 
(47) \citet{devine1999b}; 
(48) \citet{rodriguez2003};    
(49) \citet{osorio2003}; 
(50) \citet{rodriguez1998};     
(51) \citet{pfalzner2008};   
(52) \citet{mattila1989};  
(53) \citet{persi1999};  
(54) \citet{gomez2003b};     
(55) \citet{gomez2004};
(56) \citet{persi2007};  
(57) \citet{strom1993};   
(58) \citet{prusti1993}; 
(59) \citet{gredel1994}; 
(60) \citet{chavarria1981}; 
(61) \citet{reipurth1997a};
(62) \citet{carr1989};   
(63) \citet{mcmuldroch1993};    
(64) \citet{biscaya1997};    
(65) \citet{mundt1998};     
(66) \citet{henning1998};    
(67) \citet{lorenzetti2000};   
(68) \citet{hamann1992};
(69) \citet{abraham2004a};  
(70) \citet{delara1991};   
(71) \citet{hodapp1996};  
(72) \citet{heyer1990};  
(73) \citet{reipurth1997b};   
(74) \citet{stanke2000};  
(75) \citet{lee2002};    
(76) \citet{andrews2004};     
(77) \citet{briceno2004};    
(78) \citet{mcgehee2004};    
(79) \citet{reipurth2004a};     
(80) \citet{aspin2011b};     
(81) \citet{vacca2004};   
(82) \citet{walter2004}; 
(83) \citet{muzerolle2005};    
(84) \citet{acosta-pulido2007};    
(85) \citet{lis1999};
(86) \citet{miller2011};
(87) \citet{semkov2010};
(88) \citet{laugalys2006};
(89) \citet{cohen1979};
(90) \citet{green2011};
(91) \citet{dunham2012};
(92) \citet{allen2004};
(93) \citet{liu2011b};
(94) \citet{staude1992};
(95) \citet{covey2011};
(96) \citet{straizys1989};
(97) \citet{aspin2011a};
(98) \citet{reipurth2012}
(99) \citet{gregorio-hetem2008};
(100) \citet{lombardi2011};
(101) \citet{carattiogaratti2011};
(102) \citet{fischer2012};
(103) \citet{sandstrom2007};
(104) \citet{menten2007};
(105) \citet{kim2008}.
}

\end{deluxetable}

\clearpage

\begin{deluxetable}{lcc|lcc|lcc|lcc|lcc}
\rotate
\tabletypesize{\scriptsize}
\tablecolumns{12}
\tablewidth{0pt}
\tablecaption{Fluxes F$_{\nu}$(Jy) used to construct the SED of each source \label{t:fluxes}}
\tablehead{%
	\colhead{$\lambda$ ($\mu$m)} & \colhead{V1057 Cyg} & \colhead{ref} & \colhead{$\lambda$ ($\mu$m)} & \colhead{V2492 Cyg} & \colhead{ref} & \colhead{$\lambda$ ($\mu$m)} & \colhead{Z CMa} & \colhead{ref} & \colhead{$\lambda$ ($\mu$m)} & \colhead{V1331 Cyg} & \colhead{ref} & \colhead{$\lambda$ ($\mu$m)} & \colhead{L 1551 IRS5} & \colhead{ref} } %
\startdata
0.36 & 0.02 & 73 & 0.55 & $7.67\times10^{-4}$ & 17 & 0.36 & 0.04 & 28 & 0.34 & $1.90\times10^{-3}$ & 9 & 0.90 & $3.00\times10^{-4}$ & 75 \\
0.36 & $5.00\times10^{-4}$ & 33 & 0.55 & $1.08\times10^{-4}$ & 17 & 0.44 & 0.20 & 28 & 0.35 & $7.10\times10^{-3}$ & 9 & 1.20 & 0.02 & 76 \\
0.44 & 0.01 & 33 & 0.70 & $2.60\times10^{-3}$ & 17 & 0.55 & 0.57 & 28 & 0.36 & $6.10\times10^{-3}$ & 23 & 1.20 & 0.01 & 13 \\
0.44 & 0.27 & 73 & 0.70 & $4.19\times10^{-4}$ & 17 & 0.70 & 0.96 & 28 & 0.36 & $8.50\times10^{-3}$ & 10 & 1.23 & 0.01 & 13 \\
0.44 & 0.23 & 49 & 0.90 & $8.12\times10^{-3}$ & 17 & 0.90 & 1.62 & 28 & 0.38 & 0.01 & 9 & 1.60 & $7.90\times10^{-3}$ & 76 \\
0.55 & 0.03 & 33 & 0.90 & $1.64\times10^{-3}$ & 17 & 1.24 & 3.85 & 68 & 0.40 & 0.02 & 9 & 1.60 & $1.70\times10^{-3}$ & 13 \\
0.55 & 0.53 & 73 & 1.25 & $3.00\times10^{-5}$ & 89 & 1.25 & 6.08 & 37 & 0.44 & 0.03 & 10 & 1.63 & 0.02 & 57 \\
0.55 & 0.62 & 49 & 1.25 & 0.05 & 17 & 1.65 & 10.10 & 37 & 0.45 & 0.02 & 23 & 1.65 & $2.60\times10^{-3}$ & 50 \\
0.70 & 0.13 & 33 & 1.25 & 0.01 & 17 & 1.66 & 8.39 & 68 & 0.46 & 0.03 & 9 & 1.66 & 0.04 & 13 \\
1.25 & 1.91 & 37 & 1.65 & $2.50\times10^{-4}$ & 89 & 2.16 & 20.80 & 68 & 0.52 & 0.06 & 9 & 2.19 & 0.07 & 57 \\
1.65 & 2.81 & 37 & 1.65 & 0.08 & 17 & 2.20 & 19.00 & 37 & 0.55 & 0.06 & 10 & 2.20 & 0.18 & 76 \\
2.20 & 3.16 & 37 & 1.65 & 0.02 & 17 & 3.40 & 51.50 & 37 & 0.55 & 0.04 & 23 & 2.20 & 0.12 & 13 \\
3.50 & 3.74 & 37 & 2.20 & $3.80\times10^{-3}$ & 89 & 4.29 & 71.80 & 21 & 0.58 & 0.05 & 9 & 2.20 & 0.05 & 30 \\
4.63 & 6.12 & 37 & 2.20 & 0.16 & 17 & 4.35 & 72.10 & 21 & 0.64 & 0.11 & 9 & 2.22 & 0.11 & 13 \\
4.80 & 2.67 & 2 & 2.20 & 0.07 & 17 & 4.63 & 82.30 & 37 & 0.64 & 0.05 & 23 & 2.23 & $9.50\times10^{-3}$ & 50 \\
5.00 & 7.42 & 73 & 3.60 & 0.07 & 87 & 4.80 & 53.20 & 2 & 0.70 & 0.12 & 10 & 3.45 & 0.32 & 13 \\
10.20 & 8.48 & 37 & 3.60 & 0.08 & 25 & 8.28 & 828.00 & 21 & 0.79 & 0.06 & 23 & 3.50 & 0.40 & 76 \\
11.00 & 27.00 & 73 & 4.50 & 0.16 & 87 & 10.20 & 136.00 & 37 & 0.90 & 0.19 & 10 & 3.50 & 0.31 & 13 \\
12.00 & 14.90 & 1 & 4.50 & 0.23 & 25 & 11.60 & 184.00 & 72 & 1.25 & 0.37 & 10 & 3.75 & 0.71 & 50 \\
12.00 & 5.68 & 1 & 5.60 & 0.65 & 87 & 12.00 & 97.30 & 2 & 1.30 & 0.16 & 27 & 4.63 & 1.56 & 13 \\
20.00 & 96.10 & 73 & 5.80 & 0.61 & 25 & 12.00 & 135.00 & 37 & 1.65 & 0.18 & 10 & 4.80 & 0.94 & 76 \\
11.00 & 18.60 & 37 & 8.00 & 1.17 & 25 & 12.00 & 127.00 & 72 & 2.20 & 0.28 & 10 & 4.80 & 0.50 & 6 \\
25.00 & 28.70 & 1 & 8.00 & 1.20 & 87 & 14.65 & 159.00 & 21 & 3.40 & 0.41 & 10 & 8.40 & 4.28 & 13 \\
25.00 & 23.20 & 1 & 8.28 & 1.61 & 21 & 21.34 & 203.00 & 21 & 5.00 & 0.65 & 10 & 9.60 & 1.20 & 13 \\
60.00 & 53.00 & 1 & 9.00 & 1.96 & 84 & 25.00 & 215.00 & 2 & 8.00 & 0.65 & 21 & 10.00 & 2.80 & 6 \\
60.00 & 53.70 & 1 & 9.00 & 1.96 & 34 & 25.00 & 183.00 & 37 & 10.20 & 1.30 & 10 & 10.20 & 3.81 & 13 \\
65.00 & 42.30 & 1 & 12.00 & 3.39 & 86 & 60.00 & 290.00 & 2 & 12.00 & 1.12 & 2 & 10.50 & 2.30 & 6 \\
100.00 & 62.10 & 1 & 12.13 & 2.49 & 21 & 60.00 & 312.00 & 37 & 12.13 & 1.32 & 21 & 11.00 & 3.28 & 13 \\
100.00 & 47.00 & 1 & 14.65 & 3.05 & 20 & 100.00 & 369.00 & 2 & 14.65 & 1.20 & 21 & 12.50 & 5.91 & 13 \\
100.00 & 34.50 & 1 & 15.00 & 3.88 & 34 & 100.00 & 479.00 & 37 & 20.00 & 2.61 & 10 & 12.80 & 6.90 & 6 \\
350.00 & 4.92 & 80 & 18.00 & 3.88 & 84 & 350.00 & 28.80 & 19 & 21.34 & 2.23 & 21 & 18.00 & 20.00 & 6 \\
800.00 & 0.45 & 80 & 21.34 & 3.14 & 20 & 450.00 & 13.80 & 19 & 25.00 & 2.62 & 2 & 19.30 & 32.80 & 13 \\
\tablehead{ %
	\colhead{$\lambda$ ($\mu$m)} & \colhead{OO Ser} & \colhead{ref} & \colhead{$\lambda$ ($\mu$m)} & \colhead{V2492 Cyg} & \colhead{ref} & \colhead{$\lambda$ ($\mu$m)} & \colhead{Z CMa} & \colhead{ref} & \colhead{$\lambda$ ($\mu$m)} & \colhead{PP 13S} & \colhead{ref} & \colhead{$\lambda$ ($\mu$m)} & \colhead{L 1551 IRS5} & \colhead{ref} }
800.00 & 0.43 & 80 & 24.00 & 3.40 & 88 & 800.00 & 1.96 & 19 & 60.00 & 6.88 & 2 & 20.00 & 37.00 & 6 \\
850.00 & 0.27 & 71 & 24.00 & 2.22 & 63 & 1100.00 & 0.71 & 19 & 100.00 & 8.22 & 2 & 25.00 & 36.00 & 6 \\
1300.00 & 0.12 & 80 & 25.00 & 6.59 & 86 & 1300.00 & 0.60 & 72 & 850.00 & 0.51 & 71 & 37.00 & 220.00 & 18 \\
\cline{1-3}\cline{10-12}
 & OO Ser &  & 60.00 & 27.90 & 86 & 1300.00 & 0.45 & 27 &  & PP 13S &  & 40.00 & 200.00 & 14 \\
\cline{1-3}\cline{10-12}
1.24 & $1.00\times10^{-4}$ & 68 & 70.00 & 5.88 & 88 &  & RNO 1B &  & 1.24 & $9.80\times10^{-4}$ & 68 & 47.00 & 320.00 & 14 \\
1.25 & $8.00\times10^{-5}$ & 22 & 70.00 & 6.73 & 63 & 1.24 & 0.03 & 68 & 1.24 & $9.46\times10^{-4}$ & 70 & 47.00 & 390.00 & 14 \\
1.60 & $1.00\times10^{-3}$ & 31 & 100.00 & 57.40 & 86 & 1.24 & 0.19 & 38 & 1.63 & 0.01 & 70 & 47.00 & 270.00 & 14 \\
1.65 & $1.20\times10^{-4}$ & 22 & 1100.00 & 0.15 & 85 & 1.63 & 0.44 & 38 & 1.66 & $3.90\times10^{-3}$ & 68 & 52.00 & 355.00 & 14 \\
\cline{4-6}
1.66 & $1.00\times10^{-4}$ & 68 &  & Re 50 N IRS1 &  & 1.66 & $3.00\times10^{-3}$ & 68 & 2.16 & 0.04 & 68 & 58.00 & 280.00 & 18 \\
\cline{4-6}
2.10 & $3.80\times10^{-3}$ & 31 & 7.80 & 56.00 & 55 & 2.16 & 0.16 & 68 & 2.19 & 0.13 & 70 & 60.00 & 373.00 & 15 \\
2.16 & $7.00\times10^{-4}$ & 68 & 8.70 & 20.00 & 55 & 2.19 & 0.68 & 38 & 3.77 & 0.88 & 70 & 60.00 & 373.00 & 78 \\
2.17 & $7.10\times10^{-4}$ & 22 & 9.50 & 10.00 & 55 & 3.60 & 0.29 & 62 & 5.00 & 2.48 & 70 & 63.00 & 344.00 & 16 \\
2.20 & 0.01 & 35 & 10.10 & 22.00 & 55 & 3.77 & 0.89 & 38 & 10.20 & 3.99 & 70 & 63.00 & 450.00 & 16 \\
2.20 & 0.02 & 31 & 10.30 & 11.00 & 55 & 3.80 & 0.68 & 61 & 20.00 & 9.84 & 70 & 85.00 & 750.00 & 24 \\
2.20 & $1.16\times10^{-3}$ & 42 & 11.60 & 24.00 & 55 & 4.50 & 0.39 & 62 & 20.00 & 8.89 & 70 & 95.00 & 490.00 & 14 \\
2.20 & $1.54\times10^{-3}$ & 42 & 12.00 & 24.10 & 65 & 4.67 & 0.58 & 61 & 350.00 & 16.20 & 70 & 100.00 & 470.00 & 14 \\
3.60 & $7.00\times10^{-3}$ & 42 & 12.50 & 40.00 & 55 & 5.80 & 0.56 & 62 & 450.00 & 7.89 & 70 & 100.00 & 456.00 & 15 \\
3.60 & 0.01 & 22 & 20.00 & 10.00 & 55 & 8.00 & 0.64 & 62 & 750.00 & 1.88 & 70 & 100.00 & 456.00 & 78 \\
3.80 & 0.44 & 31 & 25.00 & 64.20 & 65 & 8.28 & 2.04 & 21 & 800.00 & 1.56 & 70 & 103.00 & 512.00 & 18 \\
4.50 & 0.08 & 42 & 60.00 & 147.00 & 65 & 10.80 & 1.24 & 61 & 1100.00 & 0.64 & 70 & 150.00 & 475.00 & 24 \\
4.50 & 0.07 & 22 & 100.00 & 223.00 & 65 & 12.13 & 2.36 & 21 & 1300.00 & 0.45 & 70 & 160.00 & 390.00 & 14 \\
4.80 & 1.36 & 31 & 450.00 & 37.20 & 64 & 14.65 & 2.38 & 21 & 1300.00 & 0.24 & 58 & 168.00 & 565.00 & 18 \\
\cline{10-12}
5.80 & 0.08 & 42 & 450.00 & 5.04 & 19 & 18.00 & 161.00 & 61 &  & AR 6A &  & 190.00 & 550.00 & 18 \\
\cline{10-12}
5.80 & 0.22 & 22 & 450.00 & 2.92 & 64 & 21.34 & 8.15 & 21 & 1.24 & 0.03 & 68 & 350.00 & 100.00 & 4 \\
6.70 & 2.24 & 35 & 800.00 & 0.82 & 19 & 350.00 & 111.00 & 19 & 1.24 & 0.04 & 5 & 350.00 & 164.00 & 8 \\
8.00 & 0.57 & 42 & 850.00 & 0.50 & 64 & 450.00 & 65.70 & 71 & 1.63 & 0.19 & 5 & 377.00 & 107.00 & 60 \\
8.00 & 0.60 & 22 & 850.00 & 3.70 & 64 & 450.00 & 48.90 & 19 & 1.66 & 0.19 & 68 & 400.00 & 68.00 & 18 \\
11.70 & 6.82 & 31 & 870.00 & 783.00 & 65 & 800.00 & 6.16 & 19 & 2.16 & 0.52 & 68 & 450.00 & 94.00 & 8 \\
\tablehead{ %
	\colhead{$\lambda$ ($\mu$m)} & \colhead{Par 21} & \colhead{ref} & \colhead{$\lambda$ ($\mu$m)} & \colhead{HBC 722} & \colhead{ref} & \colhead{$\lambda$ ($\mu$m)} & \colhead{V1735 Cyg} & \colhead{ref} & \colhead{$\lambda$ ($\mu$m)} & \colhead{V883 Ori} & \colhead{ref} & \colhead{$\lambda$ ($\mu$m)} & \colhead{L1551 IRS5} & \colhead{ref} }
12.00 & 0.64 & 42 & 1100.00 & 0.40 & 19 & 450.00 & 6.60 & 71 & 2.20 & 0.43 & 5 & 730.00 & 8.40 & 43 \\
14.30 & 4.46 & 35 & 1300.00 & 262.00 & 65 & 1100.00 & 2.30 & 19 & 3.80 & 0.73 & 5 & 750.00 & 18.20 & 8 \\
\cline{4-6}
20.60 & 12.30 & 31 &  & HBC 722 &  & 1300.00 & 2.51 & 27 & 4.60 & 0.63 & 5 & 800.00 & 8.05 & 52 \\
\cline{4-6}\cline{7-9}
24.00 & 13.30 & 42 & 0.44 & $5.97\times10^{-5}$ & 25 &  & V1735 Cyg &  & 8.28 & 0.83 & 21 & 811.00 & 15.00 & 60 \\
\cline{7-9}
60.00 & 14.00 & 42 & 0.55 & $2.35\times10^{-4}$ & 25 & 1.24 & 0.18 & 68 & 12.13 & 0.91 & 21 & 850.00 & 12.10 & 8 \\
70.00 & 18.00 & 22 & 0.55 & 0.01 & 42 & 1.25 & 0.25 & 37 & 14.65 & 1.35 & 21 & 850.00 & 16.90 & 43 \\
\cline{10-12}
800.00 & 0.60 & 31 & 0.55 & $7.53\times10^{-3}$ & 42 & 1.65 & 0.74 & 37 &  & V883 Ori &  & 870.00 & 2.24 & 44 \\
\cline{10-12}
1100.00 & 1.30 & 22 & 0.70 & 0.03 & 42 & 1.66 & 0.60 & 68 & 1.24 & 0.32 & 68 & 1000.00 & 5.70 & 36 \\
\cline{1-3}
 & Par 21 &  & 0.70 & 0.02 & 42 & 2.16 & 1.20 & 68 & 1.66 & 2.02 & 68 & 1100.00 & 5.10 & 43 \\
 \cline{1-3}
0.36 & $7.62\times10^{-4}$ & 29 & 0.90 & $1.96\times10^{-3}$ & 25 & 2.20 & 1.34 & 37 & 2.16 & 5.79 & 68 & 1100.00 & 2.77 & 52 \\
0.45 & $3.80\times10^{-3}$ & 29 & 0.90 & 0.05 & 42 & 3.40 & 1.75 & 37 & 3.60 & 2.06 & 54 & 1250.00 & 2.37 & 36 \\
0.55 & 0.01 & 29 & 0.90 & 0.03 & 42 & 3.63 & 1.77 & 37 & 4.50 & 3.09 & 54 & 1300.00 & 1.28 & 4 \\
1.25 & 0.05 & 82 & 1.27 & 0.05 & 45 & 4.80 & 1.29 & 37 & 5.80 & 3.95 & 54 & 1300.00 & 4.26 & 53 \\
1.67 & 0.08 & 56 & 1.25 & $8.00\times10^{-3}$ & 25 & 8.28 & 1.35 & 21 & 8.00 & 6.43 & 54 & 1300.00 & 4.26 & 78 \\
1.68 & 0.07 & 82 & 1.25 & 0.17 & 41 & 10.20 & 1.49 & 37 & 11.60 & 52.50 & 72 & 1360.00 & 0.70 & 81 \\
2.25 & 0.07 & 56 & 1.25 & 0.13 & 42 & 12.00 & 2.19 & 2 & 12.00 & 7.50 & 72 & 1650.00 & 0.17 & 36 \\
2.25 & 0.08 & 82 & 1.65 & 0.03 & 25 & 12.00 & 1.62 & 2 & 12.00 & 55.00 & 79 & 2700.00 & 0.10 & 32 \\
2.27 & 0.06 & 56 & 1.65 & 0.34 & 42 & 14.65 & 1.34 & 21 & 24.00 & 15.60 & 54 & 2700.00 & 0.30 & 46 \\
3.60 & 0.14 & 41 & 1.65 & 0.31 & 42 & 20.00 & 2.28 & 37 & 25.00 & 125.00 & 79 & 2730.00 & 3.00 & 36 \\
4.50 & 0.19 & 41 & 1.69 & 0.07 & 45 & 21.34 & 3.27 & 21 & 60.00 & 155.00 & 79 & 2730.00 & 0.13 & 36 \\
5.80 & 0.30 & 41 & 2.20 & 0.02 & 25 & 24.00 & 2.50 & 69 & 70.00 & 17.60 & 54 & 2900.00 & 0.24 & 60 \\
8.00 & 0.83 & 41 & 2.20 & 0.22 & 42 & 25.00 & 8.09 & 2 & 100.00 & 133.00 & 79 & 3000.00 & 0.24 & 78 \\
8.28 & 0.71 & 21 & 2.20 & 0.19 & 42 & 25.00 & 4.94 & 2 & 350.00 & 13.70 & 19 & 3400.00 & 0.08 & 32 \\
10.80 & 0.96 & 61 & 2.23 & 0.07 & 45 & 50.00 & 0.40 & 71 & 450.00 & 11.70 & 19 & 3410.00 & 0.09 & 36 \\
\cline{13-15}
12.00 & 0.81 & 1 & 3.60 & 0.02 & 25 & 60.00 & 40.80 & 2 & 800.00 & 1.90 & 19 &  & V1515 Cyg &  \\
\cline{13-15}
12.13 & 1.23 & 21 & 4.50 & 0.04 & 25 & 60.00 & 41.80 & 2 & 1100.00 & 0.77 & 19 & 0.36 & $4.64\times10^{-4}$ & 11 \\
14.65 & 1.83 & 21 & 5.80 & 0.04 & 25 & 60.00 & 36.00 & 69 & 1300.00 & 0.27 & 72 & 0.44 & 0.01 & 11 \\
\cline{10-12}
18.00 & 3.43 & 61 & 8.00 & 0.05 & 25 & 65.00 & 38.70 & 2 &  & V1647 Ori (post) &  & 0.55 & 0.03 & 11 \\
\cline{10-12}
21.34 & 4.59 & 21 & 10.00 & 1.44 & 12 & 100.00 & 93.00 & 2 & 0.44 & $2.20\times10^{-5}$ & 47 & 0.70 & 0.10 & 11 \\
24.00 & 5.53 & 41 & 18.00 & 3.51 & 12 & 100.00 & 77.20 & 2 & 0.48 & $3.00\times10^{-6}$ & 68 & 1.24 & 0.44 & 68 \\
25.00 & 4.07 & 1 & 65.00 & 9.97 & 83 & 100.00 & 97.50 & 69 & 0.55 & $8.20\times10^{-5}$ & 47 & 1.25 & 0.60 & 40 \\
60.00 & 11.50 & 1 & 90.00 & 16.80 & 83 & 250.00 & 187.00 & 69 & 0.63 & $3.81\times10^{-4}$ & 47 & 1.65 & 0.75 & 40 \\
65.00 & 12.60 & 41 & 140.00 & 58.00 & 83 & 350.00 & 122.00 & 69 & 0.63 & $3.89\times10^{-4}$ & 68 & 1.66 & 0.66 & 68 \\
65.00 & 18.50 & 1 & 160.00 & 56.00 & 83 & 450.00 & 2.35 & 71 & 0.77 & $2.09\times10^{-3}$ & 68 & 2.16 & 0.75 & 68 \\
\cline{4-6}
70.00 & 13.30 & 41 &  & RNO 1C &  & 500.00 & 66.90 & 69 & 0.79 & $4.43\times10^{-3}$ & 7 & 2.20 & 0.81 & 40 \\
\cline{4-6}\cline{7-9}
100.00 & 14.20 & 41 & 1.24 & 0.05 & 68 &  & V1647 Ori (pre) &  & 0.91 & $4.70\times10^{-3}$ & 47 & 3.40 & 0.88 & 40 \\
\cline{7-9}
100.00 & 20.70 & 1 & 1.24 & 0.05 & 38 & 0.63 & $2.00\times10^{-3}$ & 47 & 1.24 & 0.06 & 68 & 4.63 & 0.95 & 40 \\
100.00 & 15.60 & 1 & 1.63 & 0.29 & 38 & 0.77 & 0.02 & 47 & 1.25 & 0.06 & 22 & 4.80 & 1.57 & 2 \\
870.00 & 0.09 & 61 & 1.66 & 0.26 & 68 & 0.79 & 0.11 & 7 & 1.65 & 0.06 & 22 & 8.28 & 1.16 & 21 \\
\cline{1-3}
 & FU Ori &  & 2.16 & 0.64 & 68 & 0.91 & 0.11 & 47 & 1.66 & 0.26 & 68 & 10.20 & 2.16 & 40 \\
\cline{1-3}
0.36 & 0.04 & 39 & 2.19 & 0.64 & 38 & 1.24 & 1.99 & 2 & 2.16 & 0.73 & 68 & 12.00 & 11.60 & 2 \\
\tablehead{ %
	\colhead{$\lambda$ ($\mu$m)} & \colhead{FU Ori} & \colhead{ref} & \colhead{$\lambda$ ($\mu$m)} & \colhead{RNO 1C} & \colhead{ref} & \colhead{$\lambda$ ($\mu$m)} & \colhead{V1647 Ori (pre)} & \colhead{ref} & \colhead{$\lambda$ ($\mu$m)} & \colhead{V1647 Ori (post)} & \colhead{ref} & \colhead{$\lambda$ ($\mu$m)} & \colhead{V1515 Cyg} & \colhead{ref} }
0.36 & 0.05 & 48 & 3.77 & 0.78 & 38 & 1.66 & 13.90 & 2 & 2.17 & 0.06 & 22 & 12.00 & 3.70 & 2 \\
0.44 & 0.21 & 39 & 3.80 & 0.78 & 61 & 2.16 & 51.00 & 2 & 3.60 & 0.06 & 22 & 12.13 & 1.43 & 21 \\
0.44 & 0.32 & 48 & 4.50 & 0.07 & 21 & 6.70 & 267.00 & 2 & 3.60 & 2.06 & 54 & 14.65 & 2.23 & 21 \\
0.55 & 0.57 & 39 & 4.67 & 0.67 & 61 & 12.00 & 527.00 & 2 & 3.77 & 2.46 & 77 & 20.00 & 2.95 & 40 \\
0.55 & 1.10 & 48 & 5.80 & 0.10 & 51 & 14.30 & 559.00 & 2 & 4.50 & 0.06 & 22 & 21.34 & 2.64 & 21 \\
0.70 & 2.46 & 48 & 8.00 & 0.25 & 21 & 25.00 & $1.20\times10^{3}$ & 2 & 4.50 & 3.09 & 54 & 25.00 & 11.00 & 2 \\
0.90 & 4.76 & 48 & 8.28 & 2.04 & 21 & 60.00 & $2.00\times10^{3}$ & 2 & 4.68 & 3.63 & 77 & 25.00 & 6.80 & 2 \\
1.25 & 7.73 & 48 & 10.80 & 0.79 & 61 & 350.00 & $2.50\times10^{3}$ & 71 & 5.80 & 0.06 & 22 & 60.00 & 25.80 & 2 \\
1.65 & 3.20 & 48 & 12.00 & 2.50 & 51 & 850.00 & 180.00 & 3 & 5.80 & 3.95 & 54 & 60.00 & 25.00 & 2 \\
2.20 & 5.96 & 37 & 12.13 & 2.36 & 21 & 1300.00 & 93.00 & 71 & 8.00 & 0.06 & 22 & 100.00 & 110.00 & 2 \\
\cline{7-9}
2.20 & 9.19 & 48 & 14.65 & 2.38 & 21 &  & ISO-ChaI 192 &  & 8.00 & 6.27 & 54 & 120.00 & 78.50 & 2 \\
\cline{7-9}
2.34 & 9.28 & 48 & 18.00 & 2.34 & 61 & 2.20 & 0.04 & 59 & 24.00 & 15.60 & 54 & 450.00 & 1.10 & 40 \\
3.50 & 5.40 & 37 & 21.34 & 8.15 & 21 & 3.60 & 0.12 & 59 & 70.00 & 0.06 & 22 & 850.00 & 0.10 & 71 \\
4.63 & 6.18 & 37 & 25.00 & 24.00 & 51 & 4.50 & 0.29 & 59 & 70.00 & 17.60 & 54 & 850.00 & 0.08 & 40 \\
10.20 & 3.98 & 37 & 60.00 & 389.00 & 51 & 5.80 & 0.54 & 59 & 450.00 & 1.59 & 3 & 1300.00 & 0.03 & 40 \\
\cline{13-15}
12.00 & 5.90 & 79 & 100.00 & 829.00 & 51 & 8.00 & 0.85 & 59 & 850.00 & 0.32 & 3 &  & V2775 Ori &  \\
\cline{13-15}
25.00 & 14.00 & 79 & 143.00 & $1.62\times10^{3}$ & 51 & 8.90 & 0.80 & 59 & 1100.00 & 0.06 & 22 & 1.20 & $4.38\times 10^{-4}$ & 82 \\
\cline{10-12}
40.00 & 18.00 & 26 & 185.00 & $2.32\times10^{3}$ & 51 & 9.80 & 0.55 & 59 &  & BBW 76 &  & 1.70 & $3.95\times 10^{-3}$ & 82 \\
\cline{10-12}
50.00 & 12.00 & 26 & 450.00 & 65.70 & 71 & 12.90 & 1.15 & 59 & 0.32 & $1.75\times10^{-3}$ & 67 & 2.20 & 0.01 & 82 \\
55.50 & 0.10 & 74 & 850.00 & 6.60 & 71 & 71.00 & 12.30 & 59 & 0.36 & $3.65\times10^{-3}$ & 67 & 3.40 & 0.60 & 91 \\
\cline{4-9}
60.00 & 15.00 & 79 &  & V900 Mon &  &  & V346 Nor &  & 0.38 & $4.96\times10^{-3}$ & 67 & 3.60 & 0.12 & 87 \\
\cline{4-9}
100.00 & 8.00 & 26 & 0.10 & 0.02 & 90 & 0.55 & $1.20\times10^{-3}$ & 72 & 0.43 & 0.01 & 67 & 4.50 & 0.16 & 87 \\
100.00 & 40.10 & 79 & 0.48 & $1.64\times 10^{-4}$ & 90 & 1.25 & 0.14 & 66 & 0.54 & 0.04 & 67 & 4.60 & 1.17 & 91 \\
160.00 & 13.00 & 26 & 0.62 & $1.16\times 10^{-3}$ & 90 & 1.65 & 0.41 & 66 & 0.70 & 0.07 & 67 & 5.80 & 0.19 & 87 \\
181.00 & 12.80 & 74 & 0.76 & $4.66\times 10^{-3}$ & 90 & 2.20 & 0.96 & 66 & 0.90 & 0.13 & 67 & 8.00 & 0.22 & 87 \\
450.00 & 0.40 & 71 & 1.25 & 0.19 & 90 & 3.40 & 2.42 & 66 & 1.24 & 0.32 & 68 & 12.00 & 1.78 & 91 \\
850.00 & 0.07 & 71 & 1.65 & 0.45 & 90 & 4.80 & 4.59 & 2 & 1.24 & 0.38 & 67 & 22.00 & 4.17 & 91 \\
\cline{1-3}
 & AR 6B &  & 2.20 & 0.78 & 90 & 5.00 & 4.24 & 66 & 1.63 & 0.54 & 67 & 24.00 & 0.69 & 88 \\
 \cline{1-3}
1.24 & 0.03 & 68 & 3.40 & 0.62 & 91 & 11.60 & 7.29 & 72 & 1.66 & 0.47 & 68 & 24.00 & 5.01 & 91 \\
\tablehead{ %
	\colhead{$\lambda$ ($\mu$m)} & \colhead{AR 6B} & \colhead{ref} & \colhead{$\lambda$ ($\mu$m)} & \colhead{V900 Mon} & \colhead{ref} & \colhead{$\lambda$ ($\mu$m)} & \colhead{V346 Nor} & \colhead{ref} & \colhead{$\lambda$ ($\mu$m)} & \colhead{BBW 76} & \colhead{ref} & \colhead{$\lambda$ ($\mu$m)} & \colhead{V2775 Ori} & \colhead{ref} }
1.24 & $1.00\times10^{-3}$ & 5 & 3.80 & 0.77 & 90 & 12.00 & 9.73 & 2 & 2.16 & 0.49 & 68 & 70.00 & 3.66 & 88 \\
1.63 & $2.00\times10^{-3}$ & 5 & 4.60 & 0.97 & 91 & 12.00 & 6.61 & 2 & 2.19 & 0.57 & 67 & 70.00 & 16.20 & 92 \\
1.66 & 0.19 & 68 & 9.00 & 1.54 & 83 & 25.00 & 30.80 & 2 & 4.64 & 0.50 & 67 & 160.00 & 20.20 & 92 \\
2.16 & 0.52 & 68 & 12.00 & 2.38 & 91 & 25.00 & 30.90 & 2 & 8.28 & 0.56 & 21 & 350.00 & 3.18 & 93 \\
2.20 & 0.02 & 5 & 18.00 & 2.64 & 83 & 60.00 & 69.10 & 2 & 11.60 & 0.82 & 2 & 870.00 & 0.58 & 93 \\
3.80 & 0.09 & 5 & 22.00 & 4.04 & 91 & 60.00 & 46.50 & 2 & 12.00 & 1.00 & 67 & \nodata & \nodata & \nodata \\
4.60 & 0.15 & 5 & 65.00 & 10.07 & 83 & 100.00 & 39.50 & 2 & 12.00 & 1.03 & 72 & \nodata & \nodata & \nodata \\
8.28 & 0.83 & 21 & 90.00 & 10.93 & 83 & 100.00 & 74.90 & 2 & 25.00 & 1.70 & 37 & \nodata & \nodata & \nodata \\
12.13 & 0.91 & 21 & 140.00 & 18.52 & 83 & 450.00 & 2.25 & 2 & 60.00 & 1.70 & 37 & \nodata & \nodata & \nodata \\
14.65 & 1.35 & 21 & 160.00 & 14.88 & 83 & 1300.00 & 0.27 & 72 & 1300.00 & 0.01 & 72 & \nodata & \nodata & \nodata \\
\enddata
\tablecomments{Different flux values for the same source at the same wavelength from the same authors implies different appertures.}
\tablerefs{
(1) \citet{abraham2004a}
(2) \citet{abraham2004b}
(3) \citet{andrews2004}
(4) \citet{andrews2005}
(5) \citet{aspin2003}
(6) \citet{beichman1981}
(7) \citet{briceno2004}
(8) \citet{chandler2000}
(9) \citet{chavarria-k1981}
(10) \citet{chavarria1981}
(11) \citet{clarke2005a}
(12) \citet{cohen1974}
(13) \citet{cohen1983b}
(14) \citet{cohen1984}
(15) \citet{cohen1987}
(16) \citet{cohen1988}
(17) \citet{covey2011}
(18) \citet{davidson1984}
(19) \citet{dent1998}
(20) \citet{egan1999a}
(21) \citet{egan2003}
(22) \citet{evans2009}
(23) \citet{fernandez1995}
(24) \citet{fridlund1980}
(25) \citet{guieu2009}
(26) \citet{harvey1982}
(27) \citet{henning1998}
(28) \citet{hessman1991}
(29) \citet{hillenbrand1992}
(30) \citet{hodapp1988}
(31) \citet{hodapp1996}
(32) \citet{hogerheijde1997}
(33) \citet{ibrahimov1999}
(34) \citet{ishihara2010}
(35) \citet{kaas2004}
(36) \citet{keene1990}
(37) \citet{kenyon1991}
(38) \citet{kenyon1993a}
(39) \citet{kenyon2000}
(40) \citet{kospal2004}
(41) \citet{kospal2007}
(42) \citet{kospal2011a}
(43) \citet{ladd1995}
(44) \citet{lay1994}
(45) \citet{li1994}
(46) \citet{looney1997}
(47) \citet{mcgehee2004}
(48) \citet{mendoza1971}
(49) \citet{mendozav1971}
(50) \citet{moneti1988}
(51) \citet{mookerjea1999}
(52) \citet{moriarty-schieven1994}
(53) \citet{motte2001b}
(54) \citet{muzerolle2005}
(55) \citet{myers1987}
(56) \citet{neckel1984}
(57) \citet{park2002}
(58) \citet{perez2010}
(59) \citet{persi2007}
(60) \citet{phillips1982}
(61) \citet{polomski2005}
(62) \citet{quanz2007b}
(63) \citet{rebull2011}
(64) \citet{reipurth1986}
(65) \citet{reipurth1993}
(66) \citet{reipurth1997a}
(67) \citet{reipurth2002b}
(68) \citet{reipurth2004a}
(69) \citet{roy2011}
(70) \citet{sandell1998}
(71) \citet{sandell2001}
(72) \citet{schutz2005}
(73) \citet{simon1972}
(74) \citet{smith1982}
(75) \citet{snell1985}
(76) \citet{strom1976}
(77) \citet{vacca2004}
(78) \citet{walker1990}
(79) \citet{weaver1992}
(80) \citet{weintraub1991}
(81) \citet{woody1989}
(82) 2MASS \citep{skr2006}
(83) AKARI \citep{murakami2007}
(84) AKARI-IRC \citep{ishihara2010}
(85) BolcamGPS \citep{rosol2010}
(86) IRAS \citet{odenwald1989}
(87) Spitzer-IRAC \citep{fazio2004}
(88) Spitzer-MIPS \citep{rieke2004}
(89) UKIDSS \citep{law2007}
(90) \citet{reipurth2012}
(91) WISE \citep{wright2010}
(92) PACS \citep{pogli2010}
(93) \citet{fischer2012}.
} %
\end{deluxetable}

\clearpage

\begin{deluxetable}{lcccccccccccc}
\rotate
\tabletypesize{\scriptsize}
\tablecolumns{13}
\tablewidth{0pt}
\tablecaption{Model parameters for the class~II FU Orionis \label{t:resclass2}}
\tablehead{
\colhead{\multirow{2}{*}{Parameter}} & \colhead{V1515} & \colhead{\multirow{2}{*}{BBW 76}} & \colhead{V1735}& \colhead{V883} &\colhead{\multirow{2}{*}{RNO 1B\tablenotemark{1}}} & \colhead{\multirow{2}{*}{RNO 1C}} & \colhead{\multirow{2}{*}{PP 13S}} &\colhead{V1647} & \colhead{V1647}	&	 \colhead{\multirow{2}{*}{class~II\tablenotemark{2}}}\\
\colhead{} & \colhead{Cyg}  & \colhead{} & \colhead{Cyg} & \colhead{Ori} & \colhead{} & \colhead{} & \colhead{} & \colhead{Ori (pre)} & \colhead{Ori (post)}	& \colhead{ }
} %
\startdata																						M$_{\rm *}$ (\msun)   	M$_{\rm *}$ (\msun)   	&	0.3	&	0.5	&	0.4	&	1.5	&	0.2	&	0.2	&	0.6	&	1.0	&	1.0	&	1.61\\
R$_{\rm *}$ (R$_{\odot}$)   	&	2.0	&	3.0	&	3.0	&	2.5	&	1.8	&	1.8	&	2.5	&	1.5	&	1.5	&	  \nodata\\
T$_{\rm *}$ (K)   	&	5900	&	6500	&	5000	&	6000	&	    6000 (5600)   	&	6000	&	4800	&	4500	&	6000	&	4268\\
$\dot{\rm M}$ ($10^{-6}$\msun/yr)   	&	0.10	&	0.10	&	0.08	&	0.01	&	0.10	&	0.10	&	0.10	&	0.30	&	0.10	&	0.01\\
R$_{\rm c}$ (AU)   	&	32	&	160	&	200	&	200	&	80	&	81	&	200	&	300	&	300	&	239\\
R$_{min}\;(\rm AU)$ 	&	0.47	&	0.42	&	0.56	&	0.19	&	0.04	&	0.04	&	0.12	&	0.06	&	0.11	&	1.16\\
R$_{max}\; (\rm AU)$ 	&	8200	&	4700	&	8000	&	3800	&	1000	&	6000	&	5000	&	2840	&	2840	&	  \nodata\\
M$_{env}$ (\msun)   	&	0.050	&	0.003	&	0.900	&	0.240	&	0.370	&	0.370	&	0.120	&	0.000	&	0.000	&	  \nodata\\
M$_{disk}$ (\msun)   	&	0.13	&	0.08	&	0.20	&	0.30	&	    0.20 (0.01)   	&	0.20	&	0.24	&	0.40	&	0.40	&	  0.03\\
$\dot{\rm M}_{disk}$ ($10^{-5}$\msun/yr) 	&	1.0	&	1.0	&	1.4	&	1.0	&	   0.8 (0.3)   	&	0.8	&	8.0	&	0.01	&	0.5	&	  0.2\\
$\rho_{amb}$ ($10^{-22}$)  	&	1.0	&	23.0	&	10.0	&	1.0	&	500.0	&	500.0	&	130.0	&	2.2	&	2.2	&	  \nodata\\
$\rho_{cav}$ ($10^{-20}$)  	&	1.00	&	0.30	&	0.80	&	7.90	&	100.00	&	100.00	&	19.00	&	3.00	&	3.00	&	  \nodata\\
A   	&	2.050	&	2.005	&	2.005	&	2.213	&	2.100	&	2.103	&	2.250	&	2.200	&	2.100	&	  \nodata\\
B   	&	1.050	&	1.005	&	1.005	&	1.213	&	1.100	&	1.103	&	1.250	&	1.200	&	1.100	&	  \nodata\\
$\theta $ (\deg)   	&	25	&	48	&	55	&	5	&	10	&	5	&	20	&	7	&	7	&	  \nodata\\
$i$ (\deg)   	&	25	&	60	&	76	&	18	&	85	&	83	&	50	&	60	&	60	&	  \nodata\\
\enddata

\tablenotetext{1}{We obtained two models, corresponding to different observing periods.
Between brackets are the values that best reproduce the most recent data when they differ from those from the older data model.}
\tablenotetext{2}{Averages values from \citet{robitaille2007}.}

\end{deluxetable}


\begin{deluxetable}{l*{6}{c}}
\tabletypesize{\scriptsize}
\tablecolumns{6}
\tablewidth{0pt}
\tablecaption{Model parameters for the class~I FU Orionis \label{t:resclass1}}
\tablehead{
\colhead{Parameters} & \colhead{FU Ori} & \colhead{V1057~Cyg} & \colhead{Z CMa} & \colhead{AR 6A} & \colhead{class~I\tablenotemark{2}}
}
\startdata
M$_{\rm *}$ (M$_{\odot}$)  & 0.70       & 0.5             & 0.80            & 0.80            & 0.93\\
R$_{\rm *}$ (R$_{\odot}$)  & 5.00&3.6&2.00 & 5.48 & \nodata\\
T$_{\rm *}$ (K)  &6030&	6000& 6500 & 4100 & 3073\\
$\dot{M}$ ($10^{-6}$M$_{\odot}$/yr) & 1.0 & 0.50 & 10.0	& 30.0 & 9.73  \\
R$_{c}$ (UA)  &	70& 60 & 65 & 80 & 397\\
R$_{min}\;(\rm R_{*})$ & 0.47	& 1.00 & 0.09 &	0.19 & 5.9\\
R$_{max}\; (\rm AU)$   & 10000	& 5200 & 16170 & 7900 &	  \nodata\\
M$_{env}$ (M$_{\odot}$) & 0.138	& 0.015	& 1.470	& 0.200	&   \nodata\\
M$_{\rm disk}$ (M$_{\odot}$)  &	0.010 &	0.10 & 0.100	& 0.340	& 0.01\\
$\dot{M}_{\rm disk}$ ($10^{-6}$ M$_{\odot}$/yr) 	&	1.00	&	14.0	&	20.00	&	4.30	&	 0.6\\
$\rho_{amb}$ $(10^{-22}$) &	1.0	&	10.0	&	15.9	&	3.5	&	  \nodata\\
$\rho_{cav}$ ($10^{-20}$)  &	0.01	&	0.01	&	0.25	&	0.33	&	  \nodata\\
A &	2.090	&	2.005	&	2.064	&	2.200	&	  \nodata\\
B &	1.090	&	1.005	&	1.064	&	1.200	&	  \nodata\\
$\rm \theta $ ($^{\rm{o}}$)  	&	70	&	35	&	25	&	25	&	  \nodata\\
$i$ ($^{\rm{o}}$)  	&	75	&	30	&	32	&	73	&	  \nodata\\

\tableline\tableline\\[-2ex]
\colhead{Parameters}	&	 \colhead{AR 6B}	&	 \colhead{L1551~IRS 5}	&	 \colhead{V900~Mon}	&	 \colhead{ ISO-Cha~I~192}	& \colhead{class~I\tablenotemark{2}}\\ [+1ex]
\tableline\\[-2ex]		
M$_{\rm *}$ (M$_{\odot}$)  	&	0.87	&	1.50	&	1.00	&	1.20	&	 0.93\\
R$_{\rm *}$ (R$_{\odot}$)  	&	5.50	&	2.50	&	1.50	&	6.10	&	  \nodata\\
T$_{\rm *}$ (K)  	&	4100	&	4800	&	6400	&	5000	&	 3073\\
$\dot{M}$ ($10^{-6}$M$_{\odot}$/yr) 	&	7.0	&	10.0	&	4.0	&	5.0	&	9.73  \\
R$_{c}$ (UA)  	&	80	&	200	&	300	&	60	&	397\\
R$_{min}\;(\rm R_{*})$	&	0.18	&	0.12	&	0.13	&	0.30	&	5.9\\
R$_{max}\; (\rm AU)$	&	7900	&	5000	&	2840	&	5000	&	  \nodata\\
M$_{env}$ (M$_{\odot}$)  	&	0.010	&	0.170	&	0.027	&	0.180	&	  \nodata\\
M$_{\rm disk}$ (M$_{\odot}$)  	&	0.370	&	0.200	&	0.200	&	0.150	&	 0.01\\
$\dot{M}_{\rm disk}$ ($10^{-6}$ M$_{\odot}$/yr) 	&	1.00	&	0.30	&	2.00	&	0.10	&	 0.6\\
$\rho_{amb}$ $(10^{-22}$) 	&	3.5	&	1.3	&	2.2	&	1.3	&	  \nodata\\
$\rho_{cav}$ ($10^{-20}$)  	&	0.33	&	1.90	&	30.00	&	1.90	&	  \nodata\\
A 	&	2.191	&	2.250	&	2.050	&	2.010	&	  \nodata\\
B  	&	1.191	&	1.250	&	1.050	&	1.010	&	  \nodata\\
$\rm \theta $ ($^{\rm{o}}$)  	&	15	&	33	&	50	&	20	&	  \nodata\\
$i$ ($^{\rm{o}}$)  	&	72	&	70	&	30	&	50	&	  \nodata\\
\tableline\tableline\\[-2ex]	
\colhead{ Parameters}	&	 \colhead{ V346 Nor}	&	 \colhead{OO Ser\tablenotemark{1}}	&	 \colhead{ Re 50 N IRS1}	&	 \colhead{ V2492~Cyg\tablenotemark{1} }	&	 \colhead{class~I\tablenotemark{2}}\\ [-2ex]
\tablehead{ %
\colhead{ Parameters}	&	 \colhead{ V346 Nor}	&	 \colhead{OO Ser\tablenotemark{1}}	&	 \colhead{ Re 50 N IRS1}	&	 \colhead{ V2492~Cyg\tablenotemark{1} }	&	 \colhead{class~I\tablenotemark{2}}
}\\
\tableline\\[-2ex]	
M$_{\rm *}$ (M$_{\odot}$)  	&	0.30	&	0.70	&	1.00	&	1.20	&	 0.93\\
R$_{\rm *}$ (R$_{\odot}$)  	&	3.00	&	3.00 (2.00)	&	4.00	&	2.80/3.00 (2.50)	&	  \nodata\\
T$_{\rm *}$ (K)  	&	7000	&	6000 (5000)	&	6000	&	6100/6500 (5000)	&	 3073\\
$\dot{M}$ ($10^{-6}$M$_{\odot}$/yr) 	&	6.0	&	10.0	&	12.4	&	1.0	&	9.73  \\
R$_{c}$ (UA)  	&	90	&	200	&	30	&	500	&	397\\
R$_{min}\;(\rm R_{*})$	&	0.24	&	0.22	&	0.70	&	0.01	&	5.9\\
R$_{max}\; (\rm AU)$	&	5200	&	10000	&	6080	&	2000	&	  \nodata\\
M$_{env}$ (M$_{\odot}$)  	&	0.300	&	0.910	&	0.280	&	0.200	&	  \nodata\\
M$_{\rm disk}$ (M$_{\odot}$)  	&	0.050	&	0.010	&	0.060	&	0.030	&	 0.01\\
$\dot{M}_{\rm disk}$ ($10^{-6}$ M$_{\odot}$/yr) 	&	9.00	&	50.00 (1.00)	&	1.30	&	0.40 (0.10)	&	 0.6\\
$\rho_{amb}$ $(10^{-22}$) 	&	10.0	&	57.0	&	4.2	&	17.0	&	  \nodata\\
$\rho_{cav}$ ($10^{-20}$)  	&	1.00	&	17.00	&	1.50	&	17.00	&	  \nodata\\
A 	&	2.050	&	2.250	&	2.174	&	3.300	&	  \nodata\\
B  	&	1.050	&	1.250	&	1.174	&	1.500	&	  \nodata\\
$\rm \theta $ ($^{\rm{o}}$)  	&	20	&	70	&	40	&	70	&	  \nodata\\
$i$ ($^{\rm{o}}$)  	&	5	&	5	&	15	&	13	&	  \nodata\\

\tableline\tableline\\[-2ex]
\colhead{ Parameters}	&	 \colhead{V1331 Cyg\tablenotemark{1}}	&	 \colhead{HBC 722\tablenotemark{1}}	&	 \colhead{Par 21}	&	 \colhead{V2775 Ori\tablenotemark{1}}	&	 \colhead{class~I\tablenotemark{2}}\\ [+1ex]
\tableline\\[-2ex]	
M$_{\rm *}$ (M$_{\odot}$)  	&	0.8	&	1.00	&	1.00	&	0.50	&	 0.93\\
R$_{\rm *}$ (R$_{\odot}$)  	&	2.0	&	 1.90 (1.50) 	&	3.20	&	1.90 (1.70)	&	  \nodata\\
T$_{\rm *}$ (K)  	&	  6600 (5770) 	&	 7100 (5600) 	&	8700	&	6800 (5600)	&	 3073\\
$\dot{M}$ ($10^{-6}$M$_{\odot}$/yr) 	&	0.80	&	1.0	&	1.5	&	0.7 (3.0)	&	9.73  \\
R$_{c}$ (UA)  	&	100	&	51	&	60	&	36	&	397\\
R$_{min}\;(\rm R_{*})$	&	0.19	&	0.09	&	0.89	&	0.09	&	5.9\\
R$_{max}\; (\rm AU)$	&	8000	&	10000	&	5200	&	30200	&	  \nodata\\
M$_{env}$ (M$_{\odot}$)  	&	0.120	&	0.060	&	0.020	&	0.300	&	  \nodata\\
M$_{\rm disk}$ (M$_{\odot}$)  	&	  0.10 (0.02)  	&	0.100	&	0.300	&	0.150 (0.008)	&	 0.01\\
$\dot{M}_{\rm disk}$ ($10^{-6}$ M$_{\odot}$/yr) 	&	  2.0 (0.1)	&	 4.00 (0.40) 	&	4.00	&	10.00 (0.60)	&	 0.6\\
$\rho_{amb}$ $(10^{-22}$) 	&	4.2	&	5.0	&	10.0	&	5.0	&	  \nodata\\
$\rho_{cav}$ ($10^{-20}$)  	&	1.50	&	0.01	&	1.00	&	0.01	&	  \nodata\\
A 	&	2.250	&	2.100	&	2.050	&	2.205	&	  \nodata\\
B  	&	1.250	&	1.100	&	1.050	&	1.205	&	  \nodata\\
$\rm \theta $ ($^{\rm{o}}$)  	&	50	&	30	&	80	&	35	&	  \nodata\\
$i$ ($^{\rm{o}}$)  	&	2	&	85	&	79	&	60	&	  \nodata\\

\enddata

\tablenotetext{1}{For these objects we obtained two models associated with two different periods of observations. We indicate between brackets the best model parameters corresponding to the most recent data when they differ from those obtained form older observations (see Figures~\ref{f:ooser}, \ref{f:v2492cyg}, \ref{f:v1331cyg}, \ref{f:hbc722} and \ref{f:v2775}).}
\tablenotetext{2}{Averages values from \citet{robitaille2007}.}
\end{deluxetable}


\begin{deluxetable}{l*{4}{c}}
\tabletypesize{\small}
\tabletypesize{\scriptsize}
\tablecolumns{5}
\tablewidth{0pt}
\tablecaption{Average values for class~I and class~II FUORS and YSOs \label{t:averages} }
\tablehead{
\colhead{\multirow{2}{*}{Parameter}}& \colhead{Average values\tablenotemark{1}}& \colhead{Average values\tablenotemark{2}} & \colhead{ Average values\tablenotemark{1} }&\colhead{Average values\tablenotemark{2}} \\
\colhead{ }&\colhead{class~I FUORS}&\colhead{class~I YSOs}& \colhead{class~II FUORS} &\colhead{class~II YSOs}
}
\startdata
M$_{\rm *}$ (M$_{\odot}$)	&	0.87	&	0.93	&	0.59	&	 1.61 \\
T$_{*}$ (K)	&	6077	&	3073	&	5775	&	 4268  \\
$\dot{\rm M}$ ($10^{-6}$M$_{\odot}$/yr)	&	6.3	&	9.73	&	0.1	&	  0.01 \\
R$_{\rm c}$ (AU)	&	124	&	397	&	157	&	 239 \\
R$_{min}\;(\rm AU)$	&	0.31	&	5.93	&	0.24	&	 1.16  \\
M$_{disk}$ (M$_{\odot}$)  	&	0.14	&	0.01	&	0.22	&	  0.03  \\
$\dot{\rm M}_{disk}$ ($10^{-6}$M$_{\odot}$/yr)	&	8.3	&	0.6	&	9.1	&	0.2 \\
\enddata

\tablenotetext{1}{Averages of the results shown in the Tables~\ref{t:resclass2} and \ref{t:resclass1} for class~I and class~II, respectively.}
\tablenotetext{2}{Averages values from \citet{robitaille2007}.}

\end{deluxetable}


\begin{deluxetable}{lcccc}
\tabletypesize{\small}
\tabletypesize{\scriptsize}
\tablecolumns{5}
\tablewidth{0pt}
\tablecaption{Median values of FUORS and classical T Tauri stars and K-S analysis \label{t:median} }
\tablehead{
\colhead{\multirow{2}{*}{Parameter}}&\colhead{Class I-II}&\colhead{Class I-II from} & \colhead{\multirow{2}{*}{$D$\tablenotemark{1}}} & \colhead{\multirow{2}{*}{$s$\tablenotemark{2}}}\\
\colhead{}&\colhead{ FUORS }&\colhead{\citet{robitaille2007}} & &
} 
\startdata
$\dot{\rm M}_{disk}$ ($10^{-6}$ M$_{\odot}$/yr) & 8.0 & 0.1 & 0.79 & 3 $\times$ 10$^{-8}$ \\
${\rm M}_{disk}$ (M$_{\odot}$) & 0.15        & 0.02 & 0.83 & 3 $\times$ 10$^{-9}$\\
$\dot{\rm M}$ ($10^{-6}$ M$_{\odot}$/yr) & 1.00       & 0.03 & 0.84 & 4 $\times$ 10$^{-8}$ \\
T$_{\rm *}$ (K)                & 6000        & 3989  & 0.84 & 4 $\times$ 10$^{-9}$ \\
\enddata

\tablenotetext{1}{$D$: maximum difference.}
\tablenotetext{2}{$s$: significance or confidence level.}

\end{deluxetable}

\end{document}